\PassOptionsToPackage{unicode}{hyperref}
\PassOptionsToPackage{hyphens}{url}
\PassOptionsToPackage{dvipsnames,svgnames,x11names}{xcolor}
\documentclass[12pt]{article}

\usepackage{amsmath,amssymb,amsfonts}
\usepackage{iftex}
\ifPDFTeX
  \usepackage[T1]{fontenc}
  \usepackage[utf8]{inputenc}
  \usepackage{textcomp}
\else 
  \usepackage{unicode-math}
  \defaultfontfeatures{Scale=MatchLowercase}
  \defaultfontfeatures[\rmfamily]{Ligatures=TeX,Scale=1}
\fi
\usepackage{lmodern}
\IfFileExists{upquote.sty}{\usepackage{upquote}}{}
\IfFileExists{microtype.sty}{%
  \usepackage[]{microtype}
  \UseMicrotypeSet[protrusion]{basicmath}
}{}
\makeatletter
\@ifundefined{KOMAClassName}{%
  \IfFileExists{parskip.sty}{%
    \usepackage{parskip}
  }{%
    \setlength{\parindent}{0pt}
    \setlength{\parskip}{6pt plus 2pt minus 1pt}}
}{%
  \KOMAoptions{parskip=half}}
\makeatother
\usepackage{xcolor}
\providecommand{\tightlist}{%
  \setlength{\itemsep}{0pt}\setlength{\parskip}{0pt}}

\usepackage{longtable,booktabs,array}
\usepackage{calc}
\usepackage{etoolbox}
\makeatletter
\patchcmd\longtable{\par}{\if@noskipsec\mbox{}\fi\par}{}{}
\makeatother
\IfFileExists{footnotehyper.sty}{\usepackage{footnotehyper}}{\usepackage{footnote}}
\makesavenoteenv{longtable}
\usepackage{graphicx}
\graphicspath{{media/}}
\makeatletter
\def\maxwidth{\ifdim\Gin@nat@width>\linewidth\linewidth\else\Gin@nat@width\fi}
\def\maxheight{\ifdim\Gin@nat@height>\textheight\textheight\else\Gin@nat@height\fi}
\makeatother
\setkeys{Gin}{width=\maxwidth,height=\maxheight,keepaspectratio}
\makeatletter
\def\fps@figure{htbp}
\makeatother

\makeatletter
\@ifpackageloaded{caption}{}{\usepackage{caption}}
\@ifpackageloaded{float}{}{\usepackage{float}}
\floatstyle{ruled}
\@ifundefined{c@chapter}{\newfloat{codelisting}{h}{lop}}{\newfloat{codelisting}{h}{lop}[chapter]}
\floatname{codelisting}{Listing}

\makeatother
\makeatletter
\@ifpackageloaded{caption}{}{\usepackage{caption}}
\@ifpackageloaded{subcaption}{}{\usepackage{subcaption}}
\makeatother

\ifLuaTeX
  \usepackage{selnolig}
\fi
\usepackage[]{natbib}
\usepackage{amsthm}
\usepackage{algorithm,algorithmic}
\usepackage{underscore}
\usepackage{nicefrac}

\usepackage{hyperref}
\usepackage{bookmark}
\IfFileExists{xurl.sty}{\usepackage{xurl}}{}
\usepackage[capitalize]{cleveref}  

\hypersetup{
  pdftitle={Synthetic Data Augmentation for Functional Data},
  colorlinks=true,
  linkcolor={blue},
  citecolor={Blue},
  urlcolor={Blue},
  pdfcreator={LaTeX}}

\providecommand\texorpdfstring[2]{#1}
\providecommand\bookmarksetup[1]{}

\providecommand{\subsubsubsection}{}
\renewcommand{\subsubsubsection}{%
  \vskip5pt{\noindent\normalsize\rm\raggedright}%
}

\newtheorem{theorem}{Theorem}[section]
\newtheorem{lemma}{Lemma}[section]
\newtheorem{assumption}{Assumption}[section]
\newtheorem{remark}{Remark}[section]
\newtheorem{corollary}{Corollary}[section]

\crefname{assumption}{Assumption}{Assumptions}
\Crefname{assumption}{Assumption}{Assumptions}
\crefname{remark}{Remark}{Remarks}
\Crefname{remark}{Remark}{Remarks}
\crefname{corollary}{Corollary}{Corollaries}
\Crefname{corollary}{Corollary}{Corollaries}
\crefname{equation}{Equation}{Equations}
\Crefname{equation}{Equation}{Equations}

\newcommand{\jasaKeywords}[1]{%
  \noindent{\it Keywords:} #1}

\newcommand{\anon}{1}

\date{}

\begin{document}

\def\spacingset#1{\renewcommand{\baselinestretch}%
{#1}\small\normalsize} \spacingset{1}

\if1\anon
{
  \title{\bf Synthetic Data Augmentation for Functional Data: Wasserstein Risk Bounds and Diffusion-Based Conditional Generation}
  \author{
    Yuhao Qi \\
    Institute of Statistics and Big Data, Renmin University of China, Beijing, China \\
    \texttt{qiyh@ruc.edu.cn}
    \and
    Wenlin Dai\thanks{Corresponding author. Email: \texttt{wenlin.dai@ruc.edu.cn}} \\
    Institute of Statistics and Big Data, Renmin University of China, Beijing, China \\
    \texttt{wenlin.dai@ruc.edu.cn}
    \and
    Xuehu Zhu \\
    School of Mathematics and Statistics, Xi'an Jiaotong University, Xi'an, China \\
    \texttt{zhuxuehu@xjtu.edu.cn}
  }
  \maketitle
} \fi

\if0\anon
{
  \bigskip
  \bigskip
  \bigskip
  \begin{center}
    {\LARGE\bf Synthetic Data Augmentation for Functional Data: Wasserstein Risk Bounds and Diffusion-Based Conditional Generation}
  \end{center}
  \medskip
} \fi

\bigskip
\begin{abstract}
Functional data are costly to collect; samples of tens to hundreds of curves are common. Synthetic data augmentation can help, but its theory is developed for finite-dimensional data and relies on total variation, which is uninformative for functions. We study augmentation with a class-conditional functional diffusion generator and state guarantees in the Wasserstein-1 distance. For stable downstream estimators, we bound the risk of training on synthetic curves, from above and below, by the risk of training on as many real curves plus a generation penalty, and derive the best number of synthetic curves and a condition under which synthetic data outperform real data. For allocating synthetic curves under class imbalance, we bound the excess risk of training on real and synthetic curves together, without sample splitting. In applications to temperature, hyperspectral and gait curves, augmentation improved classification, function-to-scalar regression and function-to-function regression, and directing synthetic curves to the minority class gave the largest gain under class imbalance.
\end{abstract}

\jasaKeywords{Functional data analysis; Functional diffusion models; Generational effect; Synthetic data augmentation; Wasserstein distance.}

\vfill
\newpage
\spacingset{1} 

\section{Introduction}\label{sec:introduction}

Functional data, in which each observation is a curve, surface or trajectory, arise throughout science, from gait biomechanics and climatology to spectroscopy. The toolkit for such data is mature \citep{gertheiss2024functional}, from functional principal component analysis \citep{yao2005functional} and functional regression \citep{morris2015functional,greven2017general} to functional boosting \citep{brockhaus2015functional}, penalized function-on-function regression \citep{scheipl2015functional} and functional neural networks \citep{thind2023deep}. Yet functional samples are often small: a gait curve requires a motion-capture session and a climate station decades of monitoring, so typical studies contain tens to a few hundred curves, and biomedical ones are often subject to sharing restrictions. Small samples limit the accuracy of flexible learners.

Synthetic data offer one response \citep{jordon2022synthetic}: train a generative model on the real sample and use its draws to enlarge the training set of a downstream procedure. \citet{shen2024boosting} bound the risk of a method trained on \(m\) synthetic observations by its risk on \(m\) real ones plus a term proportional to \(m\,\mathrm{TV}(\widetilde F,F)\), the total variation (TV) distance between the generated and true distributions. This explains the generational effect: accuracy first improves with the synthetic volume and may then degrade or plateau, with an optimal volume, the reflection point. \citet{Tian2026} extend the theory to conditional generation within subpopulations such as minority classes, where a domain-adaptation index and a generation-error index govern how synthetic data should be allocated. Both works rely on generators fine-tuned from large pre-trained models, with generation-error rates available only on Euclidean spaces. Related advances use diffusion models for high-dimensional variable selection \citep{wang2025diffusion}, causal effect estimation \citep{chen2025enhancing} and multimodal prediction \citep{tian2025gdp}.

Two features of this theory fail for functional data. First, TV is uninformative in function space: by the Feldman--H\'ajek theorem, \(\mathrm{TV}\big(N(0,C),N(0,cC)\big)=1\) for every trace-class \(C\) and every \(c\neq1\), however close \(c\) is to 1, whereas \(W_1\le|\sqrt c-1|(\operatorname{Tr}C)^{1/2}\to0\), and a generator of finitely many basis coefficients is at TV distance one from any distribution without mass on that subspace, so a bound in \(m\,\mathrm{TV}\) says nothing about a practical generator of curves. Second, the conditional-generation guarantee of \citet{Tian2026} derives its rates for latent diffusion in Euclidean space with pre-trained encoder--decoders and splits the real data between the generator and the downstream fit.

A generator for curves must also respect the function-space structure. Standard diffusion models \citep{sohl2015deep,ho2020denoising,song2021scorebased} perturb data with independent Gaussian noise at each grid point. In the continuum limit this converges to white noise, which is not an \(\mathcal{H}\)-valued random variable, and the denoising objective can have unbounded variance as the grid is refined \citep{pidstrigach2024infinite,lim2025score}. Recent work resolves this by using Gaussian noise with trace-class covariance and defining the score with respect to a Gaussian reference measure: \citet{kerrigan2023diffusion} work with Gaussian measures on Hilbert spaces and neural-operator parametrizations, \citet{pidstrigach2024infinite} establish well-posedness of the forward and reverse infinite-dimensional SDEs, and \citet{lim2025score} introduce the denoising diffusion operator (DDO), which defines the score as a Fr\'echet derivative along the Cameron--Martin space and also supports conditional sampling for inverse problems. Other function-space generators, such as annealed Langevin dynamics and flow matching, are reviewed by \citet{franzese2025generative}. These works establish that diffusion generation in function space is well posed and practical, but they study the quality of the generated curves; to our knowledge, none gives a guarantee that training on them improves a downstream statistical procedure.

This paper asks three questions for functional data: when training a downstream learner on synthetic curves reduces its risk relative to real curves, how many synthetic curves to generate, and how to allocate them across imbalanced classes. Answering them raises four difficulties. First, the generator must condition on class labels and continuous responses without breaking the functional structure of the curves; we add FiLM to the DDO, which scales and shifts the network's features by the same amount at every grid point, and encode a continuous response with a value-augmented one-hot vector. Second, total variation, used by existing theory, is uninformative for curves; we measure generation error by \(W_1\). Third, the generator is trained on the same real curves that enter the downstream fit, so the synthetic and real curves are dependent. Existing bounds avoid this by splitting the real sample \citep{Tian2026}, which wastes scarce curves; we instead condition on the real sample and bound the two parts separately. Fourth, the \(W_1\) error of a trained generator cannot be computed, so the theorems cannot be checked on it directly; we instead assess their practical consequences on three real datasets.

We make four contributions.

\begin{enumerate}
\def\labelenumi{(\roman{enumi})}
\item
  \emph{Method.} Building on the DDO \citep{lim2025score}, we add feature-wise linear modulation \citep{perez2018film} to condition the score network on class labels and on continuous responses, giving a class-conditional generator for curves that is trained from scratch on the real sample.
\item
  \emph{Volume theory.} Measuring generation error by the Wasserstein-1 distance, we show that the risk of training on synthetic curves stays within a generation penalty of the raw-data risk, and that this penalty does not grow with the synthetic volume for stable estimators. We derive the best synthetic volume, a sufficient condition for synthetic data to match or beat the real sample, and conditions for a finite optimal volume.
\item
  \emph{Allocation theory.} Conditioning on the real sample, we prove an exponential-tail bound for the excess risk of the mixed real-and-synthetic estimator, in terms of approximation, estimation, domain-adaptation, allocation-mismatch, and generation-error terms, without sample splitting. For imbalanced binary classification we obtain explicit rates from the per-class generation rate and show how label imbalance and per-class generation error enter the certified level.
\item
  \emph{Evidence.} Analyses of three real datasets (daily temperature curves, grape hyperspectral reflectance, and gait biomechanics), covering classification, function-to-scalar regression, and function-to-function regression, show where augmentation helps and where it does not.
\end{enumerate}

The real-data analyses show clear gains for function-to-scalar regression and for minority-targeted allocation in classification, and consistent gains for penalized function-to-function regression.

The remainder of the paper is organized as follows. \Cref{sec:methodology} describes the conditional functional diffusion model and the generation--augmentation pipeline; details of the model are in Appendix A and implementation details in Appendix B. \Cref{sec:theoretical_results} states the volume and allocation theorems, with notation in \Cref{sec:setting} and further remarks and proofs in Appendix C. \Cref{sec:applications} analyzes the three real datasets, with details of the data in Appendix D. \Cref{sec:discussion} summarizes the findings and outlines future work.

\section{Methodology}\label{sec:methodology}

This section describes the conditional functional diffusion model and the pipeline that turns its output into training data for downstream procedures. \Cref{sec:ddo} fixes the setting and reviews diffusion in function space. \Cref{sec:film} adds label and response conditioning. \Cref{sec:synthesis} defines the generation--augmentation pipeline, the data roles, and the replication design. \Cref{sec:downstream_models} lists the downstream procedures; Appendix A gives the details of the model and Appendix B the implementation.

\subsection{Functional Data Diffusion Models}\label{sec:ddo}

\textbf{Setting.} Let \(\mathcal{H}=L^2([0,1])\). An observation is a curve \(X\in\mathcal{H}^{C_{\rm ch}}\) with \(C_{\rm ch}\) channels (\(C_{\rm ch}=1\) unless stated), recorded on a grid of \(N_{\rm grid}\) points, with a label \(Y\in\mathcal{Y}\), where \(\mathcal{Y}=\{1,\dots,K\}\) for classification and \(\mathcal{Y}\subset\mathbb{R}\) for regression. The sample \(\mathcal{D}_n=\{(X_i,Y_i)\}_{i=1}^n\) is drawn i.i.d. from the joint distribution \(F\) of \((Y,X)\); for discrete labels, \(\pi_k=\mathbb{P}(Y=k)\) and \(F_k\) is the distribution of \(X\) given \(Y=k\). The generator, trained on \(\mathcal{D}_n\), and the choice of labels for the new curves (\Cref{sec:pipeline}) induce a synthetic distribution \(\widetilde F\) with class-conditional distributions \(\widetilde F_k\). Draws from \(\widetilde F\) enlarge or replace the training set of a downstream procedure \(\widehat\theta\), and the generation error \(W_1(\widetilde F,F)\) drives the guarantees of \Cref{sec:theoretical_results}. Notation is collected in Appendix A (\Cref{tab:method_notation}).

\textbf{Noise in function space.} A standard diffusion model adds independent Gaussian noise at every grid point; as the grid is refined this noise converges to white noise, which is not a random element of \(\mathcal{H}\), and the denoising objective depends on the resolution \citep{lim2025score,pidstrigach2024infinite}. We therefore use the Denoising Diffusion Operator (DDO) of \citet{lim2025score}, which perturbs a clean curve \(x_0\) with Gaussian-process noise, \[
x_t=\alpha_t\,x_0+\sigma_t\,\varepsilon,\qquad \varepsilon\sim\mathcal{N}(0,\mathcal{C}),\qquad \alpha_t^2+\sigma_t^2=1,\quad t\in[0,1],
\] with the variance-preserving cosine schedule \citep{nichol2021improved} and independent noise in each channel. Because the self-adjoint, non-negative covariance operator \(\mathcal{C}\) is trace class, \(\mathbb{E}\|\varepsilon\|_{\mathcal{H}}^2={\rm tr}(\mathcal{C})<\infty\), so the perturbed curves stay in \(\mathcal{H}\). We take \(\mathcal{C}\) from the powered-exponential kernel of \citet[Eq. 48]{lim2025score}, \[
k(s,t)=\sigma_{\rm gp}^2\exp\!\Bigl(-\bigl(|s-t|/\ell_{\rm gp}\bigr)^{\nu}\Bigr),\qquad 0<\nu\le 2,
\] whose parameters set the amplitude, roughness and correlation length of the noise (Appendix A.1; values in Appendix B).

\textbf{Score matching.} Generation runs the process in reverse (Appendix A.1), which requires the score \(\mathcal{C}\nabla\log p_t\) of the distribution \(p_t\) of \(x_t\), defined along the Cameron--Martin space of \(\mathcal{N}(0,\mathcal{C})\) \citep{lim2025score}. Since \(\mathbb{E}[\varepsilon\mid x_t]=-\sigma_t\,\mathcal{C}\nabla\log p_t(x_t)\), a network \(\varepsilon_\psi\) that predicts the noise gives the score as \(-\varepsilon_\psi/\sigma_t\), without inverting \(\mathcal{C}\). With a conditioning vector \(v\) (\Cref{sec:film_conditioning}), we train by the conditional \(\varepsilon\)-matching objective \[
\mathcal{L}(\psi)=\mathbb{E}_{t,\,(x_0,y),\,\varepsilon}\Bigl[\tfrac12\bigl\|\varepsilon_\psi(x_t,t,v(y))-\varepsilon\bigr\|^2\Bigr],\qquad x_t=\alpha_t x_0+\sigma_t\varepsilon .
\] By the conditional score-matching result of \citet[Theorem 5]{lim2025score}, for a finite-dimensional condition, such as our encodings in \(\mathbb{R}^K\), its minimizers are those of score matching for \(F_{X\mid Y}\), so the trained network approximates the conditional score.

\subsection{Conditional Generation via Feature-wise Linear Modulation}\label{sec:film}

\citet{lim2025score} condition the DDO on observations defined on the same domain as the generated field, supplied to the network together with its input. Our conditions are instead labels or responses that take one value for the whole curve, and the classical way of conditioning on them, appending \(y\) to the discretized curve as \([x_1,\dots,x_{N_{\rm grid}},y]\), creates a jump at the end of the input and breaks its functional structure. We therefore condition with Feature-wise Linear Modulation (FiLM; \citet{perez2018film}), which modulates every residual block, as the network already does for the diffusion time. The conditioning is effective in our applications (\Cref{sec:applications}): the generated curves match the real curves of their own class or group and reproduce the differences between groups, both in the curve plots and under the quality metrics of \Cref{sec:pipeline}. We call the resulting model the \emph{Conditional Denoising Diffusion Operator} (Conditional DDO).

\subsubsection{FiLM Conditioning}\label{sec:film_conditioning}

FiLM modulates a feature map \(h\in\mathbb{R}^{C_h\times N_h}\) with a scale and a shift that depend on the condition, \[
\mathrm{FiLM}(h;\mathbf{c})=(1+s(\mathbf{c}))\odot h+b(\mathbf{c}),\qquad [s(\mathbf{c}),\,b(\mathbf{c})]=\mathrm{Linear}(\mathbf{c})\in\mathbb{R}^{2C_h},
\] where \(\odot\) multiplies channel-wise and the modulation is the same at every grid position; the parameterization \(1+s\) makes the identity map the default. The conditioning vector \(v\) is embedded by a two-layer network, \(\mathbf{c}=\mathrm{MLP}_{\rm embed}(v)\) (\(\mathrm{Linear}\to\mathrm{SiLU}\to\mathrm{Linear}\), embedding dimension \(d_c\)), and one linear layer per residual block maps \(\mathbf{c}\) to the FiLM parameters. For a class label \(y\in\{1,\dots,K\}\), \(v=e_y\in\mathbb{R}^K\) is the one-hot vector.

\textbf{Continuous conditioning (regression).} A continuous response cannot be one-hot encoded, and a raw scalar fed to the embedding network would carry no group structure. We use a \emph{value-augmented one-hot encoding}. Let \(c(y)\in\{1,\dots,K\}\) be the group of an observation, with \(K\) now the number of groups (the grape variety in \Cref{sec:regression}), and let \(y_{\rm norm}=(y-y_{\min})/(y_{\max}-y_{\min})\in[0,1]\) be the min--max normalized response, with the minimum and maximum taken over the training data. The conditioning vector replaces the unit entry of the one-hot vector by \(y_{\rm norm}\): \[
v(y)=y_{\rm norm}\,e_{c(y)}\in\mathbb{R}^K,\qquad \mathbf{c}=\mathrm{MLP}_{\rm embed}(v(y)).
\] The position of the nonzero entry encodes the group and its magnitude the response, so the embedding varies smoothly with \(y\) within each group; function-to-function regression and classification instead condition on one-hot load-condition groups and class labels.

\subsubsection{Network Architecture, Training and Sampling}\label{sec:architecture}

The denoising network \(\varepsilon_\psi(x_t,t,v)\), implemented in PyTorch, is the U-shaped Fourier neural operator of \citep{lim2025score,li2020fourier,rahman2023uno}, trained with their forward process, noise and objective (\Cref{sec:ddo}) and sampled by the ancestral sampler described below. It differs from theirs in two respects: every residual block is modulated by the condition as well as by the diffusion time, which turns the unconditional DDO into a generator of \(F_{X\mid Y=y}\); and curves of any length are accepted, by padding each curve at both ends to a length the resolution levels require and cropping the output back to the original grid (Appendix A.2).

The network is trained from scratch on the standardized real training curves with minibatches balanced across the \(K\) groups, which changes only the label marginal, not the conditional distribution of \(X\) given \(Y\); the optimizer is AdamW with cosine annealing, and the Gaussian-process noise is drawn by a fast discrete cosine approximation of the kernel (\Cref{alg:training}). Synthetic curves are drawn by the ancestral sampler of the discretized reverse process, started from \(\mathcal{N}(0,\mathcal{C})\), with the condition supplied to the network at every step (\Cref{alg:sampling}). Both algorithms are in Appendix A.3, and all settings in Appendix B.

\subsection{Conditional Data Synthesis in Function Space}\label{sec:synthesis}

\subsubsection{The Generation-Augmentation Pipeline}\label{sec:pipeline}

\begin{figure}[htbp]
\centering
\includegraphics[width=0.9\textwidth]{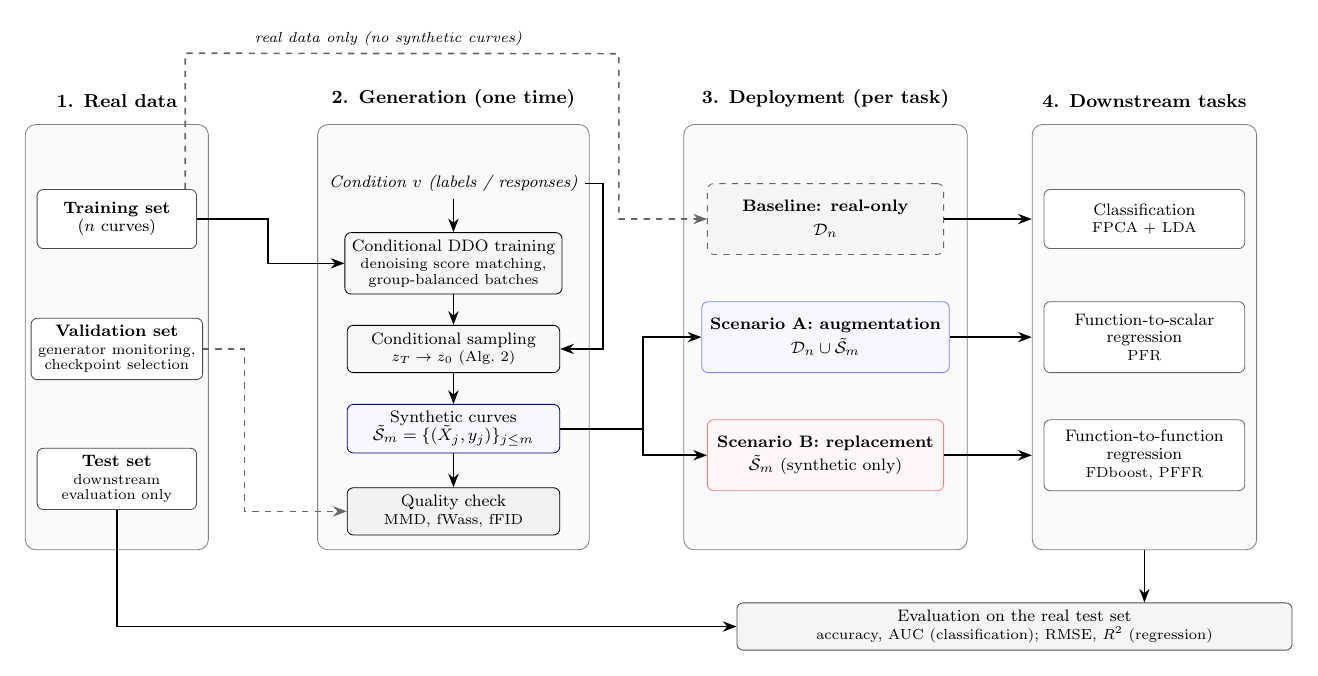}
\caption{\textbf{Generation--augmentation pipeline.} (1) The real data are split into training, validation (to monitor the generator, when the sample allows) and test sets. (2) The Conditional DDO is trained once on the training set and sampled under conditions \(v\) to give synthetic curves \(\tilde{\mathcal S}_m\), whose quality is checked by MMD, fWass and fFID. (3) Each downstream task is run on real data only (baseline), real plus synthetic data (A) and synthetic data only (B). (4) All procedures are evaluated on the real test set.}
\label{fig:framework}
\end{figure}

We follow the two-phase paradigm of \citet{shen2024boosting} and \citet{Tian2026}: a generator is trained first, and downstream procedures are then run on data that include its output. The pipeline has four stages (\Cref{fig:framework}).

In Stage 1 (data roles), the real curves are split into training, validation and test sets (\Cref{sec:datasets}): the training set fits the generator and the downstream procedures, the validation set, where available, selects the generator checkpoint, and the test set evaluates the downstream procedures and, without a validation set, generation quality. In Stage 2 (generation), the Conditional DDO trained on the training set samples \(m\) synthetic curves under conditioning vectors \(v_j\) (\Cref{sec:film_conditioning}) taken from the training data, divided among classes as in \Cref{sec:generational_effect,sec:allocation_design}; the synthetic curve--label pairs have distribution \(\widetilde F\). In function-to-scalar regression the conditions are training responses, so no new response values are added; in function-to-function regression the response is the generated moment channel. In Stage 3 (deployment), with \(\widetilde{\mathcal S}_m=\{(\widetilde X_j,y_j)\}_{j=1}^m\), a downstream procedure \(\widehat\theta\) is trained on \(\mathcal{D}_n\) (baseline), \(\mathcal{D}_n\cup\widetilde{\mathcal S}_m\) (augmentation, A) or \(\widetilde{\mathcal S}_m\) alone (replacement, B, relevant when real curves cannot be shared). In Stage 4 (evaluation), every fit is evaluated on the real test set; since the generator is fixed across deployments, differences reflect only the training data.

Because \(W_1\le W_2\), every upper bound and sufficient condition of \Cref{sec:theoretical_results} stated in \(W_1\) also holds with \(W_2\), so the theory applies to a generator with small \(W_2\) error. We therefore assess generation quality with \(W_2\)-type metrics: for Gaussian distributions, \(W_2\) has an explicit formula in the means and covariances, whereas \(W_1\) does not, and we apply it to Gaussian approximations of the synthetic and real curves. We report three such checks, defined in Appendix D.2: the maximum mean discrepancy (MMD) with an RBF kernel; fWass, the squared \(W_2\) distance between Gaussian approximations of two samples, summed over grid points and channels; and fFID, the same quantity computed on functional principal component scores. They are descriptive checks of agreement with real curves, not estimates of the generation error in the theorems.

\subsubsection{Generational Effect Analysis}\label{sec:generational_effect}

From a pool of 100 synthetic sets of \(n\) curves, each with the training class counts and its own seed, we form volumes \(m=Gn\), \(G=1,\dots,20\), by pooling \(G\) sets, with five replicates per volume that share no synthetic curve. Each analysis uses one trained generator, and a replicate is an independent run of its sampler followed by a downstream fit, so reported spreads describe the variability of sampling and of the downstream procedure. The downstream procedure is trained on each pooled set, alone or with the real data, and the trajectories in \(G\) are smoothed by penalized B-splines with pointwise 95\% confidence bands. Whether they show the interior optimum of \Cref{thm:reflection_point} is left to the data.

\subsubsection{Allocation Design}\label{sec:allocation_design}

To study how synthetic curves should be divided among classes (\Cref{thm:augmentation_risk,thm:imbalanced_classification}), we fix \(m\) and vary the share \(a_k\) of each class, adding in each replicate the integer nearest to \(a_km\) synthetic curves of class \(k\) to the real training data (\Cref{sec:targeted_allocation}).

\subsection{Downstream Models}\label{sec:downstream_models}

The training set of each deployment feeds the downstream procedures of \Cref{tab:downstream}; the models are specified with each task (\Cref{sec:classification,sec:regression,sec:ml}), and no scalar covariate other than the condition label enters any of them.

\begin{table}[htbp]
\centering
\caption{\textbf{Downstream procedures by task.}}
\label{tab:downstream}
{\renewcommand{\baselinestretch}{1}\small\setlength{\tabcolsep}{3pt}
\begin{tabular}{m{129.8pt}m{313.2pt}}
\toprule
Task & Procedure \\
\midrule
Classification & FPCA via the FACE algorithm (proportion of variance 0.95) \citep{xiao2016face}, followed by LDA on the scores \citep{ramsay2005functional} \\
Function-to-scalar regression & Penalized functional regression (PFR) \citep{goldsmith2011penalized} \\
Function-to-function regression & FDboost \citep{brockhaus2015functional}, PFFR \citep{scheipl2015functional} \\
\bottomrule
\end{tabular}
}
\end{table}

Smoothing parameters of PFR and PFFR are chosen by REML (fast REML, \texttt{bam}, for PFFR), and the FDboost stopping iteration \(m_{\rm stop}\le1000\), with step size 0.1, by \texttt{cvrisk} with five bootstrap resamples. Each procedure keeps this default tuning criterion, computed on the training set it receives, partly or wholly synthetic under augmentation and replacement, so that the training set is the only thing that changes between deployments. Classification is evaluated by accuracy, per-class precision, recall and F1, and AUC; regression by RMSE and \(R^2\), with the functional \(R^2=1-\mathrm{SSE}/\mathrm{SST}\), summed over test curves and grid points, for function-to-function regression. Generation-quality metrics and all downstream procedures are implemented in R (packages \texttt{FDboost}, \texttt{refund}, \texttt{fdapace}, \texttt{kernlab}, \texttt{fdWasserstein}).

\section{Theoretical Results}\label{sec:theoretical_results}

We state four theorems in two groups. \Cref{thm:reflection_point,thm:accuracy_gain} concern training on synthetic curves alone, the replacement deployment, and determine how many synthetic curves to use. Their statements are conditional on the generator-training sample, and their structure builds on \citet{shen2024boosting}. \Cref{thm:augmentation_risk,thm:imbalanced_classification} concern augmentation, where class-conditional synthetic curves are added to the real data, and determine how to allocate them among classes. Their statements hold under the joint distribution of the training sample and the synthetic draws, without assuming independence between real and synthetic data, and their structure builds on \citet{Tian2026}.

Further remarks and all proofs are in Appendix C, which also states and proves the auxiliary results they use.

\subsection{Setting and Choice of Metric}\label{sec:setting}

\textbf{Why not total variation.} A measure of generation error must become small when the synthetic distribution approaches the true one. In function space, total variation does not. Let \(H\) be a separable Hilbert space with \(\dim H=\infty\), and compare \(N(0,C)\) with \(N(0,cC)\) for a trace-class \(C\) and \(c\neq1\), two distributions that differ only by a scale factor.

\begin{itemize}
\tightlist
\item
  \emph{TV stays at its maximum.} By the Feldman--H\'ajek theorem \citep{da2014stochastic}, two Gaussian measures on \(H\) are either equivalent or mutually singular, and \(N(m_1,C_1)\) and \(N(m_2,C_2)\) are equivalent if and only if \(m_1-m_2\in C_1^{1/2}(H)\), \(C_1^{1/2}(H)=C_2^{1/2}(H)\), and \(C_1^{-1/2}C_2C_1^{-1/2}-I\) is Hilbert--Schmidt. For \(C_2=cC_1\) the last operator is \((c-1)I\), which is not Hilbert--Schmidt when \(\dim H=\infty\). The two measures are therefore mutually singular, and \(\mathrm{TV}\big(N(0,C),N(0,cC)\big)=1\) for every \(c\neq1\).
\item
  \emph{\(W_1\) tends to zero.} Coupling \(X\sim N(0,C)\) with \(\sqrt c\,X\sim N(0,cC)\) gives \[W_1\big(N(0,C),N(0,cC)\big)\le\mathbb E\|X-\sqrt c\,X\|_H=|\sqrt c-1|\,\mathbb E\|X\|_H\le|\sqrt c-1|\,(\operatorname{Tr}C)^{1/2}\xrightarrow[c\to1]{}0.\]
\end{itemize}

As \(c\to1\) the two distributions become indistinguishable, yet their TV distance stays at 1, so TV cannot tell a nearly exact generator from a poor one; \(W_1\) decreases continuously to zero. We therefore measure generation error by \(W_1\).

\textbf{Spaces and metrics.} \(\mathcal Y\) is a finite label set (classification) or a bounded interval (regression), and \(\mathcal Z:=\mathcal Y\times H\) with \[d_Z\big((y,x),(y',x')\big):=\|x-x'\|_H+c_Z\,\mathbf 1\{y\neq y'\}\qquad(c_Z>0),\] where \(c_Z\mathbf 1\{y\ne y'\}\) is replaced by \(|y-y'|\) for continuous responses; \((\mathcal Z,d_Z)\) is Polish. On \(\mathcal Z^m\) we use \(d_m(w,w'):=\sum_{i=1}^m d_Z(w_i,w_i')\). For \(\mu,\nu\in\mathcal P_1(\mathcal Z)\), \[W_1(\mu,\nu):=\inf_{\gamma\in\Gamma(\mu,\nu)}\int d_Z\,d\gamma=\sup_{\operatorname{Lip}(\varphi)\le1}\Big(\int\varphi\,d\mu-\int\varphi\,d\nu\Big),\] where \(\Gamma(\mu,\nu)\) is the set of couplings of \(\mu\) and \(\nu\); the second equality is the Kantorovich--Rubinstein duality \citep{panaretos2019,villani2009}.

\textbf{Data, generator, risks.} \(F\in\mathcal P_1(\mathcal Z)\) is the data law and \(Z^{(n)}=(Z_1,\dots,Z_n)\overset{\text{iid}}{\sim}F\). The generator (the Conditional DDO of \Cref{sec:methodology}) is trained on \(Z^{(n)}\); \(\widetilde F:=\mathcal L(\widetilde Z\mid Z^{(n)})\) is its output distribution, assumed to lie in \(\mathcal P_1(\mathcal Z)\) so that \(w<\infty\), and \(\widetilde Z_1,\dots,\widetilde Z_m\mid Z^{(n)}\overset{\text{iid}}{\sim}\widetilde F\). For estimators \(\widehat\theta_m:\mathcal Z^m\to\Theta\), a loss \(\ell:\Theta\times\mathcal Z\to\mathbb R\) and an independent test point \(Z\sim F\), define \[\begin{aligned}
g_m(w)&:=\mathbb E_{Z\sim F}\,\ell\big(\widehat\theta_m(w),Z\big),\qquad w\in\mathcal Z^m,\\
R_{\mathrm{raw}}(m)&:=\mathbb E\,g_m(Z^{(m)}),\quad Z^{(m)}\overset{\text{iid}}{\sim}F,\qquad
R_{\mathrm{syn}}(m):=\mathbb E\big[g_m(\widetilde Z^{(m)})\mid Z^{(n)}\big],\\
\mathrm{Gr}(m)&:=R_{\mathrm{syn}}(m)-R_{\mathrm{raw}}(m),\qquad
R^\ast_{\mathrm{syn}}:=\inf_{m\ge1}R_{\mathrm{syn}}(m),\qquad
w:=W_1(\widetilde F,F).
\end{aligned}\] We call \(\mathrm{Gr}(m)\) the generational gap. In \Cref{sec:synthetic_volume}, \(Z^{(n)}\) is fixed, so \(\widetilde F\) is a fixed measure and all quantities above are real numbers.

\textbf{Generation error.} The DDO's generation error splits into three terms: a noising term bounded by \((\operatorname{Tr}C)^{1/2}\) \citep{lim2025score}, a score-approximation term, and a discretization term \citep{pidstrigach2024infinite}.

\subsection{Training on Synthetic Data: How Many Curves}\label{sec:synthetic_volume}

\begin{assumption}\label{asm:bounded_loss}

There is \(U<\infty\) with \(0\le\ell(\theta,z)\le U\) for all \((\theta,z)\in\Theta\times\mathcal Z\).

\end{assumption}

\begin{assumption}\label{asm:generational_growth}

There exist \(m^\ast\in\mathbb N\) and \(\psi:\mathbb R_+\to\mathbb R_+\) with \(\psi(0)=0\) and \(\psi(t)>0\) for \(t>0\), such that

\begin{enumerate}
\def\labelenumi{(\alph{enumi})}
\item
  \(\mathrm{Gr}(m)\ge m\,\psi(w)\) for \(1\le m\le m^\ast\);
\item
  \(\mathrm{Gr}(m)\ge\mathrm{Gr}(m^\ast)\) for \(m>m^\ast\).
\end{enumerate}

\end{assumption}

\begin{assumption}\label{asm:level_condition}

\(m^\ast\,\psi(w)>R^\ast_{\mathrm{syn}}\).

\end{assumption}

Part (a) is the \(W_1\) analogue of the condition \(\mathrm{Gr}(m)\ge m f(\mathrm{TV})\) of \citet{shen2024boosting}; the proof uses it only at \(m=m^\ast\). Part (b) states that the bias of synthetic training does not shrink beyond \(m^\ast\).

\begin{theorem}\label{thm:reflection_point}

Fix \(Z^{(n)}\) and suppose \Cref{asm:bounded_loss,asm:generational_growth,asm:level_condition}. Then \(m^\ast\ge2\) and:

\begin{itemize}
\tightlist
\item
  \textbf{(i$'$)} there is \(m_0\in\{1,\dots,m^\ast-1\}\) with \(R_{\mathrm{syn}}(m_0)=R^\ast_{\mathrm{syn}}\);
\item
  \textbf{(ii$'$)} for every \(m\ge m^\ast\), \(R_{\mathrm{syn}}(m)\ge m^\ast\psi(w)>R_{\mathrm{syn}}(m_0)\).
\end{itemize}

\end{theorem}

The optimal volume \(m_0\) is the reflection point, beyond which more synthetic data cannot restore the minimum risk (Appendix C.1 discusses the assumptions).

To make the generation penalty quantitative, we assume that the influence of a single training curve on the downstream risk decays with the training size; we call the estimator \emph{stable} when it decays like \(m^{-1}\) (\(\kappa=1\)).

\begin{assumption}\label{asm:influence_decay}

There exist \(L>0\) and \(\kappa\in[0,1]\) such that, for every \(m\ge1\), \(g_m\) is \(Lm^{-\kappa}\)-Lipschitz with respect to \(d_m\).

\end{assumption}

\begin{lemma}\label{lem:estimator_stability}

Fix \(m\ge1\) and \(Z^{(n)}\). Assume \Cref{asm:bounded_loss}, and that \(g_m\) is Borel and \(L\)-Lipschitz with respect to \(d_m\) for some \(L>0\). Then \[|\mathrm{Gr}(m)|\le L\,m\,W_1(\widetilde F,F).\]

\end{lemma}

\begin{assumption}\label{asm:benchmark_rate}

There exist \(\alpha>0\) and \(C_\theta>0\) with \(R_{\mathrm{raw}}(m)=C_\theta m^{-\alpha}\) for all \(m\ge1\).

\end{assumption}

This is the form used by \citet{shen2024boosting}; in function space, the minimax rates of \citet{hall2007methodology} for functional linear regression motivate the value of \(\alpha\). When the risk tends to a positive Bayes risk \(R_\infty\), as for overlapping classes, the assumption is applied to the excess risk \(R_{\mathrm{raw}}(m)-R_\infty\); \Cref{rem:benchmark_rate_motivation} in Appendix C.1 shows that, for functional linear regression, these rates bound \(\alpha\) from below in terms of the eigenvalue decay of the predictor covariance and the smoothness of the slope.

\begin{assumption}\label{asm:generation_error_tail}

There exist \(\tau_{n_g}\to0\) as \(n_g\to\infty\) and \(\delta_g\in(0,1)\) with \(\mathbb P\big(W_1(\widetilde F,F)\le\tau_{n_g}\big)\ge1-\delta_g\), where \(n_g\) is the generator-training size (\(n_g=n\) in all our experiments).

\end{assumption}

\begin{theorem}\label{thm:accuracy_gain}

Fix \(Z^{(n)}\) and suppose \Cref{asm:bounded_loss,asm:influence_decay,asm:benchmark_rate}. Write \(e_\kappa:=1+\alpha-\kappa\). For \(\kappa<1\) define \[\begin{aligned}
K_{\alpha,\kappa}&:=\alpha^{-\alpha/e_\kappa}(1-\kappa)^{\alpha/e_\kappa}+\alpha^{(1-\kappa)/e_\kappa}(1-\kappa)^{-(1-\kappa)/e_\kappa},\\
D_{\alpha,\kappa}&:=(1-\kappa)\,\alpha^{-\kappa/e_\kappa}\,C_\theta^{-\kappa/e_\kappa}\,\big((1-\kappa)L\big)^{\kappa/e_\kappa}L,\qquad R_{\alpha,\kappa}(w):=D_{\alpha,\kappa}\,w^{(1+\alpha)/e_\kappa}.
\end{aligned}\]

\textbf{Part 1 (transfer bound).} For every \(m\ge1\), \(|\mathrm{Gr}(m)|\le\min\big(Lm^{1-\kappa}w,\;U\big)\). In particular \[C_\theta m^{-\alpha}-Lm^{1-\kappa}w\;\le\;R_{\mathrm{syn}}(m)\;\le\;C_\theta m^{-\alpha}+Lm^{1-\kappa}w.\]

\textbf{Part 2 (optimized synthetic risk).}

\begin{itemize}
\tightlist
\item
  If \(\kappa<1\) and \(w>0\), let \(s_0:=\big(\alpha C_\theta/((1-\kappa)Lw)\big)^{1/e_\kappa}\). Then \[R^\ast_{\mathrm{syn}}\le R_{\mathrm{syn}}(\lceil s_0\rceil)\le K_{\alpha,\kappa}\,C_\theta^{(1-\kappa)/e_\kappa}L^{\alpha/e_\kappa}w^{\alpha/e_\kappa}+R_{\alpha,\kappa}(w).\]
\item
  If \(\kappa=1\), then \(R_{\mathrm{syn}}(m)\le C_\theta m^{-\alpha}+Lw\) for every \(m\), and \(R^\ast_{\mathrm{syn}}\le Lw\).
\item
  If \(w=0\), then \(R^\ast_{\mathrm{syn}}=0\).
\end{itemize}

\textbf{Part 3 (gain condition).}

\begin{itemize}
\tightlist
\item
  If \(\kappa<1\), let \(c_1:=C_\theta\,2^{-e_\kappa/\alpha}K_{\alpha,\kappa}^{-e_\kappa/\alpha}L^{-1}\) and \(c_2:=\big(C_\theta/(2D_{\alpha,\kappa})\big)^{e_\kappa/(1+\alpha)}\). If \(w\le\min(c_1,c_2)\,n^{-e_\kappa}\), then \(R^\ast_{\mathrm{syn}}\le R_{\mathrm{raw}}(n)=C_\theta n^{-\alpha}\).
\item
  If \(\kappa=1\) and \(w\le(C_\theta/L)\,n^{-\alpha}\), then \(R^\ast_{\mathrm{syn}}\le C_\theta n^{-\alpha}\).
\end{itemize}

\end{theorem}

Part 1 is a two-sided transfer bound: synthetic training can neither lose nor gain more than \(Lm^{1-\kappa}w\) relative to training on \(m\) real curves. Part 2 optimizes the upper bound over the volume \(m\); the minimizer \(\lceil s_0\rceil\) is the volume suggested by the bound. Part 3 gives a sufficient condition on the generation error for \(m\) synthetic curves, at their best volume, to do at least as well as the \(n\) real curves.

Two features are new relative to \citet{shen2024boosting}. The lower bound of Part 1 makes the transfer bound two-sided, so it also limits how much synthetic training can gain over real training and can be tested in both directions. The dependence of the threshold exponent \(e_\kappa\) on \(\kappa\) shows that for stable estimators the exponent drops from \(1+\alpha\) to \(\alpha\), so a much larger generation error is tolerated (Appendix C.1).

Under a generation-error tail with rate \(\tau_{n_g}\le\tau_0n_g^{-\gamma}\), the gain condition holds with high probability for all sufficiently large \(n\) when \(\gamma>e_\kappa\), that is, \(\gamma>\alpha\) for stable estimators; Appendix C.1 states this gain window precisely (\Cref{rem:gain_window}), together with the probability version of \Cref{thm:accuracy_gain}.

\subsection{Augmentation: Allocation of Synthetic Curves}\label{sec:augmentation_allocation}

\textbf{Setting.} Classes are \(\mathcal K:=\{0,\dots,K-1\}\) with \(C_k:=\{Y=k\}\). Write \(\pi_k:=F(C_k)\) and \(F_k:=F(\cdot\mid C_k)\). \(Q\) is the evaluation law with \(q_k:=Q(C_k)\), and \(R_Q(\theta):=\mathbb E_Q\,\ell(\theta,\cdot)\). The theorem allows any \(Q\) that satisfies \Cref{asm:partition_validity_c}; \Cref{sec:imbalanced} takes \(Q\) balanced, while the test set of \Cref{sec:classification} follows the imbalanced real law, so that \(Q=F\) there. The class-conditional generator, trained on \(Z^{(n)}\), yields \(\widetilde F_k:=\mathcal L(\widetilde Z\mid Z^{(n)},Y=k)\). Given \(Z^{(n)}\), draw \(m_k\) curves conditionally i.i.d. from \(\widetilde F_k\), independently across \(k\). Set \[m:=\sum_km_k,\qquad a_k:=m_k/m,\qquad N:=m+n,\qquad\lambda:=(a,m),\qquad\widetilde F:=\sum_ka_k\widetilde F_k .\] The mixed estimator is \[\widehat\theta_\lambda\in\operatorname*{argmin}_{\theta\in\Theta}L_{m,n}(\theta),\qquad L_{m,n}(\theta):=\frac1N\Big[\sum_{k\in\mathcal K}\sum_{j=1}^{m_k}\ell(\theta,\widetilde Z_{kj})+\sum_{i=1}^n\ell(\theta,Z_i)\Big].\] The real part of the training set is the generator's own training sample; there is no sample splitting. Let \(\theta_0\) minimize \(R_Q\) over all measurable rules, \(\theta_Q\in\operatorname{argmin}_{\Theta}R_Q\), and \(\rho(\theta,\theta'):=R_Q(\theta)-R_Q(\theta')\).

\textbf{Indices.} With \(\widetilde a_k:=(m_k+\pi_kn)/N\), the class-\(k\) proportion of the mixed sample, \[D_\lambda:=\sum_k|\widetilde a_k-q_k|,\qquad G_\lambda:=\frac mN\sum_ka_k\,\tau_{k,\pi_kn},\qquad A_\lambda:=\frac M2\,\|a-\pi\|_1,\] where \(\|\cdot\|_1\) is the \(\ell^1\) norm on \(\mathbb R^K\), \(\tau_{k,\pi_kn}\) is the per-class generation rate of \Cref{asm:generation_error_c}, and \(M\) is the diameter of \Cref{asm:bounded_support_c}. \(D_\lambda\) is the domain-adaptation index (mismatch between the class proportions of the mixed sample and of the evaluation law), \(G_\lambda\) the generation-error index, and \(A_\lambda\) the allocation-mismatch index (cost of synthetic class proportions that differ from the real ones).

\textbf{Estimation-error index.} \(\delta_{m+n}\) is the estimation-error rate of the mixed sample, the larger of the rates of its real and synthetic blocks, and \(\bar\theta\) minimizes the mixed-sample risk \(\bar R(\theta):=\frac mN\mathbb E_{\widetilde F}\ell(\theta,\cdot)+\frac nN\mathbb E_F\ell(\theta,\cdot)\); both are defined precisely in Appendix C.1.3.

\begin{assumption}\label{asm:partition_validity_c}

\(Q(\cdot\mid C_k)=F_k\) for every \(k\in\mathcal K\).

\end{assumption}

\begin{assumption}\label{asm:loss_regularity_c}

For every \(\theta\in\Theta\), \(z\mapsto\ell(\theta,z)\) is \(\beta\)-Lipschitz in \(d_Z\), and \(0\le\ell\le U\).

\end{assumption}

For cross-entropy this requires \(\Theta\) to consist of classifiers with \(\theta(x)\in[\eta,1-\eta]\) that are Lipschitz in \(x\); then \(U=\log(1/\eta)\).

\begin{assumption}\label{asm:bounded_support_c}

There is \(M<\infty\) such that \(F\) and, almost surely, every \(\widetilde F_k\) are supported in one set of \(d_Z\)-diameter at most \(M\).

\end{assumption}

\begin{assumption}\label{asm:variance_control_c}

\(\operatorname{Var}_F\big[\ell(\theta,Z)-\ell(\theta_0,Z)\big]\le c_v\,\rho(\theta,\theta_0)\) for all \(\theta\in\Theta\).

\end{assumption}

\begin{assumption}\label{asm:generation_error_c}

For every \(k\in\mathcal K\): \(\mathbb P\big(W_1(\widetilde F_k,F_k)>\tau_{k,\pi_kn}\big)\le\exp\big(-c_g\,\pi_kn\,\tau_{k,\pi_kn}\big)\) for some \(c_g>0\).

\end{assumption}

\begin{assumption}\label{asm:benchmark_rate_c}

\(\delta^{\mathrm{real}}_v=\delta^{\mathrm{syn}}_v=C_\theta v^{-\alpha}\) for all \(v\ge1\), almost surely in \(Z^{(n)}\), in the sense that \(\mathcal S_b\supseteq[C_\theta n_b^{-\alpha},\infty)\) for both blocks. Hence \(\delta_{m+n}=C_\theta\min(m,n)^{-\alpha}\).

\end{assumption}

\begin{theorem}\label{thm:augmentation_risk}

Probabilities are under the joint distribution of \(Z^{(n)}\) and the synthetic draws. Suppose:

\begin{itemize}
\tightlist
\item
  \Cref{asm:partition_validity_c,asm:loss_regularity_c,asm:bounded_support_c,asm:variance_control_c,asm:generation_error_c,asm:benchmark_rate_c} hold;
\item
  the minimizers \(\widehat\theta_\lambda\), \(\theta_Q\), \(\theta_0\) and \(\bar\theta\) exist;
\item
  for the block shares \(\lambda_s+\lambda_r=1\), the Bernstein-range condition holds: \[\gamma_B\in(0,1),\qquad\gamma_B\ge\frac{UN}{c_v}\max\Big(\frac{\lambda_s}{m},\frac{\lambda_r}{n}\Big).\]
\end{itemize}

Define \[\begin{aligned}
c_e&:=2+2U+2\beta,\qquad c_{v2}:=2c_v(4\beta+4U+2)+8U\beta\Big(\frac Nm+1\Big),\\
c_{d,b}&:=\frac{4n_b}{\lambda_bN}(4c_v+c_{v2})+\frac U3,\qquad c_{d'}:=\frac{2\max_bc_{d,b}}{\min(\lambda_s,\lambda_r)}.
\end{aligned}\] Then for every \(\varepsilon_\lambda\) with \[\varepsilon_\lambda\ge\max\big\{\rho(\theta_Q,\theta_0),\,\delta_{m+n},\,D_\lambda,\,G_\lambda,\,A_\lambda\big\}\quad\text{and}\quad(1-\gamma_B)N\varepsilon_\lambda\ge c_{d'}\ln4,\] \[\mathbb P\big(\rho(\widehat\theta_\lambda,\theta_0)\ge c_e\varepsilon_\lambda\big)\le8\exp\Big(-\frac{(1-\gamma_B)N\varepsilon_\lambda}{c_{d'}}\Big)+\sum_{k\in\mathcal K}\exp\big(-c_g\pi_kn\,\tau_{k,\pi_kn}\big).\]

\end{theorem}

The excess risk of the mixed estimator is thus controlled, with exponential tails, by the largest of five terms: approximation \(\rho(\theta_Q,\theta_0)\), estimation \(\delta_{m+n}\), domain adaptation \(D_\lambda\), generation \(G_\lambda\), and allocation mismatch \(A_\lambda\). The second tail term is the probability that some class-conditional generator misses its rate.

\begin{remark}

\Cref{thm:augmentation_risk} requires no sample splitting: because the real and synthetic blocks are bounded separately and combined by a union bound, the generator can be trained on the same curves that enter the downstream fit (Appendix C.1.3).

\end{remark}

\subsection{Imbalanced Classification}\label{sec:imbalanced}

\Cref{thm:imbalanced_classification} applies \Cref{thm:augmentation_risk} at explicit rates. It introduces no new empirical process, following the architecture of \citet[Thm 2]{Tian2026}.

\textbf{Setting.} \(K=2\), \(\mathcal K=\{0,1\}\), the minority class is \(0\) (\(\pi_0\le\pi_1\)), \(q=(\frac12,\frac12)\), and \(\ell\) is the cross-entropy loss \(\ell(p,(y,x)):=-y\log p(x)-(1-y)\log(1-p(x))\). In binary form, \[D_\lambda=2\big|\widetilde a_1-\tfrac12\big|,\qquad A_\lambda=M|a_1-\pi_1|.\] The balancing allocation \(a^\circ_k:=q_k+\frac nm(q_k-\pi_k)\), feasible if and only if \(m\ge n(\pi_1-\pi_0)\), makes the mixed sample balanced. For \(\pi_0n>1\) define the per-class rate profile, the generation envelope and the certified level: \[\begin{aligned}
r_k(n)&:=\tau_0(\pi_kn)^{-\gamma}\big(\log(\pi_kn)\big)^{\beta_\gamma},\qquad\bar G_\lambda:=\frac mN\big(a_0r_0(n)+a_1r_1(n)+\varepsilon_N\big),\\
\varepsilon^\ast_\lambda&:=\max\big\{\rho(\theta_Q,\theta_0),\;C_\theta\min(m,n)^{-\alpha_{\mathrm{cls}}},\;D_\lambda,\;A_\lambda,\;\bar G_\lambda\big\}.
\end{aligned}\]

\begin{assumption}\label{asm:benchmark_rate_cls}

\Cref{asm:benchmark_rate_c} holds with exponent \(\alpha_{\mathrm{cls}}>0\) in place of \(\alpha\).

\end{assumption}

\begin{assumption}\label{asm:generation_rate_form}

For \(k\in\{0,1\}\): \(\tau_{k,\pi_kn}\le\varepsilon_N+r_k(n)\), with \(\gamma>0\), \(\beta_\gamma\in\mathbb R\), \(\tau_0>0\) and \(\varepsilon_N\to0\).

\end{assumption}

\begin{theorem}\label{thm:imbalanced_classification}

Suppose:

\begin{itemize}
\tightlist
\item
  \Cref{asm:partition_validity_c,asm:loss_regularity_c,asm:bounded_support_c,asm:variance_control_c,asm:generation_error_c,asm:benchmark_rate_cls,asm:generation_rate_form} hold;
\item
  the existence and Bernstein-range conditions of \Cref{thm:augmentation_risk} hold;
\item
  \(\pi_0n>1\).
\end{itemize}

Let \(c_e\) and \(c_{d'}\) be as in \Cref{thm:augmentation_risk}.

\textbf{(a) Bound.} For every \(\varepsilon\ge\varepsilon^\ast_\lambda\) with \((1-\gamma_B)N\varepsilon\ge c_{d'}\ln4\), \[\mathbb P\big(\rho(\widehat\theta_\lambda,\theta_0)\ge c_e\varepsilon\big)\le8\exp\Big(-\frac{(1-\gamma_B)N\varepsilon}{c_{d'}}\Big)+\sum_{k=0}^1\exp\big(-c_g\pi_kn\,\tau_{k,\pi_kn}\big).\]

\textbf{(b) Minority term.} With \(\Lambda_n:=\big(\log(\pi_0n)/\log(\pi_1n)\big)^{\beta_\gamma}\), \[\frac{r_0(n)}{r_1(n)}=\Big(\frac{\pi_1}{\pi_0}\Big)^{\gamma}\Lambda_n,\qquad\Lambda_n\to1,\qquad\Lambda_n\ge1\ \text{if}\ \beta_\gamma\le0 .\] Hence \(a_0r_0\ge a_1r_1\) if and only if \((a_0/a_1)(\pi_1/\pi_0)^{\gamma}\Lambda_n\ge1\), and then \(\bar G_\lambda\le\frac mN(2a_0r_0(n)+\varepsilon_N)\). When \(\beta_\gamma\le0\) this holds for every \(a\) with \(a_0\ge a_1\), which includes \(a=q\) and \(a=a^\circ\). For \(a=\pi\) the condition reads \((\pi_1/\pi_0)^{\gamma-1}\Lambda_n\ge1\).

\textbf{(c) Allocation floor.} For every allocation, \(\varepsilon^\ast_\lambda\ge\max\{D_\lambda,A_\lambda\}\ge\frac{2M}{M+2}\big(\pi_1-\frac12\big)\). Specifically:

\begin{itemize}
\tightlist
\item
  at \(a=\pi\): \(D_\lambda=2(\pi_1-\frac12)\) and \(A_\lambda=0\);
\item
  at \(a^\circ\) (feasible iff \(m\ge n(\pi_1-\pi_0)\)): \(D_\lambda=0\) and \(A_\lambda=M(\pi_1-\frac12)(1+\frac nm)\).
\end{itemize}

\textbf{(d) Generation at the benchmark scale.} If \(\gamma\ge\alpha_{\mathrm{cls}}\), let \(\ell_n:=\max_k(\log(\pi_kn))^{\beta_\gamma}\) and \(K_n:=\frac{m}{NC_\theta}\big(\tau_0\pi_0^{-\gamma}\ell_n+\varepsilon_Nn^{\alpha_{\mathrm{cls}}}\big)\). Then \[\bar G_\lambda\le K_n\,C_\theta n^{-\alpha_{\mathrm{cls}}}\le K_n\,\delta_{m+n},\qquad\varepsilon^\ast_\lambda\le\max\{1,K_n\}\max\{\rho(\theta_Q,\theta_0),\delta_{m+n},D_\lambda,A_\lambda\}.\]

\end{theorem}

\begin{remark}

\textbf{Reading \Cref{thm:imbalanced_classification}.} As in the imbalanced example of \citet{Tian2026}, the certified level is set by three terms: the raw-data rate \(C_\theta n^{-\alpha_{\mathrm{cls}}}\); the imbalance cost, which allocation can only move between \(D_\lambda\) and \(A_\lambda\) (Part (c)); and the generation term, which raises the level by at most the factor \(\max\{1,K_n\}\) when the generation rate satisfies \(\gamma\ge\alpha_{\mathrm{cls}}\) (Part (d)).

\end{remark}

\section{Real-Data Analyses}\label{sec:applications}

This section applies the generation--augmentation pipeline of \Cref{sec:methodology} to three real functional datasets, from climatology, plant science and biomechanics, which cover the three downstream tasks: classification, function-to-scalar regression and function-to-function regression. We ask the three questions of \Cref{sec:introduction} and a practical fourth, whether synthetic curves can replace real ones that cannot be shared. Does training on real plus synthetic curves reduce downstream error when the real sample is small? Can synthetic curves replace the real training data? How does downstream performance change with the synthetic volume? And, under class imbalance, how should synthetic curves be allocated across classes?

\Cref{sec:datasets} describes the datasets and \Cref{sec:quality_metrics} the common protocol and the generation-quality diagnostics. \Cref{sec:classification,sec:regression,sec:ml} report the three tasks.

\subsection{Datasets}\label{sec:datasets}

\Cref{tab:datasets} summarizes the three datasets, the conditioning of the generator and the tasks.

\begin{table}[htbp]
\centering
\caption{\textbf{Datasets, generator conditioning and tasks.} \(n\): number of real training curves; validation and test sizes in parentheses.}
\label{tab:datasets}
{\renewcommand{\baselinestretch}{1}\small\setlength{\tabcolsep}{3pt}
\begin{tabular}{m{59.8pt}m{58.2pt}m{49.4pt}m{94.2pt}m{160.0pt}}
\toprule
Dataset & Grid & \(n\) (val./test) & Generator conditioning & Task \\
\midrule
Spanish temperature & 365 days & 232 (50/50) & city (one-hot) & binary classification \\
Grape spectra & 104 wavelengths & 165 (54/55) & variety and normalized sugar content & function-to-scalar regression \\
Gait & 101 points & 75 (none/18) & load condition (one-hot) & function-to-function regression \\
\bottomrule
\end{tabular}
}
\end{table}

\textbf{Temperature.} Daily mean temperature curves of Spanish stations for 1974--2013 come from the \texttt{aemet\_temp} data of \texttt{goffda} \citep{goffda2023}, based on records of the Spanish State Meteorological Agency (AEMET). We classify Madrid (seven stations) against Valladolid (two), with a chronological split (training 1974--2001, validation 2002--2007, test 2008--2013) and 178 Madrid against 54 Valladolid training curves.

\textbf{Grape.} Reflectance spectra (397--1004 nm) of 274 grape samples of three varieties, Fer, Mauzac and Syrah, have laboratory measurements of sugar content \citep{ryckewaert2022grape}; they are split 60/20/20, stratified by variety and by tercile of sugar content.

\textbf{Gait.} Angle, angular velocity, angular acceleration and moment curves under loads of 0\%, 10\% and 20\% of body weight were assembled by \citet{gertheiss2024functional} from a public running dataset \citep{fukuchi2017publicly} and two load-carriage studies \citep{liew2016running,liew2016effects}. We analyze the ankle in the vertical axis for the slow group (ankle VT) and in the anterior--posterior axis for the moderate group (ankle AP), with whole participants assigned to training (25) or test (6); the four channels are generated jointly.

Appendix D.1 gives further details of the three datasets.

\subsection{Protocol and Generation-Quality Diagnostics}\label{sec:quality_metrics}

Generation quality is compared against the validation set, or the test set for gait. We make three comparisons: real training against reference curves of the same class (reference level), synthetic against reference curves of the same class (within-class fidelity), and synthetic curves of one class against real curves of another (cross-class confusion). The task sections summarize fWass; the full comparison scheme, the fWass tables of the two regression tasks, and the fFID and MMD values are in Appendix D.2--D.6.

\subsection{Classification}\label{sec:classification}

\subsubsection{Task and Design}\label{sec:cls_task}

\textbf{Task.} We ask whether training on Conditional DDO curves improves the classification of the city of a Spanish station-year, Madrid (class 0) or Valladolid (class 1), when the real sample is small and imbalanced (\Cref{tab:datasets}), and how the synthetic curves should be allocated between the two cities to do so. The classifier is FPCA + LDA (\Cref{sec:downstream_models}).

\textbf{Allocation design.} In addition to the generational design, the allocation design (\Cref{sec:targeted_allocation}) uses the volumes \(m/n\in\{0.5,1.0,1.5,2.0,2.5,3.0\}\) and the Valladolid shares \(a_1\in\{0,0.1,\dots,1.0\}\), with 20 replicates per configuration.

\subsubsection{Generation Quality}\label{sec:cls_gen}

\Cref{tab:mv_gen} compares synthetic and real training curves with the validation years (2002--2007). The fWass values are sums over the 365 days.

\begin{table}[htbp]
\centering
\caption{\textbf{Madrid vs.~Valladolid: generation quality (fWass) against the validation years.} Rows: validation curves of each city; columns: synthetic or real training curves of each city. Lower is better.}
\label{tab:mv_gen}
{\renewcommand{\baselinestretch}{1}\small
\begin{tabular}{lllll}
\toprule
fWass & syn c0 & train c0 & syn c1 & train c1 \\
\midrule
val c0 & 5878.6 & 5561.7 & 9840.3 & 8658.4 \\
val c1 & 7826.7 & 7983.6 & 7021.4 & 6232.0 \\
\bottomrule
\end{tabular}
}
\end{table}

Two features stand out. First, within each city the synthetic curves are nearly as close to the validation curves as the real training curves are. For Madrid the fWass is 5879 against a reference of 5562 (6\% higher); for Valladolid it is 7021 against 6232 (13\% higher). Second, each synthetic city lies closer to the real curves of its own city than to those of the other (Appendix D.4).

\begin{figure}[htbp]
\centering
\includegraphics[width=0.95\textwidth]{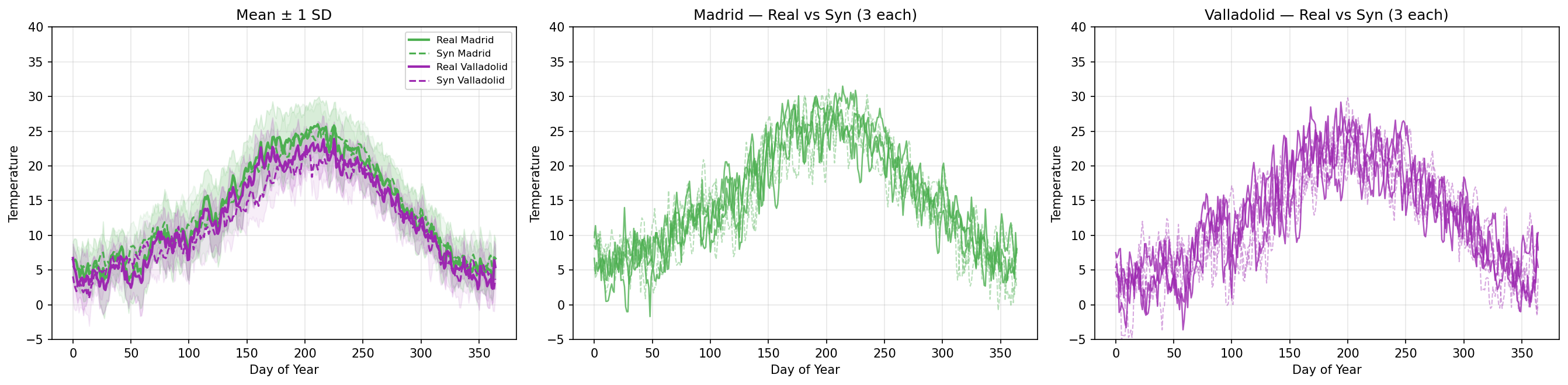}
\caption{\textbf{Madrid vs. Valladolid: real and synthetic curves.} Left: class-conditional mean curves with \(\pm1\) SD bands, real (solid) and synthetic (dashed). Middle and right: three real and three synthetic curves for Madrid and Valladolid.}
\label{fig:cls_gen}
\end{figure}

\Cref{fig:cls_gen} shows that the distribution of the synthetic curves is close to that of the real curves for both cities. The synthetic mean curves follow the real means and reproduce the seasonal amplitude and phase, and the spread of the synthetic curves around their mean is similar to that of the real curves.

\subsubsection{Augmentation and Replacement across Synthetic Volume}\label{sec:cls_volume}

The real-only classifier has accuracy 0.680 and AUC 0.636 on the test years. Its errors are mostly false Valladolid predictions: it labels 15 of the 39 Madrid test curves as Valladolid, while it recovers 10 of the 11 Valladolid curves. \Cref{fig:cls_generational} shows the generational trajectories; selected volumes are tabulated in Appendix D.4 (\Cref{tab:mv_generational}).

\begin{figure}[htbp]
\centering
\includegraphics[width=0.95\textwidth]{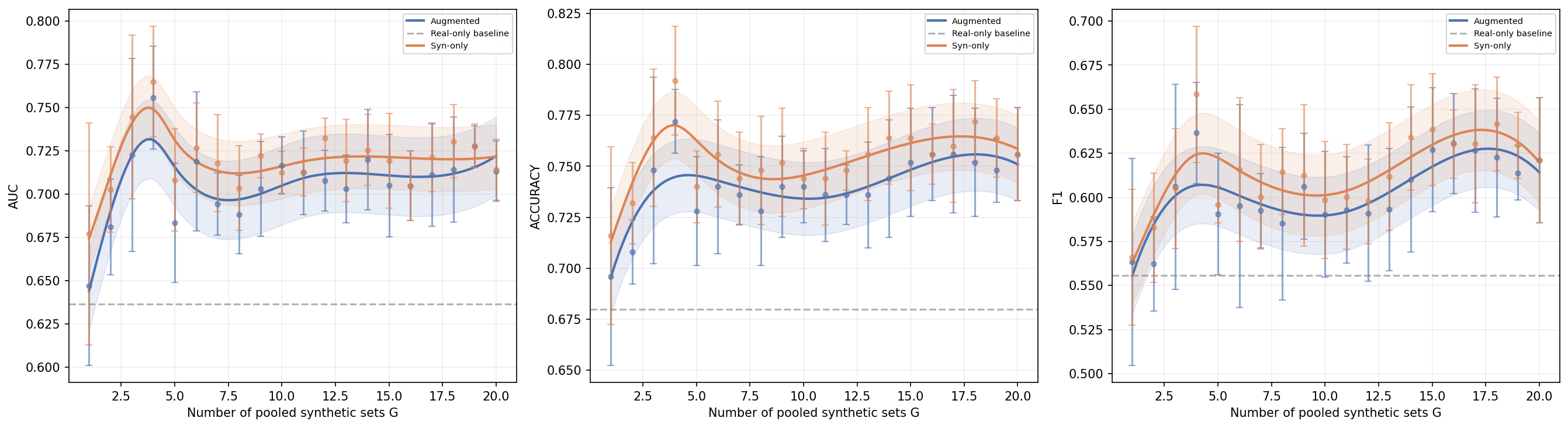}
\caption{\textbf{Madrid vs. Valladolid: performance against synthetic volume.} Test AUC, accuracy and F1 against the number \(G\) of pooled synthetic sets, \(G=1,\dots,20\) (\(m=232G\)). Augmentation (real plus pooled synthetic) and replacement (pooled synthetic only). Lines: penalized B-spline smooths of the five replicates; bands: pointwise 95\% confidence bands; dashed line: real-only baseline.}
\label{fig:cls_generational}
\end{figure}

Both deployments improve on the baseline at every volume, but not monotonically. AUC peaks at \(G=4\), at \(0.756\pm0.015\) under augmentation and \(0.765\pm0.016\) under replacement, falls to about 0.69--0.70 at \(G=8\), and then settles at about 0.71 up to \(G=20\) (\(0.713\) and \(0.714\)); accuracy and F1 also peak at \(G=4\) (accuracy \(0.772\) and \(0.792\)). The gain builds up from \(G=1\) (AUC \(0.647\) under augmentation and \(0.677\) under replacement) to its peak at \(G=4\). This rise, decline and plateau is the shape described by \Cref{thm:reflection_point}, which bounds the risk beyond the reflection point from below by a constant above its minimum.

\subsubsection{Allocation under Class Imbalance}\label{sec:targeted_allocation}

\begin{figure}[htbp]
\centering
\includegraphics[width=0.95\textwidth]{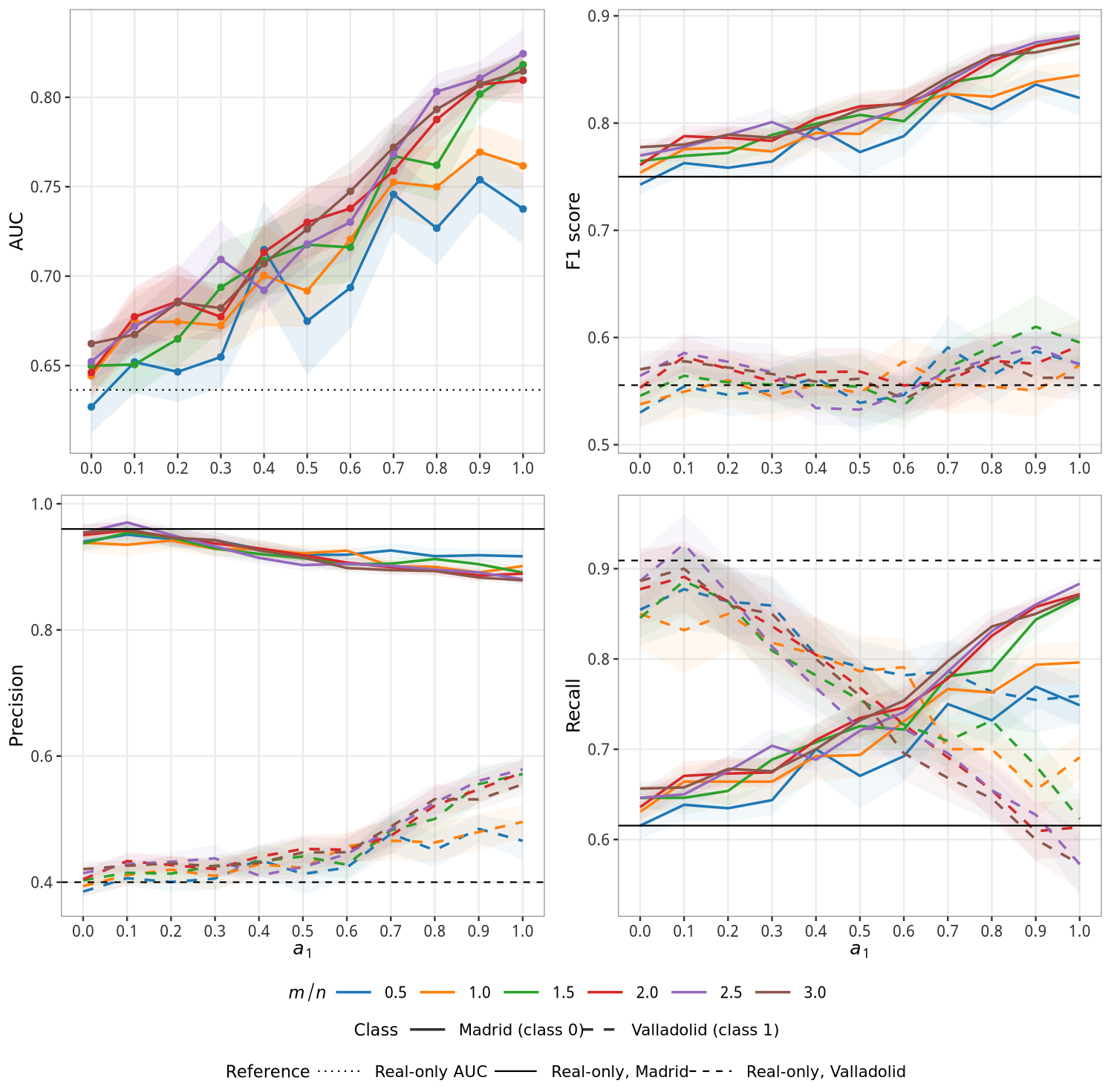}
\caption{\textbf{Madrid vs. Valladolid: AUC and per-class metrics against the minority allocation.} AUC (top left), per-class F1 (top right), per-class precision (bottom left) and per-class recall (bottom right) against the Valladolid share \(a_1\), one colour per synthetic volume \(m/n\). Coloured solid lines: Madrid (class 0); coloured dashed lines: Valladolid (class 1). Lines are means over 20 replicates; shaded bands are pointwise 95\% confidence intervals for the mean (mean \(\pm1.96\) standard errors over the 20 replicates). Black horizontal lines: the classifier trained on the 232 real training curves without synthetic curves (dotted: AUC; solid: Madrid; dashed: Valladolid).}
\label{fig:alpha_sweep}
\end{figure}

We now vary the share \(a_1\) of synthetic curves given to Valladolid at fixed volume. \Cref{fig:alpha_sweep} shows the AUC and the per-class metrics against the Valladolid share, with the real-only classifier as reference; the full AUC grid is in Appendix D.4 (\Cref{tab:appD_alpha}).

\textbf{AUC.} In the top-left panel of \Cref{fig:alpha_sweep}, AUC increases with \(a_1\) on average at every volume, with local decreases of at most 0.04 between neighbouring shares. Adding only Madrid curves (\(a_1=0\)) leaves it close to the real-only value, whereas allocating most or all synthetic curves to Valladolid (\(a_1\ge0.9\)) gives AUC between 0.74 and 0.82. The largest value, \(0.824\pm0.007\) at \(m/n=2.5\) and \(a_1=1\), is 0.19 above the real-only classifier.

\textbf{Per-class metrics.} The gain comes mainly from more reliable Valladolid predictions. Valladolid precision rises with \(a_1\) (bottom left), from the real-only value of 0.40 to about 0.57 at \(a_1=1\) for \(m/n\ge1.5\), and Madrid recall rises from its real-only value of 0.62 to 0.87 at \(m/n=2\) (bottom right): synthetic Valladolid curves remove many of the false Valladolid predictions of the real-only classifier, at the cost of some Valladolid recall. The majority class keeps its performance. Madrid precision stays close to its real-only value of 0.96, decreasing only to about 0.9, and Madrid F1 (top right) exceeds its real-only value of 0.75 at every allocation, reaching 0.88 at \(m/n=2\) and \(a_1=1\); Valladolid F1 stays near its real-only value of 0.56. Accuracy rises from 0.689 to 0.815 at \(m/n=2\), above the 0.78 of the rule that always predicts Madrid; together with the threshold-free AUC, this shows that the gain does not come from predicting mostly one city.

\textbf{Relation to the theory.} The error terms of \Cref{thm:imbalanced_classification} are indexed by the per-class sample sizes \(\pi_kn\), so the estimation error is largest in the minority class. The theorem labels the minority class 0, whereas here it is Valladolid (class 1), so \(a_1\) is the minority share. Directing synthetic curves to that class, with 54 training curves, gave the largest gain, in agreement with the theorem.

\subsection{Function-to-Scalar Regression}\label{sec:regression}

\subsubsection{Task and Models}\label{sec:reg_task}

We predict the sugar content of a grape sample from its reflectance spectrum (\Cref{sec:datasets}) by penalized functional regression \citep{goldsmith2011penalized} with \texttt{refund::pfr} in two specifications, with smoothing parameters chosen by REML:

\begin{itemize}
\tightlist
\item
  \textbf{M1:} \(Y_i=\beta_0+\int X_i(s)\,\beta(s)\,ds+\varepsilon_i\), a functional linear effect of the spectrum;
\item
  \textbf{M2:} \(Y_i=\beta_0+\alpha_{c_i}+\int X_i(s)\,\beta(s)\,ds+\varepsilon_i\), where \(c_i\) is the variety of sample \(i\) and \(\alpha_c\) is a fixed effect of variety \(c\), with one variety as the reference level (\(\alpha_c=0\)).
\end{itemize}

Performance is measured by test RMSE and \(R^2\).

As for the temperature curves, the synthetic spectra lie closest to the real spectra of their own variety, at distances of the same order as those of the real training spectra (details in Appendix D.5).

\subsubsection{Augmentation and Replacement across Synthetic Volume}\label{sec:reg_volume}

\Cref{fig:reg_generational} shows the trajectories; selected volumes are tabulated in Appendix D.5 (\Cref{tab:grape_downstream}).

\begin{figure}[htbp]
\centering
\includegraphics[width=0.95\textwidth]{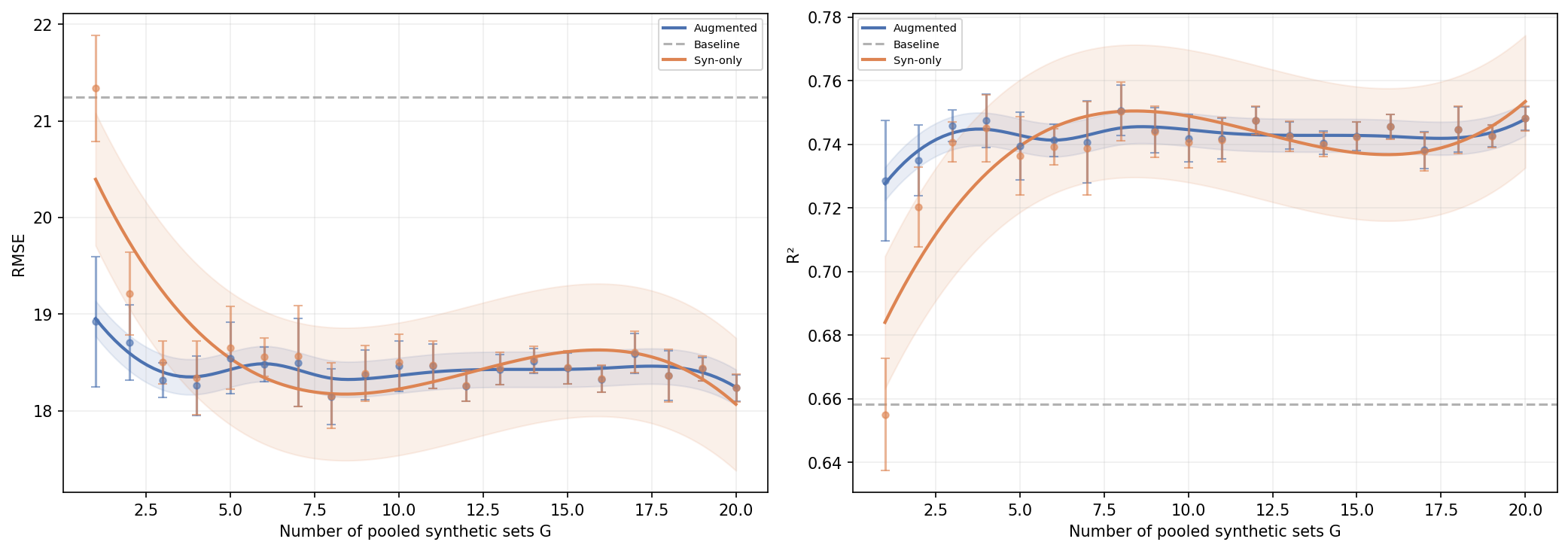}

\includegraphics[width=0.95\textwidth]{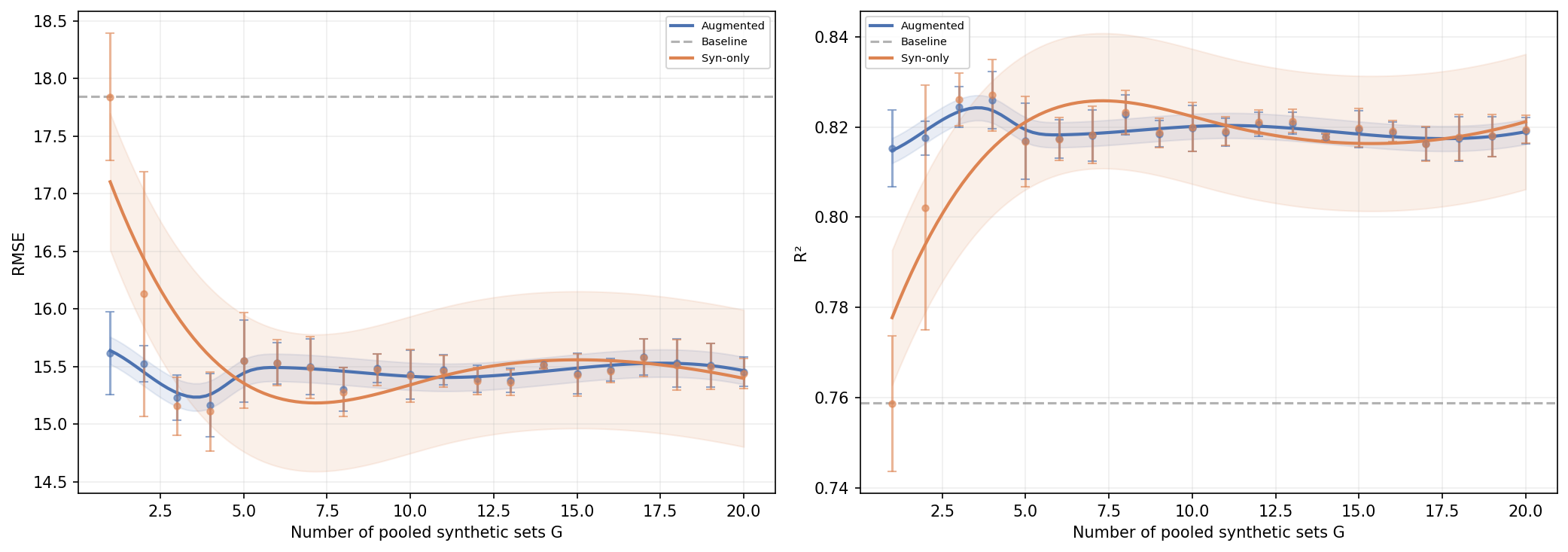}
\caption{\textbf{Grape sugar content: performance against synthetic volume.} Test RMSE and \(R^2\) against the number \(G\) of pooled synthetic sets, \(G=1,\dots,20\) (\(m=165G\)), for M1 (top) and M2 (bottom). Lines, bands and dashed line as in \Cref{fig:cls_generational}.}
\label{fig:reg_generational}
\end{figure}

Augmentation raises \(R^2\) from 0.658 to 0.747 (M1) and from 0.759 to 0.826 (M2) by \(G=4\), after which the augmented trajectories are flat up to \(G=20\); replacement, which already matches the real-data fit at \(G=1\), reaches the augmented fit by about \(G=4\). Because the synthetic responses are copies of training responses, the gain comes not from new response values but from more spectra per response value, which stabilizes the REML estimate of the coefficient function; it persists when variety is in the model (M2), so it is not a by-product of variety information carried by the generator.

\subsection{Function-to-Function Regression}\label{sec:ml}

\subsubsection{Task and Models}\label{sec:ml_task}

We predict a joint-moment curve from the angle, angular velocity and angular acceleration curves of the same joint and axis, over the gait cycle, in the two problems of \Cref{sec:datasets}: ankle VT, slow, and ankle AP, moderate. Both downstream models regress the moment curve \(Y_i(t)\) on the three predictor curves \(X_{i1},X_{i2},X_{i3}\) and the load condition \(c_i\): \[Y_i(t)=\beta_0(t)+\sum_{j=1}^3\int X_{ij}(s)\,\beta_j(s,t)\,ds+\gamma_{c_i}(t)+\varepsilon_i(t).\] \textbf{FDboost} \citep{brockhaus2015functional} fits this model by component-wise gradient boosting, and \textbf{PFFR} \citep{scheipl2015functional} by tensor-product penalized splines with REML smoothing; their settings are given in \Cref{sec:downstream_models}. Performance is measured by the test RMSE over curves and grid points and by the functional \(R^2\).

By fWass, the synthetic curves are of similar quality to the real training curves (Appendix D.6).

\subsubsection{Augmentation and Replacement across Synthetic Volume}\label{sec:ml_volume}

\begin{figure}[htbp]
\centering
\includegraphics[width=0.95\textwidth]{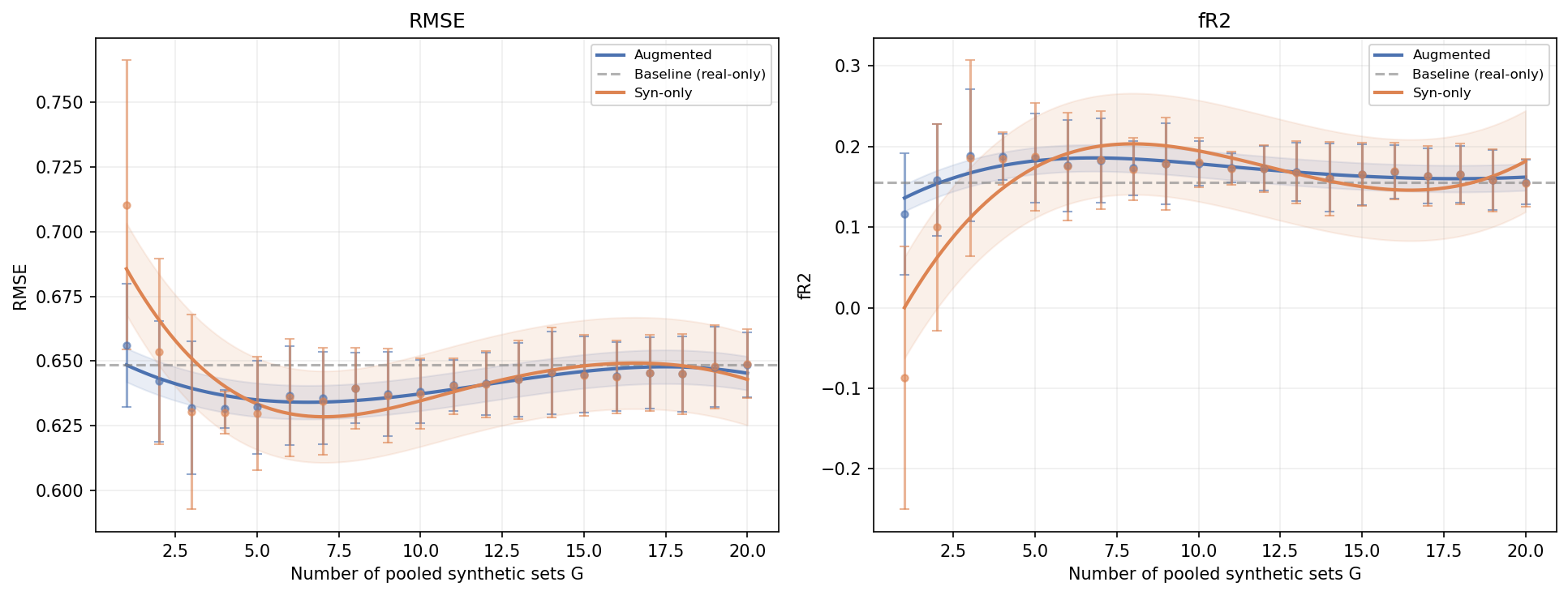}

\includegraphics[width=0.95\textwidth]{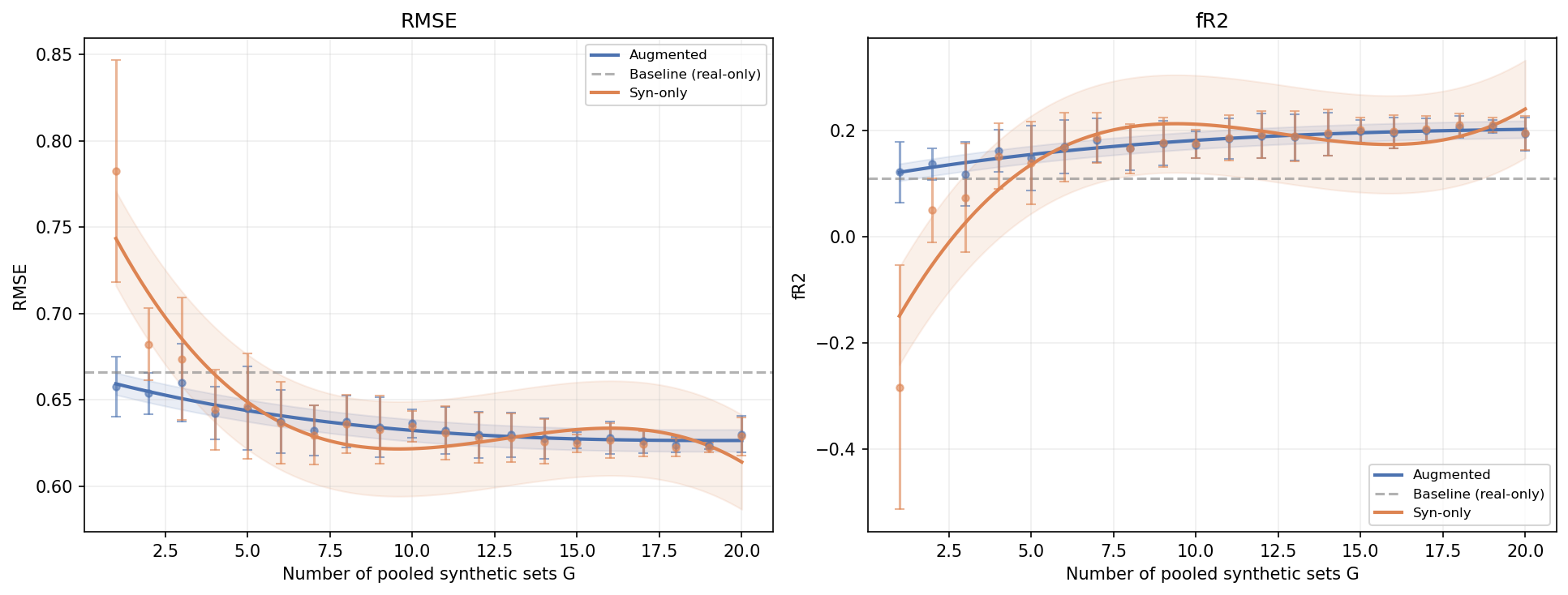}
\caption{\textbf{Ankle VT, slow: performance against synthetic volume.} Test RMSE and functional \(R^2\) against the number \(G\) of pooled synthetic sets (\(m=75G\)) for FDboost (top) and PFFR (bottom). Lines, bands and dashed line as in \Cref{fig:cls_generational}.}
\label{fig:ml_generational_vt}
\end{figure}

\Cref{fig:ml_generational_vt} shows the trajectories for ankle VT; those for ankle AP and a table of the best replicate-mean performance are in Appendix D.6 (\Cref{fig:ml_generational_ap,tab:ml_downstream}).

Augmentation improves both models in both problems, by 1.4--6.3\% in RMSE and by 0.02--0.10 in functional \(R^2\), relative to real-only fits that explain 11--30\% of the variation of the test moment curves; replacement does as well as augmentation at the best volume. PFFR improves at every volume in both problems, most on ankle VT, where the functional \(R^2\) rises from 0.109 to 0.207 and levels off after about \(G=15\). FDboost improves at intermediate volumes: on ankle VT its RMSE is smallest near \(G=4\) and returns to about the real-only level by \(G\approx15\)--20, the rise and decline described by \Cref{thm:reflection_point}.

\section{Discussion}\label{sec:discussion}

This paper developed synthetic data augmentation for functional data, with a conditional diffusion generator trained on the same small sample it augments. Measuring generation error by \(W_1\) makes the volume theory meaningful in function space: the risk of synthetic training lies within \(Lm^{1-\kappa}W_1(\widetilde F,F)\) of that of real training, in both directions, the generation penalty does not grow with the volume for stable estimators, and the allocation theory bounds the excess risk of augmentation without sample splitting. On three real datasets synthetic curves improved on real-only training, replaced the real data at moderate volumes and, under imbalance, gave their largest gain when directed to the minority class.

Future work will derive a statistical rate for conditional score matching with neural operators, which would replace the generation-error assumptions by results, and a theory for the quadratic generation penalty, which would sharpen the bounds of \Cref{thm:accuracy_gain}, whose generation penalty grows only linearly in \(W_1\). We also plan to develop a data-driven allocation rule with guarantees, to generate curves jointly with scalar covariates for richer regression models, and to study the privacy properties of synthetic functional data.

\section*{Data Availability Statement}
\addcontentsline{toc}{section}{Data Availability Statement}

The temperature data are from AEMET (Agencia Estatal de Meteorolog\'ia), obtained through the AEMET OpenData service (https://opendata.aemet.es), which permits free reuse under Spain's Law 18/2015 on the reuse of public-sector information, with attribution to AEMET (https://www.aemet.es/es/nota\_legal). The gait data are those of \citet{gertheiss2024functional}, with the demonstration data and code available at https://github.com/davidruegamer/FDA\_tutorial. Code, configuration files and seeds for every analysis will be released in a public repository in the coming months; Appendix B records all settings used.

\section{Disclosure statement}\label{disclosure-statement}
The authors report there are no competing interests to declare.

\bibliography{references}

\clearpage
\section*{Overview of the appendices}
\addcontentsline{toc}{section}{Overview of the appendices}

The appendices contain the details that the main text summarizes: the model, the settings needed to reproduce every analysis, the proofs of the theoretical results, and the full real-data results. Each appendix supports one or two sections of the main text and can be read on its own.

\begin{itemize}
\tightlist
\item
  \textbf{Appendix A (Model details; supports \Cref{sec:methodology}).} A.1 gives the diffusion process in function space: the forward and reverse stochastic differential equations, the Gaussian reference measure and the roles of the noise-kernel parameters. A.2 describes the network: the spectral convolution, the U-shaped structure, the conditional residual block with FiLM and the padding of the grid. A.3 gives the training and sampling algorithms.
\item
  \textbf{Appendix B (Implementation details; supports \Cref{sec:methodology,sec:applications}).} B.1 lists the software and computing environment, B.2 the generator configuration for each dataset, B.3 the synthetic sets and seeds of each design, and B.4 the exact specification of each downstream procedure.
\item
  \textbf{Appendix C (Theory; supports \Cref{sec:theoretical_results}).} C.1 collects remarks on each theorem: when the assumptions of \Cref{thm:reflection_point} can hold together, an estimator with \(\kappa=1\), the motivation of the benchmark rate, the probability version and gain window of \Cref{thm:accuracy_gain}, and the estimation-error indices and dependence model behind \Cref{thm:augmentation_risk,thm:imbalanced_classification}. C.2 states the auxiliary lemmas and proves all results, in the order of the main text.
\item
  \textbf{Appendix D (Real-data details; supports \Cref{sec:applications}).} D.1 describes the three datasets and their splits, D.2 defines the generation-quality metrics and D.3 the comparisons made with them. D.4--D.6 give, for the temperature, grape and gait data in turn, the generation-quality tables, the figures of real and synthetic curves, the downstream results at selected volumes, the full allocation grid for the temperature data, and the ankle AP trajectories for the gait data.
\end{itemize}

\clearpage
\appendix
\section{Model Details}\label{app:model}

This appendix gives the details of the Conditional DDO that \Cref{sec:ddo,sec:film} summarize. Settings that vary between experiments are in Appendix B.

\subsection{Diffusion in Function Space}\label{app:diffusion}

\begin{table}[htbp]
\centering
\caption{\textbf{Notation of \Cref{sec:methodology}.}}
\label{tab:method_notation}
{\renewcommand{\baselinestretch}{1}\small
\begin{tabular}{ll}
\toprule
Symbol & Meaning \\
\midrule
\(\mathcal{H},\,N_{\rm grid},\,C_{\rm ch}\) & \(L^2([0,1])\); number of grid points; number of channels \\
\(F,\ F_k,\ \widetilde F,\ \widetilde F_k\) & joint law of \((Y,X)\); class-conditional laws; their synthetic counterparts \\
\(K,\ \pi_k,\ a_k\) & number of classes; real class proportions; synthetic allocation, \(a_k=m_k/m\) \\
\(n,\ m\) & real sample size; synthetic sample size \\
\(\mathcal{C}\); \(\sigma_{\rm gp},\nu,\ell_{\rm gp}\) & noise covariance operator; kernel amplitude, exponent and length scale \\
\(\alpha_t,\ \sigma_t\) & signal scale and noise scale of the schedule \\
\(\varepsilon_\psi,\ \psi\) & denoising network and its parameters \\
\(v,\ \mathbf{c}\) & conditioning vector; its learned embedding \\
\(S,\ B,\ N_{\rm step}\) & training steps; batch size; reverse-diffusion steps \\
\(G\) & number of pooled synthetic sets, each of size \(n\), so \(m=Gn\) (\Cref{sec:generational_effect}) \\
\bottomrule
\end{tabular}
}
\end{table}

\textbf{Forward process as a stochastic differential equation.} The forward perturbation \(x_t=\alpha_tx_0+\sigma_t\varepsilon\) of \Cref{sec:ddo}, with \(\alpha_t^2+\sigma_t^2=1\) (variance preservation), is the marginal law of the linear SDE \[
dx=f(t)\,x\,dt+g(t)\,dW^{\mathcal{C}}_t,\qquad f(t)=\frac{d\log\alpha_t}{dt},\quad g(t)^2=-2f(t),
\] driven by a \(\mathcal{C}\)-Wiener process \(W^{\mathcal{C}}\). For multi-channel data the noise is drawn independently in each channel, i.e.~its covariance is \(\mathcal{C}\otimes I_{C_{\rm ch}}\).

\textbf{Reference measure and score.} The trace-class condition \({\rm tr}(\mathcal{C})<\infty\) ensures that \(\mathcal{N}(0,\mathcal{C})\) is a well-defined Gaussian measure on \(\mathcal{H}\) and that \(\mathbb{E}\|\varepsilon\|_{\mathcal{H}}^2={\rm tr}(\mathcal{C})\) is finite. By the Feldman--H\'ajek theorem, two Gaussian measures on \(\mathcal{H}\) are either equivalent or mutually singular, and equivalence requires conditions on both the means and the covariances. The score of the perturbed law is therefore defined with respect to the Gaussian reference measure along its Cameron--Martin space, rather than with respect to a Lebesgue measure \citep{lim2025score}.

\textbf{Kernel parameters.} The three parameters of the powered-exponential kernel \(k(s,t)=\sigma_{\rm gp}^2\exp(-(|s-t|/\ell_{\rm gp})^{\nu})\) play the following roles.

\begin{itemize}
\tightlist
\item
  \(\sigma_{\rm gp}>0\) is the amplitude of the noise relative to the standardized curves (\Cref{sec:architecture}).
\item
  \(\nu\) controls the roughness of the noise paths: \(\nu=1\) gives the Laplace (exponential) kernel with continuous but non-differentiable paths, and \(\nu=2\) gives the Gaussian kernel with infinitely differentiable paths.
\item
  \(\ell_{\rm gp}>0\) is the length scale on the normalized domain \citep{lim2025score}: smaller values make the noise decorrelate over shorter distances, so it varies faster along the curve.
\end{itemize}

\textbf{Reverse process.} Let \(p_t\) denote the law of \(x_t\). The reverse-time process is \[
dx=\bigl[f(t)\,x-g(t)^2\,\mathcal{C}\nabla\log p_t(x)\bigr]dt+g(t)\,d\widetilde W^{\mathcal{C}}_t ,
\] where \(\widetilde W^{\mathcal{C}}\) is a reverse-time \(\mathcal{C}\)-Wiener process. In function space, \(\mathcal{C}\nabla\log p_t\) is the derivative of the log-density along the Cameron--Martin space of \(\mathcal{N}(0,\mathcal{C})\) \citep{lim2025score}. Because \(\log p(x_t\mid x_0)\) has gradient \(-\mathcal{C}^{-1}\varepsilon/\sigma_t\), the conditional mean of the noise satisfies \(\mathbb{E}[\varepsilon\mid x_t]=-\sigma_t\,\mathcal{C}\nabla\log p_t(x_t)\), which is the identity used in \Cref{sec:ddo}.

\subsection{Network Architecture}\label{app:architecture}

\textbf{Spectral convolution.} For a feature map \(h\in\mathbb{R}^{C_h\times N_h}\), \(\mathrm{SpecConv}(h)\) proceeds in four steps: (i) \(\hat h=\mathrm{RFFT}(h)\in\mathbb{C}^{C_h\times M}\), with \(M=\lfloor N_h/2\rfloor+1\) frequency bins; (ii) keep the \(m^{\rm bin}=\lfloor m/2\rfloor+1\) lowest-frequency bins, where \(m=\min(m_{\rm modes},N_h)\) and the mode budget \(m_{\rm modes}\) counts real Fourier coefficients; (iii) multiply each kept bin by a learned complex matrix, \(W\in\mathbb{C}^{C_{\rm out}\times C_h\times m^{\rm bin}}\); (iv) apply the inverse FFT. Because \(N_h\) halves at each coarser level, the budget is capped by the grid length of the level.

\textbf{Multi-channel data.} Multi-channel curves, such as the four gait channels (joint angle, angular velocity, angular acceleration and joint moment), enter as stacked input channels; the lifting layer maps them jointly to the base width and the spectral convolutions mix them, so the channels are generated jointly.

\textbf{U-shaped structure.}

\begin{itemize}
\tightlist
\item
  \emph{Lifting:} a pointwise \(1\times1\) convolution maps the noisy curve \(x_t\), concatenated with the coordinate grid \(u\in[0,1]^{N_{\rm grid}}\), to the base channel count.
\item
  \emph{Encoder:} \(L\) resolution levels, each with \(R\) conditional spectral residual blocks followed by spectral downsampling; the resolution halves at each level and the width is multiplied by the level's channel multiplier. All reported configurations use \(L=4\) levels with multipliers \((1,2,2,4)\), and \(R=2\) (temperature, spectra) or \(R=3\) (gait); see Appendix B.
\item
  \emph{Bottleneck:} two conditional spectral residual blocks.
\item
  \emph{Decoder:} a symmetric upsampling path with skip connections, with \(R+1\) blocks per level.
\item
  \emph{Output:} a \(1\times1\) convolution to \(C\) output channels, applied directly to the last decoder block. The network of \citet{lim2025score} applies a SiLU activation before this convolution; we omit it. Either form can output any real value, and we make no claim about the effect of this choice.
\end{itemize}

\textbf{Conditional spectral residual block.} Given the block input \(h\), a timestep embedding, and the condition embedding \(\mathbf{c}\), the block computes \[
\begin{aligned}
h^{(1)}&=\mathrm{SpecConv}_1\bigl(\mathrm{Act}(\mathrm{Norm}_1(h))\bigr),\\
h^{(2)}&=\mathrm{FiLM}_{\rm time}\bigl(\mathrm{Norm}_2(h^{(1)});\,t\bigr),\\
h^{(3)}&=\mathrm{FiLM}_{\rm cond}\bigl(h^{(2)};\,\mathbf{c}\bigr),\\
h_{\rm out}&=\mathrm{Skip}(h)+\mathrm{SpecConv}_2\bigl(\mathrm{Act}(h^{(3)})\bigr),
\end{aligned}
\] where \(\mathrm{Act}\) is SiLU and \(\mathrm{Skip}\) is the residual (skip) connection, which passes the block input directly to its output: the identity when the input and output widths agree, and a pointwise linear map (\(1\times1\) convolution without bias) otherwise.

\textbf{Grid length.} Each encoder level halves the resolution, so the network needs a grid length divisible by \(2^{L-1}\); the network of \citet{lim2025score} has no padding and therefore requires the curves themselves to have such a length. We instead pad each input curve and its coordinate grid at both ends, by repeating the end values, to the next multiple of \(2^{L-1}\), and crop the output back to the original \(N\) points. With \(L=4\), the 101-point gait curves gain three points and the 365-point temperature curves three points; the 104-point spectra are not padded. The training loss and the generated curves use the original \(N\) points only.

\subsection{Training and Sampling Algorithms}\label{app:algorithms}

\begin{algorithm}[t]
\caption{Training the Conditional Denoising Diffusion Operator}
\label{alg:training}
\begin{algorithmic}[1]
\renewcommand{\algorithmicrequire}{\textbf{Input:}}
\renewcommand{\algorithmicensure}{\textbf{Output:}}
\REQUIRE training set \(\mathcal{D}_n=\{(X_i,Y_i)\}\); noise covariance \(\mathcal{C}\) (parameters \(\sigma_{\rm gp},\nu,\ell_{\rm gp}\)); schedule \((\alpha_t,\sigma_t)\); number of steps \(S\); batch size \(B\); learning rates \(\eta_{\max},\eta_{\min}\); clipping norm \(c\).
\ENSURE trained weights \(\widehat\psi\) (with checkpoints saved during training); channel means \(\bar x_c\) and standard deviations \(\hat s_c\).
\STATE Standardize each channel \(c\) of every curve: \(X\leftarrow (X-\bar x_c)/\hat s_c\).
\STATE Initialize \(\psi\) at random; set up AdamW with a cosine-annealing schedule from \(\eta_{\max}\) to \(\eta_{\min}\) over \(S\) steps (no restarts).
\FOR{\(s=1\) \textbf{to} \(S\)}
\STATE Draw a group-balanced minibatch \(\{(X_i,Y_i)\}_{i=1}^{B}\) from \(\mathcal{D}_n\).
\STATE Draw a timestep \(t_i\in[0,1]\) for each curve (low-discrepancy sampler).
\STATE Draw noise \(\varepsilon_i\sim\mathcal{N}(0,\mathcal{C})\) independently for each curve and channel.
\STATE Compute \(\alpha_{t_i},\sigma_{t_i}\) and form \(x_{t_i}=\alpha_{t_i}X_i+\sigma_{t_i}\varepsilon_i\).
\STATE Encode the condition: \(v_i=v(Y_i)\) (one-hot or value-augmented), \(\mathbf{c}_i=\mathrm{MLP}_{\rm embed}(v_i)\).
\STATE Predict \(\hat\varepsilon_i=\varepsilon_\psi(x_{t_i},t_i,\mathbf{c}_i)\).
\STATE Compute \(\mathcal{L}=\frac1{B}\sum_i \tfrac12\|\hat\varepsilon_i-\varepsilon_i\|^2\).
\STATE Clip the gradient norm to \(c=1\) and update \(\psi\) by AdamW; advance the learning-rate schedule.
\ENDFOR
\RETURN{\(\widehat\psi\), \(\bar x_c\), \(\hat s_c\)}
\end{algorithmic}
\end{algorithm}

\textbf{Sampling coefficients.} Writing \(s=t-1/N_{\rm step}\) for the previous time, the transition coefficients are \[
a_t=\frac{\alpha_t}{\alpha_s},\qquad b_t=\sigma_t^2-a_t^2\sigma_s^2,\qquad
\mu_{s\mid t}=\frac{a_t\,\sigma_s^2}{\sigma_t^2}\,z_t+\frac{\alpha_s\,b_t}{\alpha_t\,\sigma_t^2}\bigl(z_t-\sigma_t\hat\varepsilon\bigr),\qquad
\tilde\sigma^2=\frac{b_t\,\sigma_s^2}{\sigma_t^2}.
\] The second term of \(\mu_{s\mid t}\) is the weight \(\alpha_s b_t/\sigma_t^2\) on the predicted clean curve \((z_t-\sigma_t\hat\varepsilon)/\alpha_t\). \Cref{alg:sampling} gives the procedure.

\begin{algorithm}[t]
\caption{Conditional Sampling via Reverse Diffusion}
\label{alg:sampling}
\begin{algorithmic}[1]
\renewcommand{\algorithmicrequire}{\textbf{Input:}}
\renewcommand{\algorithmicensure}{\textbf{Output:}}
\REQUIRE trained \(\varepsilon_{\widehat\psi}\); constants \(\bar x_c,\hat s_c\); conditioning vectors \(v_1,\dots,v_m\); number of steps \(N_{\rm step}\).
\ENSURE synthetic curves \(\widetilde X_1,\dots,\widetilde X_m\) on the original scale.
\STATE For each \(j\): embed the condition, \(\mathbf{c}_j=\mathrm{MLP}_{\rm embed}(v_j)\).
\STATE Draw the initial state \(z\sim\mathcal{N}(0,\mathcal{C})\) for each \(j\) and channel.
\STATE Set \(t_i=1-(i-1)/N_{\rm step}\) for \(i=1,\dots,N_{\rm step}\), so \(t_1=1\) and \(t_{N_{\rm step}}=1/N_{\rm step}\).
\FOR{\(i=1\) \textbf{to} \(N_{\rm step}\)}
\STATE Set \(t=t_i\), \(s=t_i-1/N_{\rm step}\); compute \(\alpha_t,\sigma_t,\alpha_s,\sigma_s\), then \(a_t\), \(b_t\) and \(\tilde\sigma^2\).
\STATE Predict \(\hat\varepsilon=\varepsilon_{\widehat\psi}(z,t,\mathbf{c}_j)\) and compute \(\mu_{s\mid t}\).
\STATE \textbf{if} \(i<N_{\rm step}\) \textbf{then} draw \(\eta\sim\mathcal{N}(0,\mathcal{C})\) and set \(z\leftarrow\mu_{s\mid t}+\tilde\sigma\,\eta\); \textbf{else} set \(z\leftarrow\mu_{s\mid t}\).
\ENDFOR
\RETURN{\(\widetilde X_j=\hat s_c\,z+\bar x_c\) in each channel.}
\end{algorithmic}
\end{algorithm}

\section{Implementation Details}\label{app:implementation}

This appendix records the settings needed to reproduce the analyses of \Cref{sec:applications}. The code, configuration files and seeds will be released in a public repository in the coming months.

\subsection{Software and Computing Environment}\label{app:software}

The generator and its sampler are written in Python with PyTorch. Generation-quality metrics and all real-data downstream procedures are computed in R with the packages \texttt{refund} (FACE, PFR, PFFR), \texttt{FDboost}, \texttt{fdapace}, \texttt{kernlab} and \texttt{fdWasserstein}. Generators were trained on single NVIDIA A800 GPUs.

\subsection{Generator Configurations for the Real Data}\label{app:real_config}

All real-data generators use the network of \Cref{sec:architecture} with group normalization, the variance-preserving cosine schedule, \(\varepsilon\)-prediction, low-discrepancy timestep sampling, and powered-exponential Gaussian-process noise. Training uses AdamW with weight decay 0.01, cosine annealing from the initial to the final learning rate without restarts, gradient-norm clipping at 1, group-balanced minibatches, no condition dropout and no exponential moving average. Checkpoints are saved every 5,000 steps, and the one used for each dataset is listed in \Cref{tab:appB_real}; where a validation set exists, it guides this choice (\Cref{sec:pipeline}). Sampling uses the ancestral sampler of \Cref{alg:sampling} with 1,000 steps. \Cref{tab:appB_real} lists the settings that differ between datasets.

\begin{table}[htbp]
\centering
\caption{\textbf{Generator settings for the real-data analyses.} \(N_{\rm grid}\): grid points per channel; \(N_{\rm pad}\): grid length after padding to a multiple of \(2^{L-1}=8\) (\Cref{app:architecture}); \(M=\lfloor N_{\rm pad}/2\rfloor+1\): number of frequency bins at the finest level; \(m_{\rm modes}\): mode budget in real Fourier coefficients, so that \(\lfloor m_{\rm modes}/2\rfloor+1\) bins are kept at the finest level (\Cref{app:architecture}); \(R\): residual blocks per resolution level; \(d_c\): condition-embedding dimension; \(K\): number of groups in the conditioning vector. The gait settings apply to both ankle VT and ankle AP; the gait channels are angle, angular velocity, angular acceleration and moment.}
\label{tab:appB_real}
{\renewcommand{\baselinestretch}{1}\small\setlength{\tabcolsep}{3pt}
\begin{tabular}{m{101.7pt}m{62.7pt}m{138.3pt}m{128.3pt}}
\toprule
Setting & Temperature & Grape spectra & Gait \\
\midrule
Channels \(C_{\rm ch}\) & 1 & 1 & 4 \\
Grid points \(N_{\rm grid}\) & 365 & 104 & 101 \\
Padded length \(N_{\rm pad}\) & 368 & 104 & 104 \\
Frequency bins \(M\) & 185 & 53 & 53 \\
Mode budget \(m_{\rm modes}\) & 352 & 104 & 96 \\
Kept bins at finest level & 177 & 53 & 49 \\
Base width & 64 & 64 & 128 \\
Channel multipliers & (1, 2, 2, 4) & (1, 2, 2, 4) & (1, 2, 2, 4) \\
Residual blocks \(R\) & 2 & 2 & 3 \\
Embedding \(d_c\) & 512 & 512 & 512 \\
Conditioning & one-hot city & value-augmented variety & one-hot load condition \\
Groups \(K\) & 2 & 3 & 3 \\
Noise \(\sigma_{\rm gp}\) & 2.0 & 1.0 & 2.0 \\
Noise \(\nu\) & 1.0 & 1.0 & 1.5 \\
Noise \(\ell_{\rm gp}\) & 0.05 & 0.2 & 0.05 \\
Training steps \(S\) & 140,000 & 120,000 & 80,000 \\
Batch size \(B\) & 8 & 8 & 8 \\
Initial learning rate \(\eta_{\max}\) & \(10^{-4}\) & \(10^{-4}\) & \(5\times10^{-5}\) \\
Final learning rate \(\eta_{\min}\) & \(10^{-6}\) & \(10^{-6}\) & \(5\times10^{-9}\) \\
Training seed & 42 & 42 & 42 \\
Checkpoint used & 65,000 & 120,000 (final) & 80,000 (final) \\
Training curves & 232 & 165 & 75 \\
Validation curves & 50 & 54 & none \\
Test curves & 50 & 55 & 18 \\
\bottomrule
\end{tabular}
}
\end{table}

\textbf{Data preparation.} Temperature: station-years with any missing day are removed and the split is chronological (\Cref{sec:datasets}). Grape: the spectra are interpolated to a uniform grid (Appendix D.1.2), and the split is random with seed 42, stratified by variety and sugar-content tercile in proportions 6:2:2 within each of the nine strata. Gait: the participant-level partition supplied with the data is used, and the four channels of a joint and axis are stacked. In all cases the standardization constants of the generator are computed on the training split only.

\subsection{Generation and Replicates}\label{app:generation}

\textbf{Quality-assessment sets.} For the metrics of \Cref{sec:quality_metrics}, one synthetic set is generated with seed 2026 and with the class counts of the reference sample (validation set for temperature and grape, test set for gait); for the grape data the (variety, response) pairs of the reference sample are used as conditions.

\textbf{Generational pools.} For each analysis, 100 synthetic sets are generated, each with its own seed (seeds 1--100 for temperature and 200--299 for grape and gait) and each with the class counts, or for grape the (variety, response) pairs, of the training data. For a volume \(G\), the 100 set indices are shuffled with a pseudo-random generator seeded by \(G\), and replicate \(b\in\{1,\dots,5\}\) takes the \(b\)-th consecutive block of \(G\) indices. The five replicates at a given \(G\) are therefore disjoint, and at \(G=20\) they exhaust the pool.

\textbf{Allocation sweep (temperature).} For each volume \(m/n\in\{0.5,\dots,3.0\}\) and share \(a_1\in\{0,0.1,\dots,1\}\), \(m_1=\lfloor a_1m+1/2\rfloor\) synthetic Valladolid and \(m-m_1\) synthetic Madrid curves are drawn from a separate large synthetic pool, independently for each of 20 replicates, and added to the 232 real curves.

\subsection{Downstream Procedures}\label{app:downstream}

\textbf{Classification.} FACE (\texttt{refund::fpca.face}) is fitted to the training curves of each deployment, keeping the components that explain 95\% of the variance; LDA is fitted to the scores with class priors equal to the training proportions; test curves are projected onto the same basis. AUC is computed from the posterior probability of Valladolid.

\textbf{Function-to-scalar regression.} \texttt{refund::pfr} with formula \texttt{y\ \textasciitilde{}\ lf(X)} (M1) or \texttt{\spaceskip=\fontdimen2\font plus 5pt minus 3pt y \textasciitilde{} lf(X) + variety} (M2), REML smoothing, and the default basis of \texttt{lf}.

\textbf{Function-to-function regression.} FDboost with formula \texttt{\spaceskip=\fontdimen2\font plus 5pt minus 3pt moment \textasciitilde{} 1 + bsignal(angle, cycle) + bsignal(velocity, cycle) + bsignal(acceleration, cycle) + brandomc(condition)}, time formula \texttt{\spaceskip=\fontdimen2\font plus 5pt minus 3pt \textasciitilde{} bbs(cycle, df = 5)}, \texttt{\spaceskip=\fontdimen2\font plus 5pt minus 3pt boost\_control(mstop = 1000, nu = 0.1)}, and the stopping iteration chosen by \texttt{cvrisk} over the grid \(1,\dots,1000\) with five bootstrap folds. PFFR (\texttt{refund::pffr}) with a functional linear effect \texttt{ff()} of each of the three predictor curves and a functional effect of the condition, with REML smoothing. The RMSE is computed over all test curves and grid points, and the functional \(R^2\) is \(1-\mathrm{SSE}/\mathrm{SST}\) with SST taken around the mean test curve.

\section{Supplement to the Theoretical Results}\label{app:proofs}

This appendix collects further remarks on the results of \Cref{sec:theoretical_results} (C.1) and their proofs (C.2).

\subsection{Additional Remarks}\label{app:theory_remarks}

\subsubsection{\texorpdfstring{Remarks on \Cref{thm:reflection_point}}{Remarks on }}

\begin{remark}\label{rem:feasibility_window}

Under \Cref{asm:bounded_loss,asm:generational_growth,asm:level_condition}, \[R^\ast_{\mathrm{syn}}<m^\ast\psi(w)\le U .\] In particular, the three assumptions are jointly satisfiable only if \(R^\ast_{\mathrm{syn}}<U\).

\end{remark}

\begin{remark}\label{rem:generator_checkability}

Assume in addition that \(g_{m^\ast}\) is \(L(m^\ast)^{-\kappa}\)-Lipschitz with respect to \(d_{m^\ast}\) for some \(L>0\) and \(\kappa\in[0,1]\). Then:

\begin{enumerate}
\def\labelenumi{(\alph{enumi})}
\item
  \(\psi(w)\le L(m^\ast)^{-\kappa}w\);
\item
  \Cref{asm:level_condition} implies \(L(m^\ast)^{1-\kappa}w>R^\ast_{\mathrm{syn}}\);
\item
  if \(\mathrm{Gr}(m^\ast)=L(m^\ast)^{1-\kappa}w\) and \(L(m^\ast)^{1-\kappa}w>R^\ast_{\mathrm{syn}}\), then \(\psi(t):=L(m^\ast)^{-\kappa}t\) satisfies \Cref{asm:generational_growth}(a) at \(m=m^\ast\) (the only instance used in the proof of \Cref{thm:reflection_point}) and \Cref{asm:level_condition}.
\end{enumerate}

\end{remark}

\subsubsection{\texorpdfstring{Remarks on \Cref{thm:accuracy_gain}}{Remarks on }}

\begin{remark}

\textbf{Example (\(\kappa=1\)).} Let \(\Theta\) be a normed space, \(\widehat\theta_m(w)=m^{-1}\sum_{i=1}^m\phi(w_i)\) with \(\phi:\mathcal Z\to\Theta\) 1-Lipschitz, and \(\theta\mapsto\ell(\theta,z)\) \(\Lambda\)-Lipschitz. Then \(|g_m(w)-g_m(w')|\le\Lambda m^{-1}d_m(w,w')\), so \Cref{asm:influence_decay} holds with \((L,\kappa)=(\Lambda,1)\). FPCA followed by LDA and penalized functional regression are approximately of this averaging form when they are well conditioned.

\end{remark}

\begin{remark}\label{rem:benchmark_rate_motivation}

\Cref{asm:benchmark_rate} says that the risk of training on \(m\) real curves falls like a power of \(m\), \(R_{\mathrm{raw}}(m)=C_\theta m^{-\alpha}\). Power laws of this kind are standard in functional regression. For the functional linear model \(Y=\int X(t)b(t)\,dt+\varepsilon\), \citet[Thm 1]{hall2007methodology} show that the best attainable error in estimating the slope \(b\) falls like \(m^{-r}\), with \[r=\frac{2\beta_{\mathrm{HH}}-1}{\mu_{\mathrm{HH}}+2\beta_{\mathrm{HH}}},\] where \(\mu_{\mathrm{HH}}\) is how fast the variances \(\theta_j\) of the principal components of \(X\) decay (\(\theta_j\asymp j^{-\mu_{\mathrm{HH}}}\)) and \(\beta_{\mathrm{HH}}\) how fast the coefficients of \(b\) on those components decay; estimators based on functional principal components attain this rate. The prediction error is at most the largest component variance times the slope error, \(\mathbb E\big[\big(\int X(\widehat b-b)\big)^2\mid\widehat b\big]\le\theta_1\int(\widehat b-b)^2\), so it falls at least as fast, and \(\alpha\ge r\) is the range to expect. The exact form \(C_\theta m^{-\alpha}\), taken from \citet{shen2024boosting}, gives the optimal volume and the gain threshold of \Cref{thm:accuracy_gain} explicit formulas. Two further facts are used: \(C_\theta=R_{\mathrm{raw}}(1)\le U\), because the loss is bounded by \(U\) (\Cref{asm:bounded_loss}); and when even the best predictor errs, as with noise in regression or overlapping classes, the risk levels off at this Bayes risk, so the assumption is applied to the risk above it.

\end{remark}

\begin{remark}

\textbf{Probability version.} Under \Cref{asm:generation_error_tail}, Parts 1--3 hold on \(\{W_1(\widetilde F,F)\le\tau_{n_g}\}\), an event of probability at least \(1-\delta_g\). The gain conclusion holds there whenever \(\tau_{n_g}\le\min(c_1,c_2)\,n^{-e_\kappa}\) (\(\kappa<1\)), respectively \(\tau_{n_g}\le(C_\theta/L)\,n^{-\alpha}\) (\(\kappa=1\)). \Cref{rem:gain_window} makes the rate requirement explicit.

\end{remark}

\begin{remark}\label{rem:gain_window}

Let \(c^\ast:=\min(c_1,c_2)\) and \(e:=e_\kappa\) if \(\kappa<1\); let \(c^\ast:=C_\theta/L\) and \(e:=\alpha\) if \(\kappa=1\).

\begin{enumerate}
\def\labelenumi{(\roman{enumi})}
\item
  For \(\kappa<1\) and \(w>0\), the first term of the Part 2 bound is at most \(\tfrac12C_\theta n^{-\alpha}\) if and only if \(w\le c_1n^{-e_\kappa}\), and the second term \(R_{\alpha,\kappa}(w)\) is at most \(\tfrac12C_\theta n^{-\alpha}\) if \(w\le c_2n^{-\alpha e_\kappa/(1+\alpha)}\).
\item
  Under \Cref{asm:generation_error_tail}, for every \(n\ge1\) there is \(n_g(n)<\infty\) with \(\tau_{n_g(n)}\le c^\ast n^{-e}\). On the event \(\{W_1(\widetilde F,F)\le\tau_{n_g(n)}\}\), of probability at least \(1-\delta_g\), \Cref{thm:accuracy_gain} yields \(R^\ast_{\mathrm{syn}}\le C_\theta n^{-\alpha}\).
\item
  Suppose \(\tau_{n_g}\le\tau_0n_g^{-\gamma}(\log n_g)^{\beta_\gamma}\) for \(n_g\ge2\), and set \(n_g=n\). The condition \(\tau_0n^{-\gamma}(\log n)^{\beta_\gamma}\le c^\ast n^{-e}\) holds for all sufficiently large \(n\) if \(\gamma>e\), or if \(\gamma=e\) and either \(\beta_\gamma<0\) or \(\beta_\gamma=0\) and \(\tau_0\le c^\ast\). It fails for all sufficiently large \(n\) if \(\gamma<e\).
\end{enumerate}

\end{remark}

By (iii), a generator trained on the same \(n\) curves can certify a gain at all large \(n\) only if its generation rate \(\gamma\) is at least \(e_\kappa\); at \(\kappa=1\) the requirement is \(\gamma\ge\alpha\). Better-trained generators are the route to this regime.

\subsubsection{\texorpdfstring{Remarks on \Cref{thm:augmentation_risk,thm:imbalanced_classification}}{Remarks on }}

\textbf{Block estimation-error indices.} The estimation error is measured separately for the synthetic block (\(n_{\mathrm{syn}}:=m\), law \(\bar P_{\mathrm{syn}}:=\widetilde F\)) and the real block (\(n_{\mathrm{real}}:=n\), law \(\bar P_{\mathrm{real}}:=F\)). Let \(\bar R(\theta):=\frac mN\mathbb E_{\widetilde F}\ell(\theta,\cdot)+\frac nN\mathbb E_F\ell(\theta,\cdot)\) be the conditional center and \(\bar\theta\) a minimizer of \(\bar R\). Put \(f_\theta:=\ell(\theta,\cdot)-\ell(\bar\theta,\cdot)\), \(\bar{\mathcal F}:=\{f_\theta:\theta\in\Theta\}\), \(\mathcal L:=\{\ell(\theta,\cdot):\theta\in\Theta\}\) and \(\mathcal G:=\{\ell(\theta,\cdot)-\ell(\theta',\cdot):\theta,\theta'\in\Theta\}\), and let \(\mathcal C_{\mathrm{real}}:=\mathcal G\), \(\mathcal C_{\mathrm{syn}}:=\bar{\mathcal F}\). Fix block shares \(\lambda_s,\lambda_r>0\) with \(\lambda_s+\lambda_r=1\) (written \(\lambda_{\mathrm{syn}},\lambda_{\mathrm{real}}\)) and \(\gamma_B\in(0,1)\). For \(\varepsilon>0\) and \(l\in\mathbb N_0\) define the slice threshold and slice variance bound \[\varepsilon_b^{(l)}:=\lambda_b\,2^l\varepsilon\,\frac N{n_b},\qquad v^{(l)}(\varepsilon):=2c_v\,2^{l+1}\varepsilon+c_{v2}\,\varepsilon,\] with \(c_v\) from \Cref{asm:variance_control_c} and \(c_{v2}\) from \Cref{thm:augmentation_risk}. With \(H_b\) the bracketing entropy in \(L^2(\bar P_b)\), let \[\mathcal S_b:=\Big\{\delta\ge0:\ \forall\varepsilon\ge\delta\ \forall l\in\mathbb N_0,\ \int_{\gamma_B\varepsilon_b^{(l)}/8}^{\sqrt{v^{(l)}(\varepsilon)}}H_b^{1/2}(u,\mathcal C_b)\,du\le\frac{\sqrt{n_b}\,\varepsilon_b^{(l)}\gamma_B^{3/2}}{210}\Big\},\qquad\delta_{n_b}^{b}:=\inf\mathcal S_b ,\] with \(\inf\emptyset:=+\infty\), and \(\delta_{m+n}:=\max(\delta^{\mathrm{real}}_n,\delta^{\mathrm{syn}}_m)\). The set \(\mathcal S_b\) is an up-set, and ``\(\varepsilon\ge\delta^b_{n_b}\)'' is read as ``\(\varepsilon\in\mathcal S_b\)''. The index \(\delta^{\mathrm{syn}}_m\) is \(Z^{(n)}\)-measurable. By \Cref{lem:envelope_entropy}, a sufficient condition for the real-block entropy inequality is the same inequality for \(\mathcal L\) with the integration limits halved and the constant \(210\) replaced by \(420\sqrt2\).

\textbf{The term \(\varepsilon_N\).} In \Cref{asm:generation_rate_form}, \(\varepsilon_N\) is the part of the generation error that does not shrink with the real sample size, such as the reconstruction error of the pretrained encoder--decoder in \citet{Tian2026}; it is zero for our generators, which are trained from scratch.

\begin{remark}

\textbf{Dependence model.} \(\bar\theta\), the slices of the proof and \(\widetilde F\) are \(Z^{(n)}\)-measurable. The two blocks are handled differently:

\begin{itemize}
\tightlist
\item
  \emph{Synthetic block.} Conditionally on \(Z^{(n)}\) its draws are independent (not identically distributed across classes). Its class \(\{\ell(\theta,\cdot)-\ell(\bar\theta,\cdot)\}\) is therefore fixed, and the conditional tail is integrated over the generation event \(S:=\bigcap_k\{W_1(\widetilde F_k,F_k)\le\tau_{k,\pi_kn}\}\).
\item
  \emph{Real block.} Its randomness is \(Z^{(n)}\) itself, so it is analysed over the non-random class \(\mathcal G\).
\end{itemize}

The two bounds are then combined by a union bound (\Cref{sec:augmentation_allocation}).

\end{remark}

\subsection{Proofs}\label{app:proof_details}

Notation is that of \Cref{sec:theoretical_results}. In addition, \(\sigma'=\sigma(1-\sigma)\) for the logistic function \(\sigma\), and \(\mathcal K=\{0,\dots,K-1\}\). Events that are not known to be measurable are assigned outer probability \(\mathbb P^\ast\); only monotonicity and countable subadditivity of outer probability are used.

\subsubsection{Auxiliary Results}

The following results are used in the proofs of \Cref{sec:theoretical_results}. Throughout, \((\mathcal Z,d)\) is a Polish space.

\begin{lemma}\label{lem:w1_tensorization}

Let \(\mu,\nu\in\mathcal P_1(\mathcal Z)\) and \(m\ge1\). Equip \(\mathcal Z^m\) with \(d_m(z,z'):=\sum_{i=1}^md(z_i,z_i')\). Then \(W_1\big(\mu^{\otimes m},\nu^{\otimes m}\big)\le m\,W_1(\mu,\nu)\).

\end{lemma}

\subsubsubsection{Proof of \Cref{lem:w1_tensorization}}

By \citep[Lemma 1(4)]{mariucci2018wasserstein} there is \(\gamma^\ast\in\Gamma(\mu,\nu)\) with \(\int d\,d\gamma^\ast = W_1(\mu,\nu)\). Define \[T:(\mathcal Z\times\mathcal Z)^m\to\mathcal Z^m\times\mathcal Z^m,\qquad T\big((z_i,z_i')_{i\le m}\big) := \big((z_i)_{i\le m},(z_i')_{i\le m}\big),\] and \(\gamma_m := T_\#(\gamma^{\ast\otimes m})\). The marginals of \(\gamma_m\) are \(\mu^{\otimes m}\) and \(\nu^{\otimes m}\), so \(\gamma_m\in\Gamma(\mu^{\otimes m},\nu^{\otimes m})\) and \[W_1\big(\mu^{\otimes m},\nu^{\otimes m}\big)\le\int d_m\,d\gamma_m = \sum_{i=1}^m\int d(z_i,z_i')\,d\gamma^{\ast\otimes m} = m\int d\,d\gamma^\ast = m\,W_1(\mu,\nu). \qquad\blacksquare\]

\begin{lemma}\label{lem:mixture_coupling}

Let \(B\subseteq\mathcal Z\) be Borel with \(\operatorname{diam}(B)\le M<\infty\). For \(k\in\mathcal K\) let \(\mu_k,\widetilde\mu_k\) be Borel probability measures concentrated on \(B\). Let \(\pi,\widetilde\pi\) be probability vectors on \(\mathcal K\), and set \(\mu:=\sum_k\pi_k\mu_k\) and \(\widetilde\mu:=\sum_k\widetilde\pi_k\widetilde\mu_k\). Then \[W_1(\widetilde\mu,\mu)\le\sum_k\min(\pi_k,\widetilde\pi_k)\,W_1(\widetilde\mu_k,\mu_k)+\frac M2\,\|\widetilde\pi-\pi\|_1 .\] In particular: (i) \(W_1(\widetilde\mu,\mu)\le\sum_k\widetilde\pi_kW_1(\widetilde\mu_k,\mu_k)+\frac M2\|\widetilde\pi-\pi\|_1\); (ii) if \(\widetilde\mu_k=\mu_k\) for all \(k\), then \(W_1(\widetilde\mu,\mu)\le\frac M2\|\widetilde\pi-\pi\|_1\).

\end{lemma}

\subsubsubsection{Proof of \Cref{lem:mixture_coupling}}

Let \(p_k := \min(\pi_k,\widetilde\pi_k)\) and \(r := 1-\sum_kp_k\). Since \(\min(a,b) = \frac12(a+b-|a-b|)\), we have \(r = \frac12\|\pi-\widetilde\pi\|_1\). Let \(\gamma_k\in\Gamma(\widetilde\mu_k,\mu_k)\) be optimal \citep[Lemma 1(4)]{mariucci2018wasserstein}.

If \(r=0\), then \(\pi=\widetilde\pi\) and \(\gamma := \sum_k\pi_k\gamma_k\in\Gamma(\widetilde\mu,\mu)\).

If \(r>0\), then \(\sum_k(\widetilde\pi_k-p_k) = \sum_k(\pi_k-p_k) = r\), so \[\widetilde\mu^{r} := r^{-1}\sum_k(\widetilde\pi_k-p_k)\,\widetilde\mu_k,\qquad \mu^{r} := r^{-1}\sum_k(\pi_k-p_k)\,\mu_k\] are probability measures concentrated on \(B\). Put \(\gamma := \sum_kp_k\gamma_k+r\,(\widetilde\mu^r\otimes\mu^r)\). Its first marginal is \(\sum_kp_k\widetilde\mu_k+\sum_k(\widetilde\pi_k-p_k)\widetilde\mu_k = \widetilde\mu\), and its second marginal is \(\mu\) by the same computation. Since \(\widetilde\mu^r\otimes\mu^r\) is concentrated on \(B\times B\), where \(d\le M\), \[W_1(\widetilde\mu,\mu)\le\int d\,d\gamma = \sum_kp_kW_1(\widetilde\mu_k,\mu_k)+r\int d\,d(\widetilde\mu^r\otimes\mu^r)\le\sum_kp_kW_1(\widetilde\mu_k,\mu_k)+\frac M2\|\widetilde\pi-\pi\|_1 .\] Statement (i) follows from \(p_k\le\widetilde\pi_k\), and (ii) from \(W_1(\mu_k,\mu_k)=0\). \(\blacksquare\)

\begin{corollary}\label{cor:generation_mixture_bound}

Under \Cref{asm:bounded_support_c}, on the generation event \(S\), \[W_1(\widetilde F,F)\le\sum_ka_k\tau_{k,\pi_kn}+\frac M2\|a-\pi\|_1=\frac NmG_\lambda+A_\lambda .\]

\end{corollary}

\subsubsubsection{Proof of \Cref{cor:generation_mixture_bound}}

Apply \Cref{lem:mixture_coupling}(i) with \(\widetilde\mu_k := \widetilde F_k\), \(\mu_k := F_k\), \(\widetilde\pi := a\) and \(\pi := \pi\). On \(S\), \(W_1(\widetilde F_k,F_k)\le\tau_{k,\pi_kn}\). The equality is the definition of \(G_\lambda\) and \(A_\lambda\). \(\blacksquare\)

\begin{lemma}\label{lem:variance_correction}

Let \(g:\mathcal Z\to\mathbb R\) be \(L_g\)-Lipschitz with respect to \(d_Z\), with \(|g|\le U\), and let \(\mu,\widetilde\mu\in\mathcal P_1(\mathcal Z)\). Then \(\big|\operatorname{Var}_{\widetilde\mu}(g)-\operatorname{Var}_\mu(g)\big|\le4UL_g\,W_1(\mu,\widetilde\mu)\).

\end{lemma}

\subsubsubsection{Proof of \Cref{lem:variance_correction}}

For \(z,z'\in\mathcal Z\), \(|g^2(z)-g^2(z')| = |g(z)+g(z')|\,|g(z)-g(z')|\le2UL_g\,d_Z(z,z')\). By Kantorovich--Rubinstein duality, \[|\mathbb E_{\widetilde\mu}g^2-\mathbb E_\mu g^2|\le2UL_gW_1(\mu,\widetilde\mu)\quad\text{and}\quad|\mathbb E_{\widetilde\mu}g-\mathbb E_\mu g|\le L_gW_1(\mu,\widetilde\mu).\] Since \(|\mathbb E_{\widetilde\mu}g+\mathbb E_\mu g|\le2U\), \[|(\mathbb E_{\widetilde\mu}g)^2-(\mathbb E_\mu g)^2|\le2UL_gW_1(\mu,\widetilde\mu).\] Subtracting the two identities \(\operatorname{Var}(g) = \mathbb Eg^2-(\mathbb Eg)^2\) and applying the triangle inequality gives the claim. \(\blacksquare\)

\begin{lemma}\label{lem:bounded_bernstein}

Let \(X\) be a real random variable with \(|X|\le U\). Then \(\mathbb E|X-\mathbb EX|^k\le\frac12\,k!\operatorname{Var}(X)\,U^{k-2}\) for every \(k\ge2\).

\end{lemma}

\subsubsubsection{Proof of \Cref{lem:bounded_bernstein}}

Since \(|X-\mathbb EX|\le2U\), \(\mathbb E|X-\mathbb EX|^k\le(2U)^{k-2}\operatorname{Var}(X)\). It remains to show \(2^{k-2}\le\frac12k!\), i.e.~\(2^{k-1}\le k!\). This holds for \(k=2\), and if \(2^{k-1}\le k!\) then \(2^k\le2\cdot k!\le(k+1)!\). \(\blacksquare\)

\begin{lemma}\label{lem:envelope_entropy}

For every \(u>0\) and every probability measure \(P\), \(H(u,\mathcal G,L^2(P))\le2H(u/2,\mathcal L,L^2(P))\). Consequently, for \(0<a<b\), \[\int_a^bH^{1/2}(u,\mathcal G)\,du\le2\sqrt2\int_{a/2}^{b/2}H^{1/2}(t,\mathcal L)\,dt .\]

\end{lemma}

\subsubsubsection{Proof of \Cref{lem:envelope_entropy}}

Let \(\{[l_i,u_i]\}_{i\le N_0}\) be \(u/2\)-brackets covering \(\mathcal L\), i.e.~\(\|u_i-l_i\|_{L^2(P)}\le u/2\). If \(\ell(\theta,\cdot)\in[l_i,u_i]\) and \(\ell(\theta',\cdot)\in[l_j,u_j]\), then \(l_i-u_j\le\ell(\theta,\cdot)-\ell(\theta',\cdot)\le u_i-l_j\), and by Minkowski's inequality \[\|(u_i-l_j)-(l_i-u_j)\|_{L^2(P)}\le\|u_i-l_i\|+\|u_j-l_j\|\le u .\] Hence the \(N_0^2\) brackets \([l_i-u_j,\,u_i-l_j]\) cover \(\mathcal G\), which gives the entropy bound. The integral bound follows by the substitution \(u=2t\). \(\blacksquare\)

\begin{lemma}\label{lem:block_empirical_process}

Let \(Z_1,\dots,Z_{n_b}\) be independent, \(P_j:=\mathcal L(Z_j)\) and \(\bar P:=n_b^{-1}\sum_jP_j\). Let \(\mathcal F\) be a class of measurable functions such that, for some \(U>0\), \(v_j\ge0\) and every \(f\in\mathcal F\), \[\mathbb E\big|f(Z_j)-\mathbb Ef(Z_j)\big|^k\le\tfrac12\,k!\,v_jU^{k-2}\qquad(1\le j\le n_b,\ k\ge2),\] and let \(v\ge n_b^{-1}\sum_jv_j\). Let \(H(u,\mathcal F)\) be the \(L^2(\bar P)\)-bracketing entropy of \(\mathcal F\). Fix \(\gamma\in(0,1)\) and \(\varepsilon>0\) with \[\text{(i)}\ \ \varepsilon\le\frac{\gamma v}{4U},\qquad\qquad\text{(ii)}\ \ \int_{\gamma\varepsilon/8}^{\sqrt v}H^{1/2}(u,\mathcal F)\,du\le\frac{\sqrt{n_b}\,\varepsilon\,\gamma^{3/2}}{210}.\] Then, with \(\mathbb P^\ast\) the outer probability, \[\mathbb P^\ast\Big(\sup_{f\in\mathcal F}\frac1{n_b}\sum_{j=1}^{n_b}\big(f(Z_j)-\mathbb Ef(Z_j)\big)\ge\varepsilon\Big)\le3\exp\Big(-(1-\gamma)\frac{n_b\varepsilon^2}{2[4v+\varepsilon U/3]}\Big).\]

\end{lemma}

\Cref{lem:block_empirical_process} is \citet[Appendix B, Lemma 2]{Tian2026}, which restates \citet[Lemma 11]{tian2024knowledge} for independent, not necessarily identically distributed variables, with \((m,\varepsilon_m):=(n_b,\varepsilon)\). For i.i.d. variables it coincides with \citet[Lemma 11]{tian2024knowledge} under \(M:=\sqrt{n_b}\,\varepsilon\) and \(v^2:=v\). The norm defining \(H\) is not specified in the source; we adopt \(L^2(\bar P)\). Applying the lemma to \(-\mathcal F\), which satisfies the same hypotheses, bounds the lower tail by the same quantity. It is the only result used without proof.

\subsubsection{\texorpdfstring{Proof of \Cref{thm:reflection_point} and of \Cref{rem:feasibility_window,rem:generator_checkability}}{Proof of  and of }}

\subsubsubsection{Proof of \Cref{thm:reflection_point}}

Fix \(Z^{(n)}\) and write \(w := W_1(\widetilde F,F)\) and \(I := R^\ast_{\mathrm{syn}}\). By \Cref{asm:bounded_loss}, \(R_{\mathrm{raw}}(m),R_{\mathrm{syn}}(m)\in[0,U]\) for all \(m\), so \(I\in[0,U]\). If \(0\in\mathbb N\) were admitted as a value of \(m^\ast\), \Cref{asm:level_condition} would read \(0>I\), contradicting \(I\ge0\); hence \(m^\ast\ge1\).

\emph{Step 1.} For every \(m\ge m^\ast\), \[R_{\mathrm{syn}}(m)\ge m^\ast\psi(w)>I. \tag{A.1}\] Indeed, \(R_{\mathrm{syn}}(m) = R_{\mathrm{raw}}(m)+\mathrm{Gr}(m)\ge\mathrm{Gr}(m)\). Next, \(\mathrm{Gr}(m)\ge\mathrm{Gr}(m^\ast)\) for all \(m\ge m^\ast\) (by \Cref{asm:generational_growth}(b) for \(m>m^\ast\), trivially for \(m=m^\ast\)). Then \(\mathrm{Gr}(m^\ast)\ge m^\ast\psi(w)\) by \Cref{asm:generational_growth}(a) at \(m=m^\ast\). Finally \(m^\ast\psi(w)>I\) by \Cref{asm:level_condition}.

\emph{Step 2 (\(m^\ast\ge2\)).} If \(m^\ast=1\), then (A.1) holds for all \(m\ge1\), so \(\psi(w)\) is a lower bound of \(\{R_{\mathrm{syn}}(m):m\ge1\}\) and \(I\ge\psi(w)>I\), a contradiction.

\emph{Step 3 (i$'$).} By Step 2, \(\{1,\dots,m^\ast-1\}\neq\emptyset\). Let \(r^\ast := \min_{1\le m\le m^\ast-1}R_{\mathrm{syn}}(m)\) and \(s^\ast := \inf_{m\ge m^\ast}R_{\mathrm{syn}}(m)\), so that \(I = \min(r^\ast,s^\ast)\). By (A.1), \(s^\ast\ge m^\ast\psi(w)>I\), hence \(I\neq s^\ast\) and \(I = r^\ast\). Any \(m_0\in\operatorname{argmin}_{1\le m\le m^\ast-1}R_{\mathrm{syn}}(m)\) satisfies (i$'$).

\emph{Step 4 (ii$'$).} For \(m\ge m^\ast\), (A.1) and Step 3 give \(R_{\mathrm{syn}}(m)\ge m^\ast\psi(w)>I = R_{\mathrm{syn}}(m_0)\). \(\blacksquare\)

\subsubsubsection{Proof of \Cref{rem:feasibility_window}}

The left inequality is \Cref{asm:level_condition}. For the right, \(m^\ast\psi(w)\le\mathrm{Gr}(m^\ast) = R_{\mathrm{syn}}(m^\ast)-R_{\mathrm{raw}}(m^\ast)\le R_{\mathrm{syn}}(m^\ast)\le U\). \(\blacksquare\)

\subsubsubsection{Proof of \Cref{rem:generator_checkability}}

\Cref{lem:estimator_stability} with constant \(L(m^\ast)^{-\kappa}\) gives \(\mathrm{Gr}(m^\ast)\le L(m^\ast)^{1-\kappa}w\). Combined with \(m^\ast\psi(w)\le\mathrm{Gr}(m^\ast)\), this gives (a). Then (b) follows from \(R^\ast_{\mathrm{syn}}<m^\ast\psi(w)\le L(m^\ast)^{1-\kappa}w\). For (c), \(m^\ast\psi(w) = L(m^\ast)^{1-\kappa}w = \mathrm{Gr}(m^\ast)\). \(\blacksquare\)

\subsubsection{\texorpdfstring{Proof of \Cref{thm:accuracy_gain}, \Cref{lem:estimator_stability} and \Cref{rem:gain_window}}{Proof of ,  and }}

\subsubsubsection{Proof of \Cref{lem:estimator_stability}}

The test point \(Z\sim F\) is independent of \(\big(Z^{(n)},\widetilde Z^{(m)},Z^{(m)}\big)\), and \(0\le\ell\le U\). By Fubini's theorem, \[R_{\mathrm{syn}}(m) = \int_{\mathcal Z^m}g_m\,d\widetilde F^{\otimes m},\qquad R_{\mathrm{raw}}(m) = \int_{\mathcal Z^m}g_m\,dF^{\otimes m}.\] The space \((\mathcal Z^m,d_m)\) is Polish and \(\widetilde F^{\otimes m},F^{\otimes m}\in\mathcal P_1(\mathcal Z^m)\). The Kantorovich--Rubinstein duality \citep[Thm 1.4.2]{panaretos2019} applied to the 1-Lipschitz function \(g_m/L\), followed by \Cref{lem:w1_tensorization}, gives \[|\mathrm{Gr}(m)| = L\,\Big|\int\frac{g_m}{L}\,d\big(\widetilde F^{\otimes m}-F^{\otimes m}\big)\Big|\le L\,W_1\big(\widetilde F^{\otimes m},F^{\otimes m}\big)\le L\,m\,W_1(\widetilde F,F). \qquad\blacksquare\]

\subsubsubsection{Proof of \Cref{thm:accuracy_gain}}

Fix \(Z^{(n)}\) and assume \Cref{asm:bounded_loss,asm:influence_decay,asm:benchmark_rate}. Write \(e_\kappa := 1+\alpha-\kappa\).

\emph{Step 1 (Part 1).} Fix \(m\ge1\). By \Cref{asm:influence_decay}, \(g_m\) is \(Lm^{-\kappa}\)-Lipschitz, so \Cref{lem:estimator_stability} gives \(|\mathrm{Gr}(m)|\le Lm^{-\kappa}\cdot m\cdot w = Lm^{1-\kappa}w\). Since \(R_{\mathrm{syn}}(m),R_{\mathrm{raw}}(m)\in[0,U]\), also \(|\mathrm{Gr}(m)|\le U\). The two-sided bound follows from \(R_{\mathrm{syn}}(m) = C_\theta m^{-\alpha}+\mathrm{Gr}(m)\) (\Cref{asm:benchmark_rate}).

\emph{Step 2 (Part 2, \(\kappa<1\)).}

\emph{Case \(w=0\).} \(W_1\) is a metric on \(\mathcal P_1(\mathcal Z)\) \citep[Lemma 1(1)]{mariucci2018wasserstein}, so \(\widetilde F=F\). Then \(R_{\mathrm{syn}}(m) = R_{\mathrm{raw}}(m) = C_\theta m^{-\alpha}\) for all \(m\), and \(R^\ast_{\mathrm{syn}}=0\).

\emph{Case \(w>0\).} Then \(s_0\in(0,\infty)\). Let \(m_0 := \lceil s_0\rceil\ge1\), so \(s_0\le m_0<s_0+1\). By Part 1, and since \(t\mapsto t^{-\alpha}\) is decreasing and \(t\mapsto t^{1-\kappa}\) is nondecreasing, \[R_{\mathrm{syn}}(m_0)\le C_\theta s_0^{-\alpha}+Lw\,(s_0+1)^{1-\kappa}.\] The map \(t\mapsto t^{1-\kappa}\) is concave and differentiable on \((0,\infty)\), so \((s_0+1)^{1-\kappa}\le s_0^{1-\kappa}+(1-\kappa)s_0^{-\kappa}\). Hence \[R_{\mathrm{syn}}(m_0)\le C_\theta s_0^{-\alpha}+Lw\,s_0^{1-\kappa}+(1-\kappa)Lw\,s_0^{-\kappa}. \tag{A.2}\] Substituting \(s_0^{e_\kappa} = \alpha C_\theta/((1-\kappa)Lw)\) gives \[\begin{aligned}
C_\theta s_0^{-\alpha} &= \alpha^{-\alpha/e_\kappa}(1-\kappa)^{\alpha/e_\kappa}\,C_\theta^{(1-\kappa)/e_\kappa}L^{\alpha/e_\kappa}w^{\alpha/e_\kappa},\\
Lw\,s_0^{1-\kappa} &= \alpha^{(1-\kappa)/e_\kappa}(1-\kappa)^{-(1-\kappa)/e_\kappa}\,C_\theta^{(1-\kappa)/e_\kappa}L^{\alpha/e_\kappa}w^{\alpha/e_\kappa},\\
(1-\kappa)Lw\,s_0^{-\kappa} &= D_{\alpha,\kappa}\,w^{(1+\alpha)/e_\kappa} = R_{\alpha,\kappa}(w),
\end{aligned}\] where we used \(1-(1-\kappa)/e_\kappa = \alpha/e_\kappa\) and \(1+\kappa/e_\kappa = (1+\alpha)/e_\kappa\). The first two terms sum to \(K_{\alpha,\kappa}C_\theta^{(1-\kappa)/e_\kappa}L^{\alpha/e_\kappa}w^{\alpha/e_\kappa}\), and \(R^\ast_{\mathrm{syn}}\le R_{\mathrm{syn}}(m_0)\).

\emph{Step 3 (Part 2, \(\kappa=1\)).} By Part 1, \(R_{\mathrm{syn}}(m)\le C_\theta m^{-\alpha}+Lw\) for all \(m\). Since \(C_\theta m^{-\alpha}\to0\), \(R^\ast_{\mathrm{syn}}\le Lw\). If \(w=0\), this and \(R^\ast_{\mathrm{syn}}\ge0\) give \(R^\ast_{\mathrm{syn}}=0\).

\emph{Step 4 (Part 3, \(\kappa<1\)).} Let \(c := \min(c_1,c_2)\) and assume \(w\le c\,n^{-e_\kappa}\).

\emph{Case \(w=0\).} By Part 1, \(R^\ast_{\mathrm{syn}}\le R_{\mathrm{syn}}(n)\le C_\theta n^{-\alpha}\).

\emph{Case \(w>0\).} The bound \(w\le c_1n^{-e_\kappa}\) gives \(w^{\alpha/e_\kappa}\le c_1^{\alpha/e_\kappa}n^{-\alpha}\), with \(c_1^{\alpha/e_\kappa} = \tfrac12\,C_\theta^{\alpha/e_\kappa}K_{\alpha,\kappa}^{-1}L^{-\alpha/e_\kappa}\). Hence \[K_{\alpha,\kappa}C_\theta^{(1-\kappa)/e_\kappa}L^{\alpha/e_\kappa}w^{\alpha/e_\kappa}\le\tfrac12C_\theta n^{-\alpha}. \tag{A.3}\] Since \(n\ge1\) and \(e_\kappa\ge\alpha e_\kappa/(1+\alpha)\), we have \(w\le c_2n^{-e_\kappa}\le c_2n^{-\alpha e_\kappa/(1+\alpha)}\). Raising to the power \((1+\alpha)/e_\kappa\) gives \(w^{(1+\alpha)/e_\kappa}\le\frac{C_\theta}{2D_{\alpha,\kappa}}n^{-\alpha}\), that is, \[R_{\alpha,\kappa}(w)\le\tfrac12C_\theta n^{-\alpha}. \tag{A.4}\] Part 2 with (A.3) and (A.4) gives \(R^\ast_{\mathrm{syn}}\le C_\theta n^{-\alpha} = R_{\mathrm{raw}}(n)\).

\emph{Step 5 (Part 3, \(\kappa=1\)).} By Step 3, \(R^\ast_{\mathrm{syn}}\le Lw\le C_\theta n^{-\alpha}\). \(\blacksquare\)

\subsubsubsection{Proof of \Cref{rem:gain_window}}

\emph{(i)} Let \(\kappa<1\) and \(w>0\). Since \(1-(1-\kappa)/e_\kappa=\alpha/e_\kappa\), the inequality \(K_{\alpha,\kappa}C_\theta^{(1-\kappa)/e_\kappa}L^{\alpha/e_\kappa}w^{\alpha/e_\kappa}\le\tfrac12C_\theta n^{-\alpha}\) is equivalent to \(w^{\alpha/e_\kappa}\le\tfrac12K_{\alpha,\kappa}^{-1}C_\theta^{\alpha/e_\kappa}L^{-\alpha/e_\kappa}n^{-\alpha}\). Both sides are positive and \(t\mapsto t^{e_\kappa/\alpha}\) is increasing, so this is equivalent to \(w\le2^{-e_\kappa/\alpha}K_{\alpha,\kappa}^{-e_\kappa/\alpha}C_\theta L^{-1}n^{-e_\kappa}=c_1n^{-e_\kappa}\). If \(w\le c_2n^{-\alpha e_\kappa/(1+\alpha)}\), then raising to the power \((1+\alpha)/e_\kappa\) gives \(w^{(1+\alpha)/e_\kappa}\le c_2^{(1+\alpha)/e_\kappa}n^{-\alpha}=\frac{C_\theta}{2D_{\alpha,\kappa}}n^{-\alpha}\), that is, \(R_{\alpha,\kappa}(w)\le\tfrac12C_\theta n^{-\alpha}\).

\emph{(ii)} Fix \(n\ge1\). Since \(c^\ast n^{-e}>0\) and \(\tau_{n_g}\to0\), there is \(n_g(n)<\infty\) with \(\tau_{n_g(n)}\le c^\ast n^{-e}\). On the event \(\{W_1(\widetilde F,F)\le\tau_{n_g(n)}\}\) we have \(w\le c^\ast n^{-e}\), which is the hypothesis of \Cref{thm:accuracy_gain}, Part 3, for \(\kappa<1\) (\(c^\ast=\min(c_1,c_2)\), \(e=e_\kappa\)) and for \(\kappa=1\) (\(c^\ast=C_\theta/L\), \(e=\alpha\)). Hence \(R^\ast_{\mathrm{syn}}\le C_\theta n^{-\alpha}\) on this event, whose probability is at least \(1-\delta_g\) by \Cref{asm:generation_error_tail}.

\emph{(iii)} Dividing by \(n^{-\gamma}>0\), the condition \(\tau_0n^{-\gamma}(\log n)^{\beta_\gamma}\le c^\ast n^{-e}\) is equivalent to \(\tau_0(\log n)^{\beta_\gamma}\le c^\ast n^{\gamma-e}\). If \(\gamma>e\), the right-hand side grows polynomially and the left-hand side at most polylogarithmically, so the condition holds for all large \(n\). If \(\gamma=e\), the right-hand side equals \(c^\ast\); the left-hand side tends to \(0\) if \(\beta_\gamma<0\) and equals \(\tau_0\) if \(\beta_\gamma=0\), which gives the two stated cases. If \(\gamma<e\), the right-hand side tends to \(0\) as \(n^{\gamma-e}\), while the left-hand side is bounded below by \(\tau_0\) if \(\beta_\gamma\ge0\) and decays only polylogarithmically if \(\beta_\gamma<0\); in both cases the ratio of the left-hand side to the right-hand side tends to infinity, so the condition fails for all large \(n\). \(\blacksquare\)

\subsubsection{\texorpdfstring{Proof of \Cref{thm:augmentation_risk}}{Proof of }}

\subsubsubsection{Notation}

\begin{itemize}
\tightlist
\item
  \emph{Synthetic sample.} Conditionally on \(Z^{(n)}\), the variables \(\widetilde Z_{kj}\) (\(k\in\mathcal K\), \(1\le j\le m_k\)) are independent with \(\widetilde Z_{kj}\sim\widetilde F_k\).
\item
  \emph{Blocks.} Put \(n_{\mathrm{syn}} := m\), \(n_{\mathrm{real}} := n\), \(\bar P_{\mathrm{syn}} := \widetilde F = \sum_ka_k\widetilde F_k\) and \(\bar P_{\mathrm{real}} := F\).
\item
  \emph{Centering.} Let \(\bar\theta\in\operatorname{argmin}_\Theta\bar R\) and \(\bar\rho(\theta,\theta') := \bar R(\theta)-\bar R(\theta')\).
\item
  \emph{Function classes.} \(f_\theta := \ell(\theta,\cdot)-\ell(\bar\theta,\cdot)\) and \(g_\theta := \ell(\theta,\cdot)-\ell(\theta_0,\cdot)\); \[\bar{\mathcal F} := \{f_\theta:\theta\in\Theta\},\qquad \mathcal L := \{\ell(\theta,\cdot):\theta\in\Theta\},\qquad \mathcal G := \{\ell(\theta,\cdot)-\ell(\theta',\cdot):\theta,\theta'\in\Theta\}.\]
\item
  \emph{Generation event.} \(S := \bigcap_{k\in\mathcal K}\{W_1(\widetilde F_k,F_k)\le\tau_{k,\pi_kn}\}\).
\end{itemize}

The random objects \(\widetilde F\), \(\bar R\), \(\bar\theta\), \(\bar{\mathcal F}\) and \(S\) are \(Z^{(n)}\)-measurable.

The slice thresholds \(\varepsilon_b^{(l)}\), the slice variance bounds \(v^{(l)}(\varepsilon)\) and the block indices \(\delta^b_{n_b}\) are those of \Cref{sec:augmentation_allocation}.

\subsubsubsection{Proof of \Cref{thm:augmentation_risk}}

Fix \(\varepsilon\) satisfying the hypotheses of \Cref{thm:augmentation_risk}, and set \(\Delta := \beta G_\lambda+UD_\lambda\).

\emph{Step 1 (risk discrepancy).} For every \(\theta\in\Theta\), by \Cref{asm:partition_validity_c}, \[R_Q(\theta) = \sum_kq_k\,\mathbb E_{F_k}\ell(\theta,\cdot),\qquad \bar R(\theta) = \sum_k\Big[\frac mNa_k\,\mathbb E_{\widetilde F_k}\ell(\theta,\cdot)+\frac nN\pi_k\,\mathbb E_{F_k}\ell(\theta,\cdot)\Big].\] With \(\widetilde a_k = \frac mNa_k+\frac nN\pi_k\) this gives \[R_Q(\theta)-\bar R(\theta) = \sum_k(q_k-\widetilde a_k)\,\mathbb E_{F_k}\ell(\theta,\cdot)-\frac mN\sum_ka_k\big(\mathbb E_{\widetilde F_k}\ell(\theta,\cdot)-\mathbb E_{F_k}\ell(\theta,\cdot)\big).\] By \Cref{asm:loss_regularity_c}, \(0\le\mathbb E_{F_k}\ell\le U\) and, by Kantorovich--Rubinstein duality, \(|\mathbb E_{\widetilde F_k}\ell(\theta,\cdot)-\mathbb E_{F_k}\ell(\theta,\cdot)|\le\beta W_1(\widetilde F_k,F_k)\). Hence \[|\bar R(\theta)-R_Q(\theta)|\le UD_\lambda+\beta\frac mN\sum_ka_kW_1(\widetilde F_k,F_k),\] and on \(S\), \[\sup_{\theta\in\Theta}|\bar R(\theta)-R_Q(\theta)|\le\Delta\le(\beta+U)\varepsilon, \tag{A.5}\] the last inequality by \(\varepsilon\ge\max\{G_\lambda,D_\lambda\}\).

\emph{Step 2 (reduction to \(\bar\rho\)).} On \(S\), for every \(\theta\in\Theta\), \[\rho(\theta,\theta_0) = [R_Q(\theta)-\bar R(\theta)]+\bar\rho(\theta,\bar\theta)+[\bar R(\bar\theta)-\bar R(\theta_Q)]+[\bar R(\theta_Q)-R_Q(\theta_Q)]+\rho(\theta_Q,\theta_0).\] The third bracket is \(\le0\) because \(\bar\theta\) minimizes \(\bar R\). By (A.5) and \(\rho(\theta_Q,\theta_0)\le\varepsilon\), \[\rho(\theta,\theta_0)\le\bar\rho(\theta,\bar\theta)+2\Delta+\rho(\theta_Q,\theta_0)\le\bar\rho(\theta,\bar\theta)+(2\beta+2U+1)\varepsilon. \tag{A.6}\] Two consequences follow. First, with \(\theta=\bar\theta\), \[\rho(\bar\theta,\theta_0)\le(2\beta+2U+1)\varepsilon. \tag{A.7}\] Second, with \(c_e = 2+2U+2\beta\), \[\{\rho(\widehat\theta_\lambda,\theta_0)\ge c_e\varepsilon\}\cap S\subseteq\{\bar\rho(\widehat\theta_\lambda,\bar\theta)\ge\varepsilon\}\cap S. \tag{A.8}\]

\emph{Step 3 (slicing and block split).} Define \[\hat\nu(\theta) := [L_{m,n}(\theta)-\bar R(\theta)]-[L_{m,n}(\bar\theta)-\bar R(\bar\theta)] = \frac mN\nu_{\mathrm{syn}}(\theta)+\frac nN\nu_{\mathrm{real}}(\theta),\] \[\nu_{\mathrm{syn}}(\theta) := \frac1m\sum_{k\in\mathcal K}\sum_{j=1}^{m_k}\big(f_\theta(\widetilde Z_{kj})-\mathbb E_{\widetilde F_k}f_\theta\big),\qquad \nu_{\mathrm{real}}(\theta) := \frac1n\sum_{i=1}^n\big(f_\theta(Z_i)-\mathbb E_Ff_\theta\big).\] The decomposition holds because \(\sum_k(m_k/m)\,\mathbb E_{\widetilde F_k}f_\theta = \mathbb E_{\widetilde F}f_\theta\). Since \(L_{m,n}(\widehat\theta_\lambda)\le L_{m,n}(\bar\theta)\) and \(\hat\nu(\bar\theta)=0\), \[\bar\rho(\widehat\theta_\lambda,\bar\theta)\le-\hat\nu(\widehat\theta_\lambda). \tag{A.9}\] Let \(A_l := \{\theta\in\Theta:2^l\varepsilon\le\bar\rho(\theta,\bar\theta)<2^{l+1}\varepsilon\}\) for \(l\in\mathbb N_0\). Since \(\bar R\in[0,U]\), \(A_l=\emptyset\) whenever \(2^l\varepsilon>U\). If \(\bar\rho(\widehat\theta_\lambda,\bar\theta)\ge\varepsilon\), then \(\widehat\theta_\lambda\in A_l\) for \(l := \lfloor\log_2(\bar\rho(\widehat\theta_\lambda,\bar\theta)/\varepsilon)\rfloor\), and by (A.9), \(-\hat\nu(\widehat\theta_\lambda)\ge2^l\varepsilon\). If \(-\hat\nu(\theta)\ge2^l\varepsilon\), then \(-\frac mN\nu_{\mathrm{syn}}(\theta)\ge\lambda_s2^l\varepsilon\) or \(-\frac nN\nu_{\mathrm{real}}(\theta)\ge\lambda_r2^l\varepsilon\). Therefore \[\{\bar\rho(\widehat\theta_\lambda,\bar\theta)\ge\varepsilon\}\subseteq\bigcup_{l\ge0}\Big(E^{\mathrm{syn}}_l\cup E^{\mathrm{real}}_l\Big),\qquad E^b_l := \Big\{\sup_{\theta\in A_l}\big(-\nu_b(\theta)\big)\ge\varepsilon_b^{(l)}\Big\}. \tag{A.10}\]

\emph{Step 4 (variance bounds on the slices).} Fix \(l\) and \(\theta\in A_l\), and work on \(S\). By (A.6) and (A.7), \[\rho(\theta,\theta_0)+\rho(\bar\theta,\theta_0)\le2^{l+1}\varepsilon+(4\beta+4U+2)\varepsilon .\] Since \(f_\theta = g_\theta-g_{\bar\theta}\) and \(\operatorname{Var}(X-Y)\le2\operatorname{Var}X+2\operatorname{Var}Y\), \Cref{asm:variance_control_c} gives \[\operatorname{Var}_F(f_\theta)\le2c_v\big(\rho(\theta,\theta_0)+\rho(\bar\theta,\theta_0)\big)\le2c_v2^{l+1}\varepsilon+2c_v(4\beta+4U+2)\varepsilon. \tag{A.11}\] The function \(f_\theta\) is \(2\beta\)-Lipschitz with \(|f_\theta|\le U\). \Cref{lem:variance_correction} with \(L_g=2\beta\), followed by \Cref{cor:generation_mixture_bound} and \(\max\{G_\lambda,A_\lambda\}\le\varepsilon\), gives \[\operatorname{Var}_{\widetilde F}(f_\theta)\le\operatorname{Var}_F(f_\theta)+8U\beta\,W_1(\widetilde F,F)\le\operatorname{Var}_F(f_\theta)+8U\beta\Big(\frac Nm+1\Big)\varepsilon .\] With (A.11) and \(c_{v2} = 2c_v(4\beta+4U+2)+8U\beta(N/m+1)\), on \(S\) both blocks satisfy \[\operatorname{Var}_{\bar P_b}(f_\theta)\le v^{(l)}(\varepsilon)\qquad(\theta\in A_l,\ b\in\{\mathrm{syn},\mathrm{real}\}). \tag{A.12}\] For the synthetic block, the law of total variance gives \(\sum_ka_k\operatorname{Var}_{\widetilde F_k}(f_\theta)\le\operatorname{Var}_{\widetilde F}(f_\theta)\le v^{(l)}(\varepsilon)\).

\emph{Step 5 (per-block tails).} Two conditions of \Cref{lem:block_empirical_process} are verified first.

\begin{itemize}
\tightlist
\item
  \emph{Condition (i).} Since \(v^{(l)}(\varepsilon)\ge4c_v2^l\varepsilon\), the inequality \(\varepsilon_b^{(l)}\le\gamma_Bv^{(l)}(\varepsilon)/(4U)\) holds whenever \(\lambda_bN/n_b\le\gamma_Bc_v/U\). For \(b\in\{\mathrm{syn},\mathrm{real}\}\) this is the Bernstein-range condition of \Cref{thm:augmentation_risk}.
\item
  \emph{Exponent.} Since \(v^{(l)}(\varepsilon)/\varepsilon_b^{(l)} = \frac{n_b}{\lambda_bN}(4c_v+2^{-l}c_{v2})\le\frac{n_b}{\lambda_bN}(4c_v+c_{v2})\), we have \(4v^{(l)}(\varepsilon)+\varepsilon_b^{(l)}U/3\le c_{d,b}\,\varepsilon_b^{(l)}\). Hence \[\frac{n_b(\varepsilon_b^{(l)})^2}{2[4v^{(l)}(\varepsilon)+\varepsilon_b^{(l)}U/3]}\ge\frac{n_b\varepsilon_b^{(l)}}{2c_{d,b}} = \frac{\lambda_b2^l\varepsilon N}{2c_{d,b}}\ge\frac{2^l\varepsilon N}{c_{d'}} . \tag{A.13}\]
\end{itemize}

\emph{Real block.} Let \(\mathcal G^{(l)} := \{g\in\mathcal G:\operatorname{Var}_F(g)\le v^{(l)}(\varepsilon)\}\). This class is non-random, and it is symmetric: \(g\in\mathcal G^{(l)}\Rightarrow-g\in\mathcal G^{(l)}\). On \(S\), (A.12) gives \(f_\theta\in\mathcal G^{(l)}\) for all \(\theta\in A_l\), hence \[E^{\mathrm{real}}_l\cap S\subseteq\Big\{\sup_{g\in\mathcal G^{(l)}}\frac1n\sum_{i=1}^n\big(g(Z_i)-\mathbb E_Fg\big)\ge\varepsilon_{\mathrm{real}}^{(l)}\Big\}.\] Apply \Cref{lem:block_empirical_process} to \(\mathcal G^{(l)}\) with \(Z_1,\dots,Z_n\) i.i.d. \(F\) and \(v := v^{(l)}(\varepsilon)\):

\begin{itemize}
\tightlist
\item
  the moment condition holds with \(v_j := \operatorname{Var}_F(g)\le v\) by \Cref{lem:bounded_bernstein}, since \(|g|\le U\) for \(g\in\mathcal G\);
\item
  condition (i) is verified above;
\item
  condition (ii) holds because \(H(u,\mathcal G^{(l)})\le H(u,\mathcal G)\) and \(\varepsilon\ge\delta_n^{\mathrm{real}}\).
\end{itemize}

With (A.13), \[\mathbb P^\ast\big(E^{\mathrm{real}}_l\cap S\big)\le3\exp\Big(-\frac{(1-\gamma_B)2^l\varepsilon N}{c_{d'}}\Big). \tag{A.14}\]

\emph{Synthetic block.} Condition on \(Z^{(n)}\) and assume \(S\). The class \(\bar{\mathcal F}_l := \{f_\theta:\theta\in A_l\}\subseteq\bar{\mathcal F}\) is then fixed, and \(\widetilde Z_{kj}\) are independent with \(\widetilde Z_{kj}\sim\widetilde F_k\). Apply \Cref{lem:block_empirical_process} to \(-\bar{\mathcal F}_l\) with \(v := v^{(l)}(\varepsilon)\) and \(v_j := \operatorname{Var}_{\widetilde F_{k(j)}}(f_\theta)\), where \(k(j)\) is the class of the \(j\)-th synthetic draw:

\begin{itemize}
\tightlist
\item
  the moment condition holds by \Cref{lem:bounded_bernstein};
\item
  \(v\ge m^{-1}\sum_jv_j\) holds by (A.12);
\item
  condition (ii) holds because \(\varepsilon\ge\delta^{\mathrm{syn}}_m\) almost surely.
\end{itemize}

This gives \(\mathbb P^\ast\big(E^{\mathrm{syn}}_l\mid Z^{(n)}\big)\mathbf 1_S\le3\exp\big(-(1-\gamma_B)2^l\varepsilon N/c_{d'}\big)\). Integrating over \(Z^{(n)}\), \[\mathbb P^\ast\big(E^{\mathrm{syn}}_l\cap S\big)\le3\exp\Big(-\frac{(1-\gamma_B)2^l\varepsilon N}{c_{d'}}\Big). \tag{A.15}\]

\emph{Step 6 (assembly).} By (A.8), (A.10), (A.14), (A.15) and countable subadditivity, with \(C := (1-\gamma_B)N\varepsilon/c_{d'}\ge\ln4\), \[\mathbb P^\ast\big(\{\rho(\widehat\theta_\lambda,\theta_0)\ge c_e\varepsilon\}\cap S\big)\le6\sum_{l\ge0}e^{-C2^l}\le6\sum_{l\ge0}e^{-C(l+1)} = \frac{6e^{-C}}{1-e^{-C}}\le8e^{-C},\] using \(2^l\ge l+1\) and \(e^{-C}\le\frac14\). By \Cref{asm:generation_error_c} and the union bound, \[\mathbb P(S^c)\le\sum_{k\in\mathcal K}\exp\big(-c_g\pi_kn\,\tau_{k,\pi_kn}\big).\] The theorem follows from \(\mathbb P^\ast(A)\le\mathbb P^\ast(A\cap S)+\mathbb P(S^c)\). No step uses independence between \(Z^{(n)}\) and the synthetic draws beyond their conditional law given \(Z^{(n)}\). \(\blacksquare\)

\subsubsection{\texorpdfstring{Proofs for \Cref{sec:imbalanced}}{Proofs for }}

Here \(\mathcal Y=\{0,1\}\), \(\eta_F(x):=\mathbb P_F(Y=1\mid X=x)\), and the cross-entropy loss is \(\ell(q,(y,x)) := -y\log q(x)-(1-y)\log(1-q(x))\). The Q-target \(\theta_0\) of \Cref{thm:augmentation_risk} is distinct from \(\eta_F\) unless \(Q=F\).

\subsubsubsection{Proof of \Cref{thm:imbalanced_classification}}

\emph{Step 1 (binary indices).} Since \(\widetilde a_0+\widetilde a_1=1\) and \(q=(\frac12,\frac12)\), \(D_\lambda = 2|\widetilde a_1-\frac12|\). Since \(a_0+a_1 = \pi_0+\pi_1 = 1\), \(A_\lambda = M|a_1-\pi_1|\). By \Cref{asm:benchmark_rate_cls}, \(\delta_{m+n} = C_\theta\max(n^{-\alpha_{\mathrm{cls}}},m^{-\alpha_{\mathrm{cls}}}) = C_\theta\min(m,n)^{-\alpha_{\mathrm{cls}}}\). By \Cref{asm:generation_rate_form} and \(a_0+a_1=1\), \[G_\lambda = \frac mN\sum_ka_k\tau_{k,\pi_kn}\le\frac mN\Big(\sum_ka_kr_k(n)+\varepsilon_N\Big) = \bar G_\lambda .\]

\emph{Step 2 (a).} By Step 1, \(\varepsilon\ge\varepsilon^\ast_\lambda\ge\max\{\rho(\theta_Q,\theta_0),\delta_{m+n},D_\lambda,G_\lambda,A_\lambda\}\). Part (a) is \Cref{thm:augmentation_risk} with \(K=2\) and \(\varepsilon_\lambda := \varepsilon\).

\emph{Step 3 (b).} Since \((\pi_0n)^{-\gamma} = (\pi_1/\pi_0)^{\gamma}(\pi_1n)^{-\gamma}\), we have \(r_0(n)/r_1(n) = (\pi_1/\pi_0)^{\gamma}\Lambda_n\).

Since \(\log(\pi_0n)/\log(\pi_1n)\to1\), \(\Lambda_n\to1\), and if \(\beta_\gamma\le0\), then \(0<\log(\pi_0n)\le\log(\pi_1n)\) gives \(\Lambda_n\ge1\). For \(a_1>0\), \(a_0r_0\ge a_1r_1\) is equivalent to \((a_0/a_1)(\pi_1/\pi_0)^{\gamma}\Lambda_n\ge1\) (for \(a_1=0\) it holds trivially), and then \(a_0r_0+a_1r_1\le2a_0r_0\). If \(\beta_\gamma\le0\) and \(a_0\ge a_1\), all three factors are at least \(1\); this covers \(a^\circ\), since \(a^\circ_0-a^\circ_1=\frac nm(\pi_1-\pi_0)\ge0\), while for \(a=\pi\) the condition reads \((\pi_1/\pi_0)^{\gamma-1}\Lambda_n\ge1\).

\emph{Step 4 (c).} From \(\widetilde a_1 = \frac mNa_1+\frac nN\pi_1\) we get \(\widetilde a_1-\frac12 = \frac mN(a_1-\pi_1)+(\pi_1-\frac12)\). Since \(m/N\le1\), \[\pi_1-\tfrac12\le|\widetilde a_1-\tfrac12|+|a_1-\pi_1| = \frac{D_\lambda}2+\frac{A_\lambda}M\le\Big(\frac12+\frac1M\Big)\max\{D_\lambda,A_\lambda\}.\] The two special allocations are computed as follows.

\begin{itemize}
\tightlist
\item
  At \(a=\pi\): \(\widetilde a=\pi\), so \(D_\lambda = 2(\pi_1-\frac12)\) and \(A_\lambda=0\).
\item
  At \(a^\circ\): \(a^\circ_1 = \frac12-\frac n{2m}(\pi_1-\pi_0)\ge0\) if and only if \(m\ge n(\pi_1-\pi_0)\). Using \(\pi_1-\pi_0 = 2\pi_1-1\), we get \(\widetilde a_1 = \frac1N\big(\frac m2-\frac n2(2\pi_1-1)+n\pi_1\big) = \frac12\), so \(D_\lambda=0\). Also \(a^\circ_1-\pi_1 = -(\pi_1-\frac12)(1+\frac nm)\), so \(A_\lambda = M(\pi_1-\frac12)(1+\frac nm)\).
\end{itemize}

\emph{Step 5 (d).} Let \(\gamma\ge\alpha_{\mathrm{cls}}\). For \(n\ge1\) and \(k\in\{0,1\}\), \((\pi_kn)^{-\gamma}\le\pi_0^{-\gamma}n^{-\gamma}\le\pi_0^{-\gamma}n^{-\alpha_{\mathrm{cls}}}\) and \((\log(\pi_kn))^{\beta_\gamma}\le\ell_n\). Hence \(\sum_ka_kr_k(n)\le\tau_0\pi_0^{-\gamma}\ell_nn^{-\alpha_{\mathrm{cls}}}\), and \[\bar G_\lambda\le\frac mN\big(\tau_0\pi_0^{-\gamma}\ell_n+\varepsilon_Nn^{\alpha_{\mathrm{cls}}}\big)n^{-\alpha_{\mathrm{cls}}} = K_nC_\theta n^{-\alpha_{\mathrm{cls}}}\le K_n\delta_{m+n},\] since \(\delta_{m+n}\ge C_\theta n^{-\alpha_{\mathrm{cls}}}\). The bound on \(\varepsilon^\ast_\lambda\) follows by replacing \(\bar G_\lambda\) in its definition. \(\blacksquare\)

\section{Real-Data Analyses: Supplementary Details}\label{app:data}

This appendix gives details of the real-data analyses of \Cref{sec:applications}: the datasets (D.1), the generation-quality metrics (D.2) and comparisons (D.3), and further results for each dataset (D.4--D.6).

\subsection{Datasets}\label{app:datasets}

\subsubsection{Spanish Station Temperature}\label{app:data_temperature}

The \texttt{aemet\_temp} dataset of \texttt{goffda} \citep{goffda2023} contains the daily average temperatures of 73 Spanish weather stations for 1974--2013. The records were retrieved from the FTP server of the Spanish State Meteorological Agency (AEMET, Agencia Estatal de Meteorolog\'ia) in 2014 with a script of the \texttt{fda.usc} authors \citep{febrero2012statistical}, and \texttt{goffda} keeps 73 of 115 stations with consistent records. In leap years the values of 28 and 29 February are averaged, so every curve has 365 values. Curves are used on the original temperature scale. Station-years with any missing day are removed, without imputation.

The Madrid class contains seven stations (Retiro, Barajas, Cuatro Vientos, Getafe, Torrej\'on, Navacerrada and Colmenar Viejo), several of them in the surroundings of the city, and the Valladolid class contains two (Valladolid and Villanubla). The chronological split is approximately 70/15/15 and gives 232 training curves (178 Madrid, 54 Valladolid), 50 validation curves (38 and 12) and 50 test curves (39 and 11).

\subsubsection{Grape Hyperspectral Spectra}\label{app:data_grape}

Each spectrum is measured at 204 bands between 397.3 and 1003.5 nm with slightly uneven spacing (2.9--3.1 nm). Because the Fourier layers act on a uniform grid, the spectra are linearly interpolated onto 104 equally spaced wavelengths (spacing 5.9 nm), about half the raw resolution and a length divisible by \(2^{L-1}=8\), so no padding is needed (\Cref{app:architecture}). Reflectance lies between 0.03 and 0.54, and the sugar content of the training samples ranges from 101 to 283 g/L. The split is random, with seed 42, and stratified so that each of the nine strata (variety × tercile of sugar content within variety) is split 6:2:2. This gives 165 training samples (38 Fer, 51 Mauzac, 76 Syrah), 54 validation samples (12, 17, 25) and 55 test samples (13, 17, 25). The variety is the group of the value-augmented encoding of \Cref{sec:film_conditioning}.

\subsubsection{Gait Biomechanics}\label{app:data_gait}

The data were assembled by \citet{gertheiss2024functional} as a demonstration dataset for functional data analysis. They contain functional variables indexed by joint (ankle, knee, hip), axis (anterior--posterior, AP; medial--lateral; vertical, VT) and measure (angle, angular velocity, angular acceleration, joint moment), each time-normalized to 101 points over the gait cycle. In the slow, moderate and fast groups the three load conditions are recorded at one running speed. Each of our two problems has 75 training curves (25 participants under three conditions) and 18 test curves (6 participants); the data are too small to hold out validation participants. A curve holds the four channels (angle, angular velocity, angular acceleration and joint moment) of one participant under one load condition; the channels are generated jointly, without constraints linking velocity and acceleration to the derivatives of the angle. Scalar covariates such as age, height, weight and sex are available but are not used in any downstream model (\Cref{sec:downstream_models}).

\subsection{Generation-Quality Metrics}\label{app:metrics}

The three metrics of \Cref{sec:quality_metrics} are computed in R as follows.

\begin{itemize}
\tightlist
\item
  \textbf{MMD} is the maximum mean discrepancy \citep{gretton2012kernel} with an RBF kernel, computed with \texttt{kernlab} \citep{karatzoglou2004kernlab}. The bandwidth is set by the median heuristic. The statistic is the square root of the biased estimate of MMD\(^2\), so it is positive in finite samples even for identical laws.
\item
  \textbf{fWass} is the squared \(W_2\) distance between Gaussian approximations of two samples. With empirical means \(\hat\mu,\hat{\tilde\mu}\) and covariance matrices \(\hat\Sigma,\hat{\tilde\Sigma}\) of the discretized curves, \[
  \text{fWass}=\|\hat\mu-\hat{\tilde\mu}\|^2+\mathrm{tr}\Bigl(\hat\Sigma+\hat{\tilde\Sigma}-2\bigl(\hat\Sigma^{1/2}\hat{\tilde\Sigma}\hat\Sigma^{1/2}\bigr)^{1/2}\Bigr)
  \] \citep{gelbrich1990wasserstein}, where the covariance term is computed with \texttt{fdWasserstein} \citep{fdWasserstein}. The first term measures the shift of the mean function and the second is the Bures--Wasserstein distance between the covariances. Both are sums over grid points (and channels), so fWass depends on the grid and on the units of the curves; we compare it only within a dataset.
\item
  \textbf{fFID} applies the fWass formula, including the squared mean difference, to functional principal component scores from an FPCA fitted with \texttt{fdapace} \citep{fdapace} on the real reference curves \citep{heusel2017gans}.
\end{itemize}

\subsection{Comparison Scheme}\label{app:comparisons}

We compute the metrics under three comparisons, with the validation curves as reference sample (the test curves for gait).

\begin{enumerate}
\def\labelenumi{\arabic{enumi}.}
\tightlist
\item
  \textbf{Reference level.} Real training curves against the reference curves of the same class. This shows the discrepancy expected from sampling and, for the temperature data, from year-to-year variation alone. It is a reference, not a lower bound: synthetic curves can lie closer to a reference sample than another real sample does, for example when they are smoother than real curves.
\item
  \textbf{Within-class fidelity.} For each class \(k\), synthetic curves of class \(k\) against reference curves of class \(k\).
\item
  \textbf{Cross-class confusion.} Synthetic class \(k\) should lie closer to real class \(k\) than to real class \(\ell\ne k\); otherwise the generator mixes features across classes.
\end{enumerate}

For the grape data, where the generator is also conditioned on a continuous response, we add a visual check: real and synthetic spectra are stratified by tercile of sugar content within variety, and the mean and \(\pm1\) standard deviation envelopes are overlaid within each tertile. Misaligned envelopes would show that the generator has not captured the relation between spectrum and response, even when MMD and fWass look acceptable.

\subsection{Temperature: Further Results}\label{app:res_temperature}

\textbf{Generation quality.} \Cref{tab:appD_mv_gen} gives fFID and MMD for the comparisons of \Cref{tab:mv_gen}. They order the comparisons as fWass does, with one exception: by MMD, the Valladolid validation curves lie slightly closer to real than to synthetic Madrid curves (0.477 against 0.510). The cities stay distinguishable in the synthetic data: validation Madrid curves lie closer to synthetic Madrid (5879) than to synthetic Valladolid (9840), and validation Valladolid curves lie closer to synthetic Valladolid (7021) than to synthetic Madrid (7827).

\begin{table}[htbp]
\centering
\caption{\textbf{Madrid vs.~Valladolid: fFID and MMD against the validation years.} Layout as in \Cref{tab:mv_gen}. Lower is better.}
\label{tab:appD_mv_gen}
{\renewcommand{\baselinestretch}{1}\small
\begin{tabular}{lllllllll}
\toprule
 & \multicolumn{4}{l}{fFID} & \multicolumn{4}{l}{MMD} \\
 & syn c0 & train c0 & syn c1 & train c1 & syn c0 & train c0 & syn c1 & train c1 \\
\midrule
val c0 & 2258.2 & 1929.3 & 5521.1 & 4944.8 & 0.336 & 0.306 & 0.603 & 0.544 \\
val c1 & 4154.1 & 4218.0 & 2713.3 & 2408.2 & 0.510 & 0.477 & 0.447 & 0.365 \\
\bottomrule
\end{tabular}
}
\end{table}

\textbf{Checkpoint choice.} For the temperature data, generation-quality metrics were computed on the validation years at every checkpoint, and the checkpoint at 65,000 steps was used for all analyses. For the grape and gait data, the final checkpoint was used.

\textbf{Generational design.} \Cref{tab:mv_generational} gives the trajectories of \Cref{fig:cls_generational} at their end points and at the peak, \(G=4\).

\begin{table}[htbp]
\centering
\caption{\textbf{Madrid vs.~Valladolid: test accuracy and AUC at selected volumes of the generational design.} Mean \(\pm\) standard error over five replicates with disjoint synthetic data; the real-only classifier is deterministic. \(G\): number of pooled synthetic sets of 232 curves; \(G=1\) and \(G=20\) are the end points of \Cref{fig:cls_generational}, and \(G=4\) is the volume at which its trajectories peak.}
\label{tab:mv_generational}
{\renewcommand{\baselinestretch}{1}\small
\begin{tabular}{lllll}
\toprule
Deployment & \(G\) & Training curves & Accuracy & AUC \\
\midrule
Real only & --- & 232 & 0.680 & 0.636 \\
Augmentation & 1 & 464 & 0.696 \(\pm\) 0.022 & 0.647 \(\pm\) 0.024 \\
Replacement & 1 & 232 & 0.716 \(\pm\) 0.022 & 0.677 \(\pm\) 0.033 \\
Augmentation & 4 & 1160 & 0.772 \(\pm\) 0.008 & 0.756 \(\pm\) 0.015 \\
Replacement & 4 & 928 & 0.792 \(\pm\) 0.014 & 0.765 \(\pm\) 0.016 \\
Augmentation & 20 & 4872 & 0.756 \(\pm\) 0.012 & 0.713 \(\pm\) 0.009 \\
Replacement & 20 & 4640 & 0.756 \(\pm\) 0.012 & 0.714 \(\pm\) 0.009 \\
\bottomrule
\end{tabular}
}
\end{table}

\textbf{Allocation.} \Cref{tab:appD_alpha} gives the full AUC grid of \Cref{sec:targeted_allocation}. Between neighbouring shares, AUC decreases by at most 0.04 and increases by up to 0.06. Adding only Madrid curves (\(a_1=0\)) changes the AUC by at most 0.026. At small volumes the best share is \(a_1=0.9\) rather than 1, but the difference is within one standard error.

\begin{table}[htbp]
\centering
\caption{\textbf{Madrid vs.~Valladolid: test AUC over the full allocation grid.} Mean over 20 replicates, with the standard error \(\times10^3\) in parentheses. Rows: synthetic volume \(m/n\); columns: share \(a_1\) of synthetic curves allocated to Valladolid. Real-only AUC: 0.636.}
\label{tab:appD_alpha}
{\renewcommand{\baselinestretch}{1}\small\setlength{\tabcolsep}{3pt}
\begin{tabular}{m{21.7pt}m{30.0pt}m{30.0pt}m{33.9pt}m{33.9pt}m{33.9pt}m{33.9pt}m{33.9pt}m{33.9pt}m{33.9pt}m{30.0pt}m{33.9pt}}
\toprule
\(m/n\) & 0.0 & 0.1 & 0.2 & 0.3 & 0.4 & 0.5 & 0.6 & 0.7 & 0.8 & 0.9 & 1.0 \\
\midrule
0.5 & 0.627 (8) & 0.652 (9) & 0.647 (9) & 0.655 (9) & 0.715 (14) & 0.675 (15) & 0.694 (12) & 0.746 (11) & 0.727 (10) & 0.754 (9) & 0.738 (10) \\
1.0 & 0.644 (6) & 0.674 (9) & 0.674 (13) & 0.672 (9) & 0.700 (14) & 0.692 (10) & 0.721 (11) & 0.753 (9) & 0.750 (12) & 0.769 (8) & 0.762 (7) \\
1.5 & 0.650 (5) & 0.651 (8) & 0.665 (8) & 0.694 (13) & 0.709 (10) & 0.718 (12) & 0.716 (10) & 0.767 (9) & 0.762 (10) & 0.802 (6) & 0.818 (4) \\
2.0 & 0.646 (7) & 0.677 (8) & 0.686 (11) & 0.677 (7) & 0.713 (9) & 0.730 (10) & 0.738 (10) & 0.759 (10) & 0.788 (8) & 0.807 (4) & 0.810 (7) \\
2.5 & 0.652 (6) & 0.672 (8) & 0.685 (10) & 0.709 (11) & 0.692 (7) & 0.718 (9) & 0.730 (11) & 0.768 (9) & 0.803 (5) & 0.811 (5) & 0.824 (7) \\
3.0 & 0.662 (3) & 0.667 (6) & 0.685 (7) & 0.682 (7) & 0.707 (10) & 0.726 (10) & 0.747 (9) & 0.772 (8) & 0.793 (7) & 0.807 (4) & 0.815 (6) \\
\bottomrule
\end{tabular}
}
\end{table}

\subsection{Grape: Further Results}\label{app:res_grape}

\textbf{Generation quality.} \Cref{tab:grape_gen} compares synthetic and real training spectra with the validation spectra by fWass, on the reflectance scale and summed over the 104 wavelengths.

\begin{table}[htbp]
\centering
\caption{\textbf{Grape spectra: generation quality (fWass) against the validation set.} Rows: validation spectra of each variety; columns: synthetic (syn) or real training (train) spectra of each variety. fWass is multiplied by \(10^3\). Lower is better.}
\label{tab:grape_gen}
{\renewcommand{\baselinestretch}{1}\small
\begin{tabular}{lllllll}
\toprule
fWass & syn Fer & train Fer & syn Mauzac & train Mauzac & syn Syrah & train Syrah \\
\midrule
val Fer & 0.8 & 1.1 & 516.5 & 524.9 & 8.8 & 9.8 \\
val Mauzac & 525.8 & 524.7 & 4.1 & 6.4 & 479.5 & 469.1 \\
val Syrah & 7.1 & 8.0 & 470.8 & 479.8 & 7.8 & 2.8 \\
\bottomrule
\end{tabular}
}
\end{table}

Two features stand out. First, within each variety the synthetic spectra are about as close to the validation spectra as the real training spectra are: closer for Fer (0.8 against 1.1) and Mauzac (4.1 against 6.4), farther for Syrah (7.8 against 2.8). Second, each synthetic variety lies closest to the validation spectra of its own variety (columns of the table): synthetic Fer at 0.8 against at least 7.1, synthetic Mauzac at 4.1 against at least 470.8, and synthetic Syrah at 7.8 against at least 8.8.

\begin{figure}[htbp]
\centering
\includegraphics[width=0.95\textwidth]{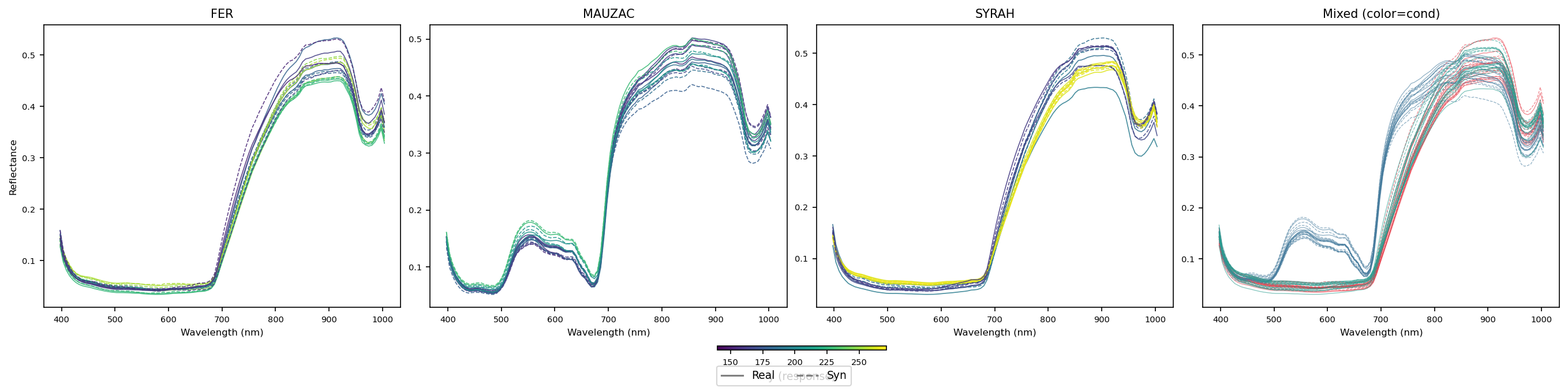}
\caption{\textbf{Grape spectra: real and synthetic reflectance.} Individual spectra for nine validation spectra per variety (solid), each paired with the synthetic spectrum of closest sugar content (dashed). First three panels: one variety each, colored by sugar content; right: all varieties together, colored by variety.}
\label{fig:reg_gen}
\end{figure}

\begin{figure}[htbp]
\centering
\includegraphics[width=0.95\textwidth]{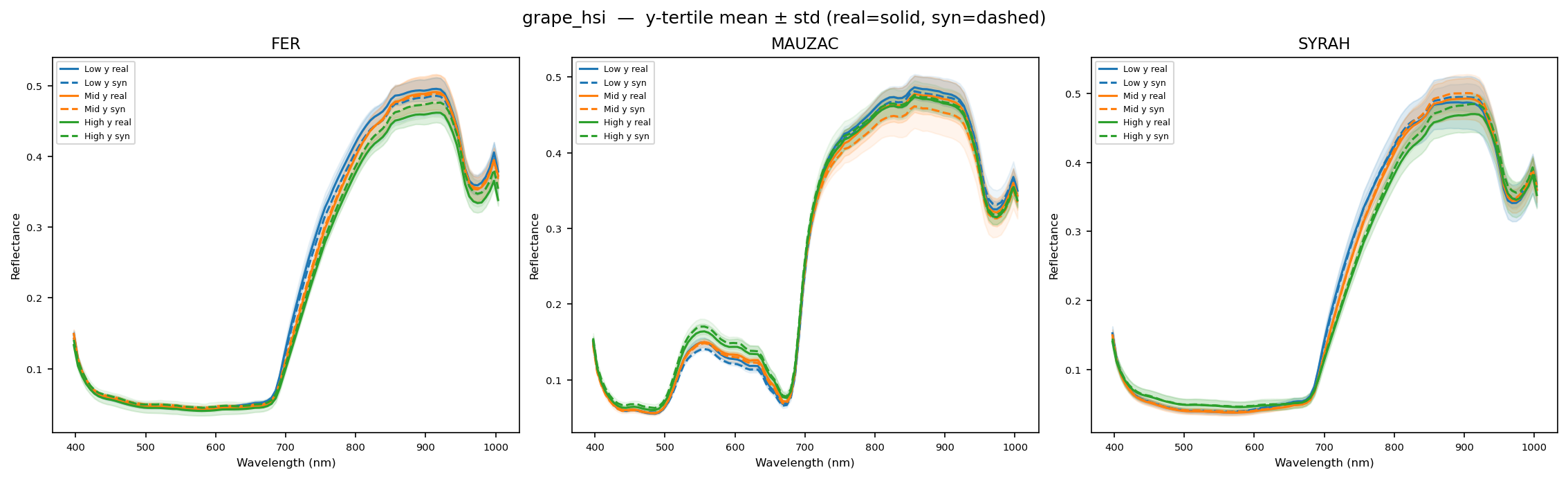}
\caption{\textbf{Grape spectra: conditional fidelity by sugar content.} Within each variety (panel), the low, middle and high terciles of sugar content are overlaid, real (solid) and synthetic (dashed).}
\label{fig:reg_tertile}
\end{figure}

\Cref{tab:appD_grape_gen} gives the fFID and MMD values within each variety. For Fer and Mauzac the synthetic spectra are closer to the validation spectra than the training spectra are by fFID, and by MMD for Mauzac; for Syrah they are farther by both. \Cref{fig:reg_gen} shows real and synthetic spectra, and \Cref{fig:reg_tertile} checks the relation between spectrum and sugar content, which these metrics ignore, by tercile of sugar content.

\begin{table}[htbp]
\centering
\caption{\textbf{Grape spectra: fFID and MMD against the validation set.} Synthetic (syn) or real training (train) spectra of a variety against validation spectra of the same variety. fFID is multiplied by \(10^3\). Lower is better.}
\label{tab:appD_grape_gen}
{\renewcommand{\baselinestretch}{1}\small
\begin{tabular}{lllll}
\toprule
Variety & fFID syn & fFID train & MMD syn & MMD train \\
\midrule
Fer & 0.5 & 0.8 & 0.162 & 0.137 \\
Mauzac & 3.6 & 6.1 & 0.245 & 0.267 \\
Syrah & 7.6 & 2.3 & 0.244 & 0.162 \\
\bottomrule
\end{tabular}
}
\end{table}

\textbf{Generational design.} \Cref{tab:grape_downstream} gives the end points of the trajectories in \Cref{fig:reg_generational} and the volume \(G=4\) at which the gain is essentially complete.

\begin{table}[htbp]
\centering
\caption{\textbf{Grape sugar content: test RMSE and \(R^2\) of penalized functional regression.} Mean \(\pm\) standard error over five replicates with disjoint synthetic data; the real-only fit is deterministic. \(G\): number of pooled synthetic sets of 165 spectra. M1: functional effect of the spectrum; M2: M1 plus a variety effect.}
\label{tab:grape_downstream}
{\renewcommand{\baselinestretch}{1}\small\setlength{\tabcolsep}{3pt}
\begin{tabular}{m{67.8pt}m{10.9pt}m{78.6pt}m{60.2pt}m{65.7pt}m{60.2pt}m{65.7pt}}
\toprule
Deployment & \(G\) & Training spectra & M1 RMSE & M1 \(R^2\) & M2 RMSE & M2 \(R^2\) \\
\midrule
Real only & --- & 165 & 21.24 & 0.658 & 17.85 & 0.759 \\
Augmentation & 1 & 330 & 18.92 \(\pm\) 0.34 & 0.729 \(\pm\) 0.010 & 15.62 \(\pm\) 0.18 & 0.815 \(\pm\) 0.004 \\
Replacement & 1 & 165 & 21.34 \(\pm\) 0.28 & 0.655 \(\pm\) 0.009 & 17.84 \(\pm\) 0.28 & 0.759 \(\pm\) 0.008 \\
Augmentation & 4 & 825 & 18.26 \(\pm\) 0.16 & 0.747 \(\pm\) 0.004 & 15.16 \(\pm\) 0.14 & 0.826 \(\pm\) 0.003 \\
Replacement & 4 & 660 & 18.34 \(\pm\) 0.19 & 0.745 \(\pm\) 0.005 & 15.11 \(\pm\) 0.18 & 0.827 \(\pm\) 0.004 \\
Augmentation & 20 & 3465 & 18.24 \(\pm\) 0.07 & 0.748 \(\pm\) 0.002 & 15.45 \(\pm\) 0.06 & 0.819 \(\pm\) 0.002 \\
Replacement & 20 & 3300 & 18.24 \(\pm\) 0.07 & 0.748 \(\pm\) 0.002 & 15.44 \(\pm\) 0.07 & 0.820 \(\pm\) 0.002 \\
\bottomrule
\end{tabular}
}
\end{table}

\subsection{Gait: Further Results}\label{app:res_gait}

\textbf{Generation quality.} Without a validation split, generation quality is described against the 18 test curves (6 per condition). \Cref{tab:ml_gen} compares synthetic and real training curves of each load condition with the test curves of each condition for ankle VT, and \Cref{tab:appD_ml_gen_ap} does so for ankle AP. The fWass values are summed over 101 grid points and four channels on the original scales, where the angular acceleration channel dominates; they are shown in units of \(10^8\).

\begin{table}[htbp]
\centering
\caption{\textbf{Ankle VT, slow: generation quality (fWass) against the test curves.} Rows: test curves of each load condition (0\%, 10\% and 20\% of body weight); columns: synthetic (syn) or real training (train) curves of each condition. fWass in units of \(10^8\). Lower is better.}
\label{tab:ml_gen}
{\renewcommand{\baselinestretch}{1}\small
\begin{tabular}{lllllll}
\toprule
Test & syn 0\% & train 0\% & syn 10\% & train 10\% & syn 20\% & train 20\% \\
\midrule
0\% & 9.4 & 11.6 & 12.7 & 10.7 & 11.1 & 11.4 \\
10\% & 12.3 & 14.7 & 12.7 & 12.4 & 12.5 & 13.8 \\
20\% & 11.2 & 13.1 & 13.6 & 11.6 & 12.0 & 12.7 \\
\bottomrule
\end{tabular}
}
\end{table}

\begin{table}[htbp]
\centering
\caption{\textbf{Ankle AP, moderate: generation quality (fWass) against the test curves.} Layout and units as in \Cref{tab:ml_gen}.}
\label{tab:appD_ml_gen_ap}
{\renewcommand{\baselinestretch}{1}\small
\begin{tabular}{lllllll}
\toprule
Test & syn 0\% & train 0\% & syn 10\% & train 10\% & syn 20\% & train 20\% \\
\midrule
0\% & 21.4 & 8.8 & 12.6 & 12.1 & 24.5 & 16.8 \\
10\% & 24.2 & 9.2 & 10.1 & 9.6 & 22.7 & 13.2 \\
20\% & 20.4 & 9.5 & 10.8 & 9.5 & 16.6 & 10.7 \\
\bottomrule
\end{tabular}
}
\end{table}

Within conditions, the ankle VT synthetic curves are as close to the test curves as the real training curves are (fWass 9.4--12.7 against 11.6--12.7); for ankle AP, the synthetic 0\% and 20\% curves are 2.4 and 1.6 times as far as the training curves. Across conditions, the load conditions are only weakly separated even in the real data (the 0\% test curves of ankle VT are closer to the 10\% than to the 0\% training curves, 10.7 against 11.6), and the synthetic curves reproduce this weak separation.

\Cref{tab:appD_ml_gen} gives the MMD values within each load condition for both gait problems. By MMD the synthetic curves are somewhat farther from the test curves than the training curves are in five of the six cases, including two of the three ankle VT conditions, where fWass finds them as close.

\begin{table}[htbp]
\centering
\caption{\textbf{Gait: MMD against the test curves.} Synthetic (syn) or real training (train) curves of a load condition against test curves of the same condition. Lower is better.}
\label{tab:appD_ml_gen}
{\renewcommand{\baselinestretch}{1}\small
\begin{tabular}{llll}
\toprule
Problem & Condition & MMD syn & MMD train \\
\midrule
Ankle VT, slow & 0\% & 0.552 & 0.493 \\
 & 10\% & 0.551 & 0.512 \\
 & 20\% & 0.422 & 0.492 \\
Ankle AP, moderate & 0\% & 0.484 & 0.376 \\
 & 10\% & 0.434 & 0.404 \\
 & 20\% & 0.401 & 0.381 \\
\bottomrule
\end{tabular}
}
\end{table}

\begin{figure}[htbp]
\centering
\includegraphics[width=0.95\textwidth]{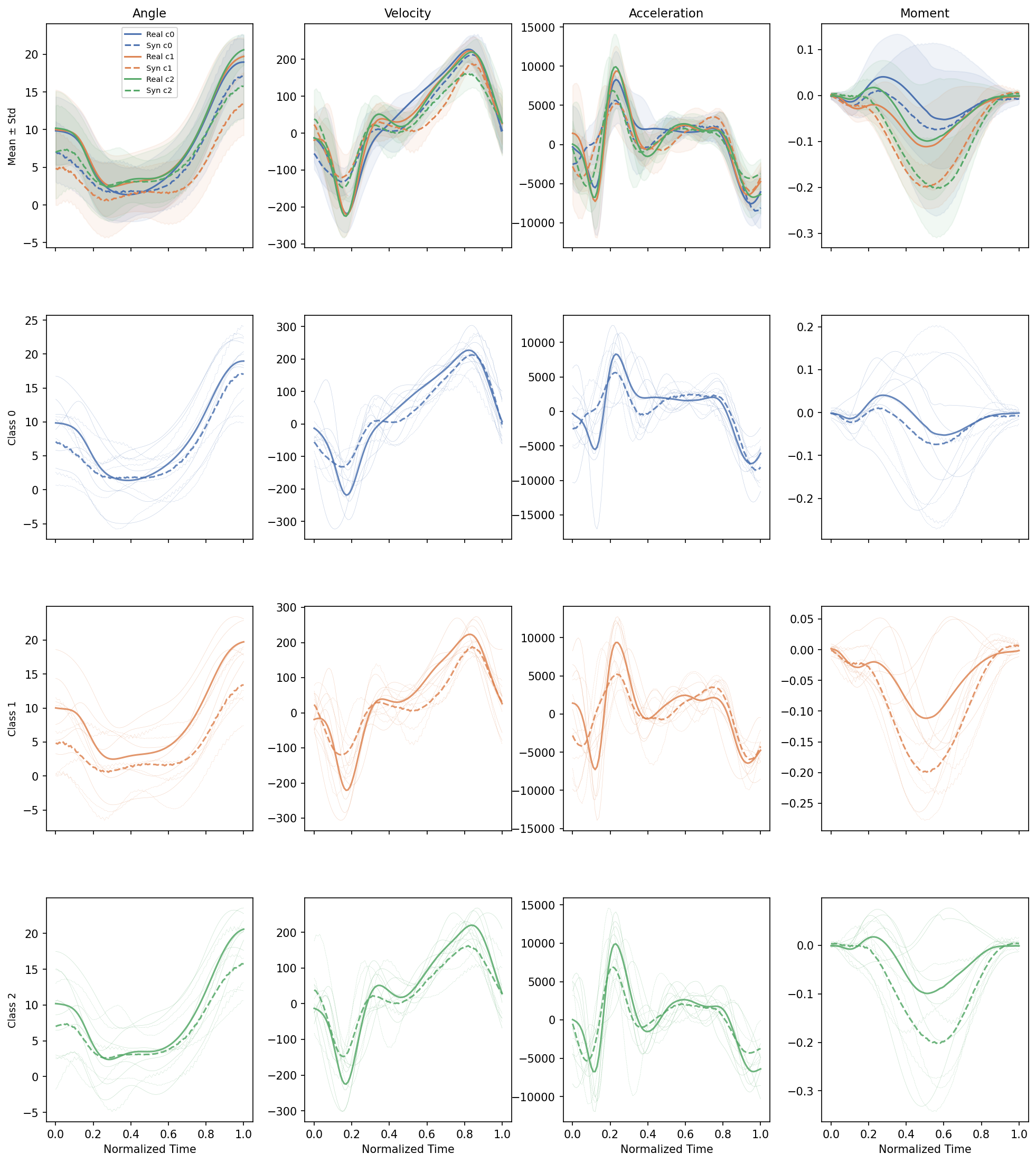}
\caption{\textbf{Ankle VT, slow: real and synthetic curves.} Columns: angle, angular velocity, angular acceleration and moment. Top row: condition-wise mean \(\pm1\) SD of the test curves of the six held-out participants (solid) and of synthetic curves (dashed); rows 2--4: individual curves and their means for the 0\%, 10\% and 20\% loads. Horizontal axis: percentage of the gait cycle.}
\label{fig:ml_gen_vt}
\end{figure}

\begin{figure}[htbp]
\centering
\includegraphics[width=0.95\textwidth]{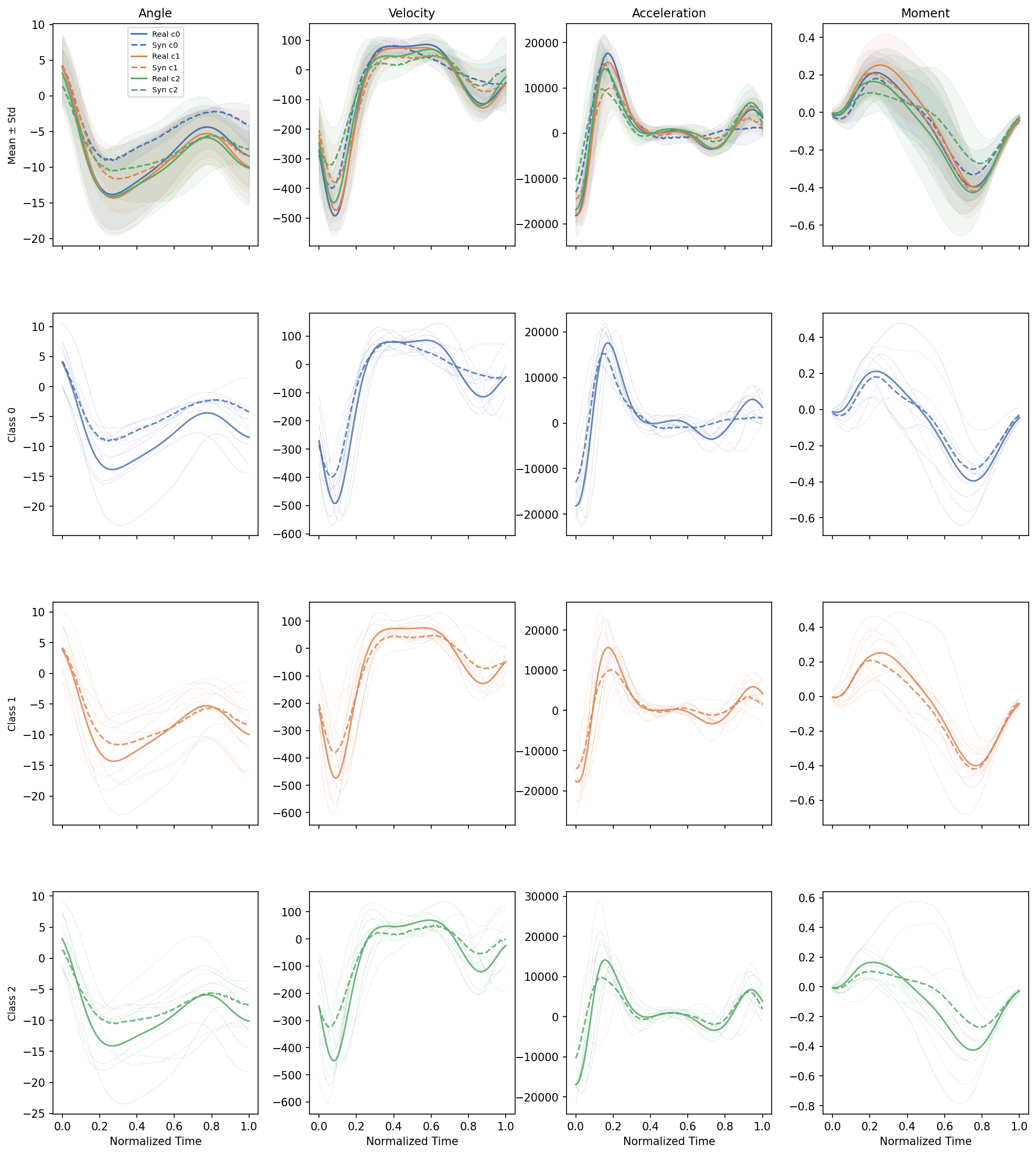}
\caption{\textbf{Ankle AP, moderate: real and synthetic curves.} Layout as in \Cref{fig:ml_gen_vt}.}
\label{fig:ml_gen_ap}
\end{figure}

\begin{figure}[htbp]
\centering
\includegraphics[width=0.95\textwidth]{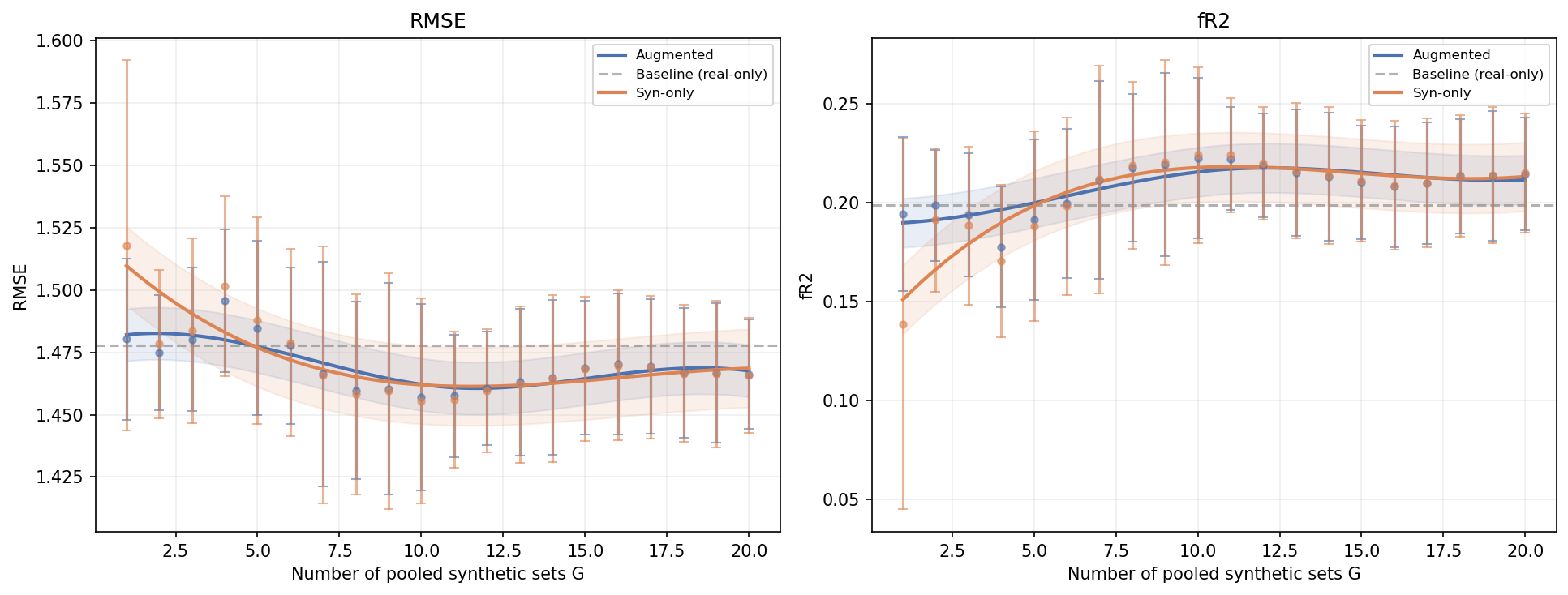}

\includegraphics[width=0.95\textwidth]{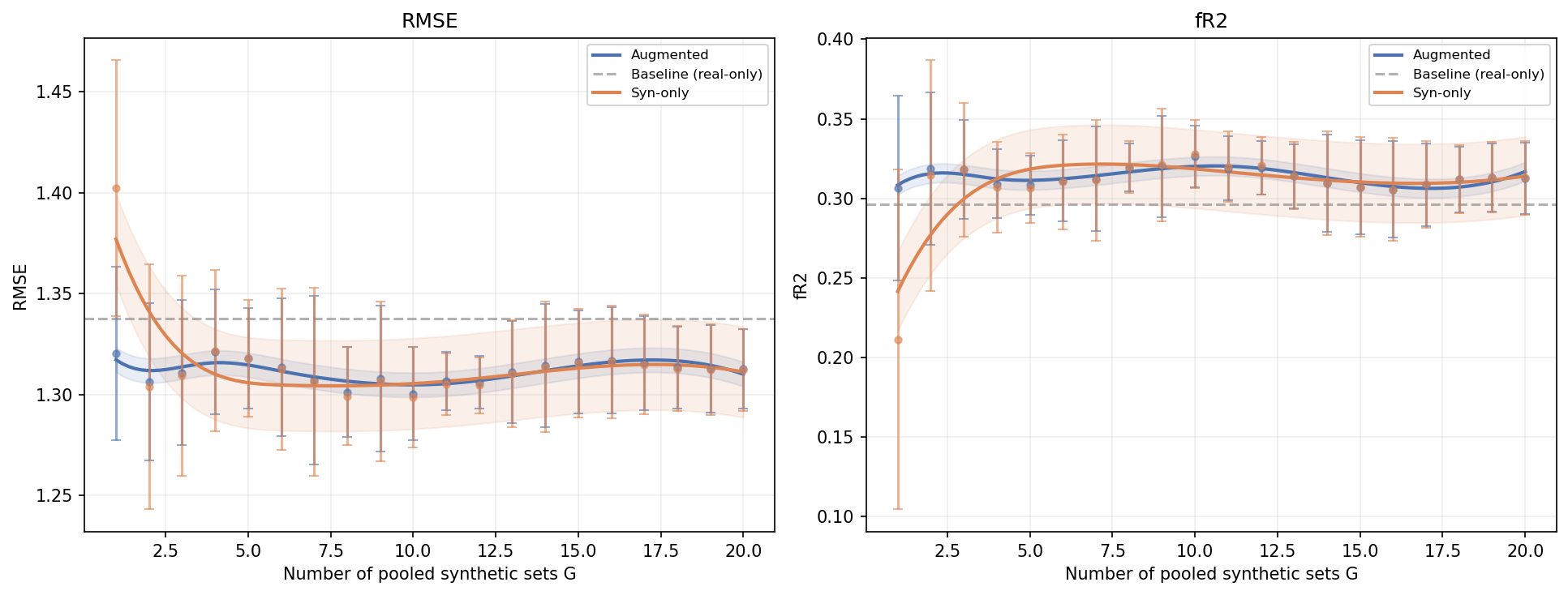}
\caption{\textbf{Ankle AP, moderate: performance against synthetic volume.} Layout as in \Cref{fig:ml_generational_vt}.}
\label{fig:ml_generational_ap}
\end{figure}

\Cref{tab:ml_downstream} gives, for each model and deployment, the real-only performance and the best replicate-mean performance over the 20 volumes of \Cref{fig:ml_generational_vt,fig:ml_generational_ap}.

\begin{table}[htbp]
\centering
\caption{\textbf{Gait moment prediction: test RMSE and functional \(R^2\).} Real only: the deterministic real-only fit. Augmentation and replacement: best value over \(G=1,\dots,20\) of the mean over five replicates, with the minimizing (RMSE) or maximizing (\(R^2\)) \(G\) in parentheses; \(m=75G\). Entries are the best value over the 20 volumes.}
\label{tab:ml_downstream}
{\renewcommand{\baselinestretch}{1}\small\setlength{\tabcolsep}{3pt}
\begin{tabular}{m{74.6pt}m{41.7pt}m{61.6pt}m{57.8pt}m{46.9pt}m{47.5pt}m{40.3pt}m{36.6pt}}
\toprule
Problem & Model & Real-only RMSE & Aug.~RMSE & Repl. RMSE & Real-only \(R^2\) & Aug.~\(R^2\) & Repl. \(R^2\) \\
\midrule
Ankle VT, slow & FDboost & 0.649 & 0.632 (4) & 0.630 (5) & 0.155 & 0.189 (3) & 0.188 (5) \\
 & PFFR & 0.666 & 0.624 (19) & 0.622 (19) & 0.109 & 0.207 (19) & 0.210 (19) \\
Ankle AP, moderate & FDboost & 1.478 & 1.457 (10) & 1.456 (10) & 0.199 & 0.223 (10) & 0.224 (10) \\
 & PFFR & 1.337 & 1.301 (10) & 1.299 (10) & 0.296 & 0.326 (10) & 0.328 (10) \\
\bottomrule
\end{tabular}
}
\end{table}

\end{document}